\documentclass[a4paper,11pt]{article}
\pdfoutput=1 

\usepackage{jheppub} 
            
\usepackage{bm,amssymb,slashed,graphicx,multirow,soul,mathtools,xspace,array}  
\usepackage{float}                   
\allowdisplaybreaks
\usepackage{ bbold }
\usepackage{subfigure}

\newcommand{\todo}[1]{{\color{red} \ifmmode\else[todo]\fi #1}}
\usepackage[usenames,dvipsnames]{xcolor}
     \definecolor{hgreen}{rgb}{0,.3,0}
      \definecolor{darkgreen}{rgb}{0.3,.8,0.2}
     \definecolor{hred}{rgb}{.3,0,0}
     \definecolor{hblue}{rgb}{0,0,.3}
     \definecolor{LightGray}{gray}{0.95}
     
\usepackage{hyperref}

\newcommand{\GeV}{\text{GeV}}
\newcommand{\MeV}{\text{MeV}}
\newcommand{\Br}{\text{Br}}

\newcommand{\re}[0]{\mathrm{Re}\,}

\newcommand{\beq}{\begin{equation} }
\newcommand{\eeq}{\end{equation}} 
\newcommand{\bi}{\begin{itemize} }
\newcommand{\ei}{\end{itemize} }

\definecolor{Red}{rgb}{1.,0.,0.}
\definecolor{Grn}{rgb}{0.,0.75,0.}
\definecolor{Blu}{rgb}{0.,0.,1.}

\newcommand{\FLt}{{\cal F}^{(2)}_L}
\newcommand{\Fpt}{{\cal F}^{(2)}_+}

\newcommand{\FL}{{\cal F}_L}
\newcommand{\Fp}{{\cal F}_+}

\newcommand{\mL}{m_{\phi_1}}
\newcommand{\mH}{m_{\phi_2}}

\DeclareMathOperator{\diag}{diag}   
\let\Re\relax
\DeclareMathOperator{\Re}{Re}
\let\Im\relax
\DeclareMathOperator{\Im}{Im}
\DeclareMathOperator{\Tr}{Tr}

\usepackage[T1]{fontenc} 

\usepackage{amsmath,amssymb,epsfig,ulem,color,slashed}
\allowdisplaybreaks  

\title{\boldmath Three Exceptions to the Grossman-Nir Bound}

\author[1]{Robert Ziegler,}
\author[2]{Jure Zupan,}
\author[3]{Roman Zwicky}

\affiliation[1]{Institute for Theoretical Particle Physics (TTP),
Karlsruhe Institute of Technology, Engesserstrasse 7, D-76128 Karlsruhe, Germany}
\affiliation[2]{Department of Physics, University of Cincinnati, Cincinnati, Ohio 45221,USA}
\affiliation[3]{Higgs Centre for Theoretical Physics, School of Physics and Astronomy, University of Edinburgh, Edinburgh EH9 3JZ, Scotland}

\emailAdd{robert.ziegler@kit.edu}
\emailAdd{zupanje@ucmail.uc.edu}
\emailAdd{roman.zwicky@ed.ac.uk}

\abstract{We show that the Grossman-Nir (GN) bound, $\Br(K_L\to \pi^0\nu \bar\nu)\leq 4.3 \, 
\Br(K^+\to \pi^+\nu \bar\nu)$, can be violated in the presence of light new physics with flavor violating couplings. We construct three sample models in which the GN bound can be violated by orders of magnitude, while 
satisfying all  other experimental bounds. In the three models the enhanced  branching ratio 
$\Br(K_L\to \pi^0+{\rm inv})$  is due to $K_L\to \pi^0\phi_1$, $K_L\to \pi^0\phi_1\phi_1$, 
$K_L\to \pi^0\psi_1\bar \psi_1$ transitions, respectively, where $\phi_1 (\psi_1)$ is a light scalar (fermion) that escapes the detector. In the three models $\Br(K^+\to \pi^++{\rm inv})$ remains very close to the SM value, while $\Br(K_L\to \pi^0+{\rm inv})$ can saturate the present KOTO bound.  Besides  invisible particles in the final state (which may account for dark matter) the models require 
additional light mediators around the GeV-scale.
}

\begin{document} 
\preprint{CP3-Origins-2020-06 DNRF90 \begin{flushright}  TTP20-019 \\ P3H-20-017 \end{flushright} }

\maketitle

\flushbottom

\section{Introduction}
\label{sec:intro}
In the SM, the $K_L\to \pi^0\nu \bar\nu$ and $K^+\to \pi^+\nu \bar\nu$ decays proceed  
through the same short distance operator, involving 
the fields  of the quark level transition ($s\to d \nu \bar\nu$).
The matrix elements for the $K_L\to \pi^0\nu \bar\nu$ and $K^+\to \pi^+\nu \bar\nu$ transitions are thus trivially related through isospin, leading to the Grossman-Nir (GN) bound \cite{Grossman:1997sk}
\beq
\Br(K_L\to \pi^0\nu \bar\nu)\leq 4.3 \, \Br(K^+\to \pi^+\nu \bar\nu).
\eeq
The bound remains valid in the presence of heavy New Physics (NP), i.e., for NP modification due to new particles with masses well above the kaon mass. The bound is saturated for the case of maximal 
CP violation, if lepton flavor violation can be neglected (see Ref.~\cite{Grossman:2003rw} for counter-examples).

\begin{figure}[t]
 \begin{center}
\includegraphics[width=7cm]{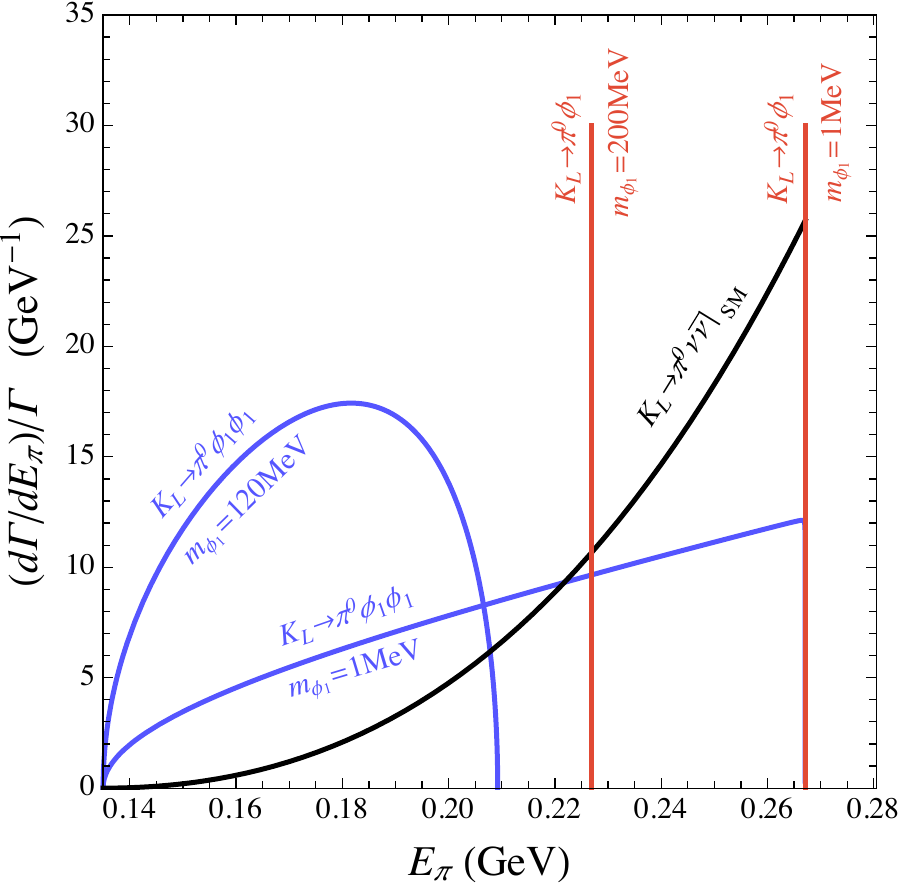}~~~~ 
\includegraphics[width=7.6cm]{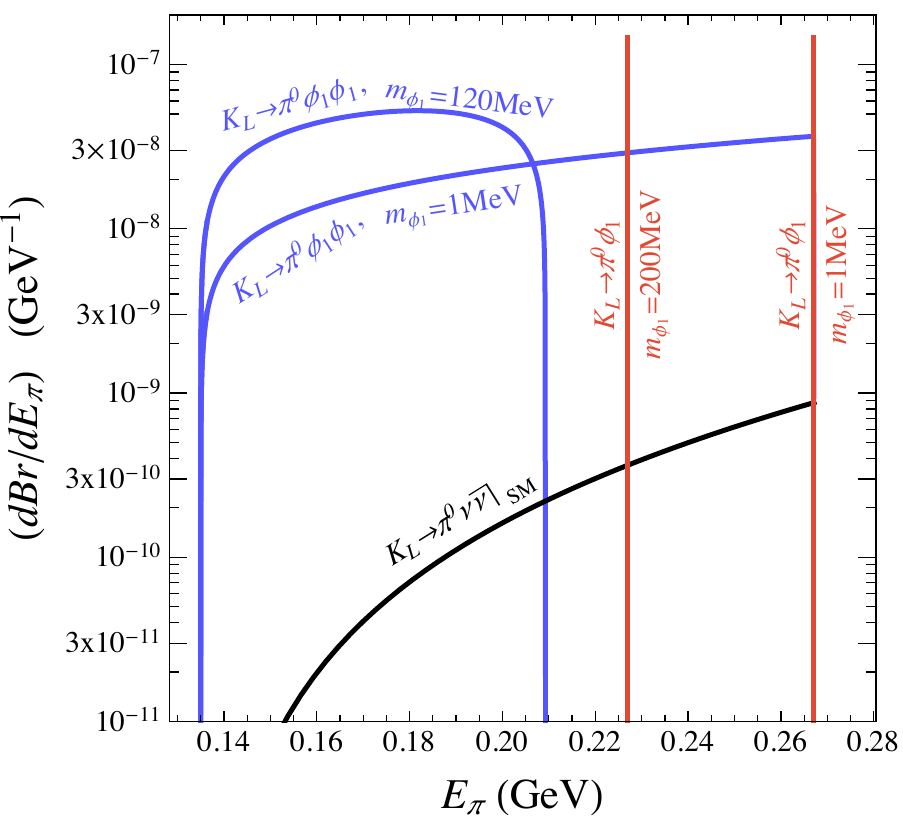}~~~~ 
\end{center}
 \caption{ \label{fig:dGamma} {\it Left:} The normalized decay width distributions as functions of the pion energy, $E_\pi$, for the SM (black line), for the decay dominated by the two body NP transition (Model 1), $K_L\to \pi^0\phi_1$, for two choices of invisible scalar masses, $m_{\phi_1}=1$ MeV, 200 MeV (red lines) and for three body NP decay (Model 2),  $K_L\to \pi^0\phi_1\phi_1$, with $m_{\phi_1}=1$ MeV, 120 MeV (blue lines). {\it Right:} The branching ratio distributions, where NP contributions saturate the present KOTO bound  \cite{Ahn:2018mvc}.
 At the kinematic endpoint, $E_\pi\to m_\pi$,  we have $d\Gamma_{\textrm{SM}} \propto p_\pi ^3 dE_\pi$ while for NP Model 2 $d\Gamma_{K_L\to \pi^0\phi_1\phi_1} \propto p_\pi d E_\pi$, where $p_\pi = (E_\pi^2 - m_\pi^2)^{1/2}$ (similarly for Model 3, Section \ref{sec:Model3}, $d\Gamma_{K_L\to \pi^0\psi_1\bar \psi_1}\sim p_\pi dE_\pi$ unless the Yukawa couplings $y_{ij}$ are purely real).
 This follows from the partial wave expansion, $d\Gamma/d E_\pi \sim p_\pi^{2l + 1}$, adapted to EFTs \cite{Hiller:2013cza}.
 In the $K \to \pi \nu\bar \nu$ rate the $V$-$A$ SM interaction   induces  a negligible  $S$-wave contribution proportional to the neutrino mass,  whereas the scalar interactions in our model induce a non-suppressed $S$-wave. The maximum recoil, $E_\pi \to E_\pi\big|_{\rm max}$, in contrast, is controlled by a single power of 
 the  $\nu$/$\phi_1$-velocity in the $q^\mu$ rest frame,
 $\beta_{\nu,\phi_1}  =  (1 -  4 m^2_{\nu/\phi_1}/q^2)^{1/2}$ (where $q^2 = m_K^2+ m_\pi^2 - 2 m_K E_\pi)$. 
 For small $m_\nu/\phi_1$ this velocity goes to $1$ for most values of $q^2$, leading to a sharp cut-off at $E_\pi\big|_{\rm max}$.}
 \end{figure}

In this paper we investigate to what extent NP contributions to $K\to \pi +{\rm inv}$ decays can violate the GN bound. Simple dimensional counting shows that for large violations of the GN bound the NP needs to be light, of order of a few GeV at most (see Section \ref{sec:EFT:analysis} and Refs.~\cite{He:2020jzn,Li:2019fhz}). Such light NP faces stringent experimental constraints from rare meson decays and collider/beam dump searches as well as from astrophysics and cosmology. Nevertheless, the couplings needed to modify the rare $K\to \pi+{\rm inv}$ decays are small enough that interesting modifications of the GN bound are indeed possible. We identify three sample models that achieve this through the following decays:
\begin{itemize}
\item Model 1: $K_L\to \pi^0\phi_1$, where the mass of the light scalar, $\phi_1$, can be anywhere from $m_{\phi_1}\lesssim m_K-m_\pi$ to a few MeV or even less,
\item Model 2: $K_L\to \pi^0\phi_1 \phi_1$, where the mass of the light scalar, $\phi_1$, is required in a large part of the parameter space to be  $m_{\phi_1}\gtrsim m_\pi/2$ in order to avoid constraints 
from  invisible pion decays, 
\item Model 3: $K_L\to \pi^0\psi_1 \bar \psi_1$, with $ \psi_1$ a light fermion whose mass is required to be  $m_{\psi_1}\gtrsim m_\pi/2$ in most of the phenomenologically viable parameter space. 
\end{itemize}
The $\phi_1$ and $\psi_1$ particles are feebly interacting and escape the detector, resulting in the $K_L\to \pi^0+{\rm inv}$ signature, as does the SM transition, $K_L\to \pi^0\nu \bar \nu$. The NP is thus detected through 
an enhanced $\Gamma(K_L\to \pi^0+{\rm inv})$ rate. Furthermore,  the three models can be distinguished from the SM and each other  by measuring the energy distribution of the neutral pion,  $d\Gamma(K_L\to \pi^0+{\rm inv})/dE_\pi$, see Fig.~\ref{fig:dGamma} for several sample distributions. While the two body decay in Model 1 results in a fixed pion energy, the three body decays in Model 2 and 3 can be close to the SM distribution for light $\phi_1$ and 
$\psi_1$ masses  and differ from it for non-negligible masses. 
Let us mention in passing that the lightness of the scalars 
could be due to them being a pseudo Goldstone boson of a broken global symmetry
whereas for fermions light masses are natural due to chiral symmetry.

In all three models the branching ratio $\Br(K^+\to \pi^++{\rm inv})$ remains close to the SM value, $\Br(K^+\to \pi^+\nu \bar \nu)_{\rm SM}=(8.4 \pm 1.0) \times 10^{-11}$ \cite{Buras:2005gr,Brod:2010hi,Buras:2015qea}, 
and thus below the preliminary NA62 bound  $\Br(K^+\to \pi^+\nu\bar \nu)_{\rm exp}<  1.85\times 10^{-10}$ \cite{NA62:2020}, 
while $\Br(K_L \to \pi^0 + {\rm inv})$ can be enhanced well above its SM value, $\Br(K_L \to \pi^0\nu \bar \nu)_{\rm SM}=(3.4 \pm 0.6) \times  10^{-11}$~ \cite{Buras:2005gr,Brod:2010hi,Buras:2015qea}. 
The NP induced $K_L \to \pi^0+ {\rm inv}$ transitions, on the other hand, can saturate the present experimental upper  bounds. The exact experimental bounds depend on the assumed NP decay channel. 
For instance, for the SM decay kinematics KOTO obtains $\Br(K_L \to \pi^0 \nu \bar \nu)_{\rm exp}<3.0 \times  10^{-9}$ 
\cite{Ahn:2018mvc}, while  for two body decays the bound is somewhat stronger, $\Br(K_L \to \pi^0\nu \bar \nu)_{\rm exp}<2.4 \times  10^{-9}$, for $m_{\phi_1}\lesssim m_\pi$ \cite{Ahn:2018mvc}. Recently, KOTO unblinded its 2016-18 data and found four events in the signal region, while only $0.05 \pm 0.02$ background events were expected \cite{KOTO:2019} (under additional scrutiny this has been revised to $0.34 \pm 0.08$ expected background events \cite{Nomura:2020}). If the preliminary data are interpreted as a signal, they correspond to a rate $\Br(K_L \to \pi^0+{\rm inv})_{\rm KOTO}=\big(2.1^{+2.0}_{-1.1}\big)\times 10^{-9}$ \cite{Kitahara:2019lws}. Note, that while some of the observed events may be due to yet unidentified backgrounds -- according to KOTO the four events do have some suspicious features -- they cannot be conclusively rejected \cite{Nomura:2020}. As a useful benchmark we will thus compare our results also with $\Br(K_L \to \pi^0+{\rm inv})_{\rm KOTO}$ as though these events are due to NP. Furthermore, in the numerics we quote the experimental bounds on three body decays, $K\to \pi \phi_1\phi_1$  $K\to \pi \psi_1\bar \psi_1$, assuming the experimental efficiencies are the same as for the SM $K \to \pi \nu \bar \nu$ transition. In reality, we expect the bounds to be weaker, since the experimental efficiencies are highest for larger values of $E_\pi$, while NP decays considered here are less peaked towards maximal $E_\pi$  (as compared to the SM). 

The three models considered in this work differ from the other proposed NP solutions to the KOTO anomaly in that they allow for large violations of the GN bound at the level of the amplitudes already. 
In contrast, Ref. \cite{Fabbrichesi:2019bmo} relies on the fact that the available phase space is larger for  neutral kaon decays due to $m_{K_L} -m_{\pi^0}>m_{K^+} -m_{\pi^+}$ and thus $K^+\to \pi^+ X_{\rm inv}$ decays can be forbidden by a finely tuned choice for the mass of the invisible final state $X_{\rm inv}$. Ref. \cite{Kitahara:2019lws} instead obtains, in one of the models, an apparent violation of the GN bound from the experimental set-up; the produced light NP particles decay on experimental length-scales, and are not observed in NA62 but are observed in KOTO due to the geometry of the experiments. Finally, the NP models of Refs. \cite{Kitahara:2019lws,Fuyuto:2014cya,Hou:2016den,Egana-Ugrinovic:2019wzj,Dev:2019hho,Jho:2020jsa,Liu:2020qgx,Cline:2020mdt,liao2020imprint} do not violate the GN bound, but can allow for a large signal in KOTO since NA62 is not sensitive to $X_{\rm inv}$ with a mass close to the pion mass. 

The paper is organized as follows. In section \ref{sec:EFT:analysis} a general 
Effective Field Theory  analysis is presented. 
The three models are discussed consecutively in Sections \ref{sec:Model1}, \ref{sec:Model2} and
\ref{sec:Model3} with the main plots collected in Figs.~\ref{fig:benchmark1}, \ref{fig:benchmark2},  in Figs.
\ref{fig:Model2:BM1}, \ref{fig:Model2:BM2}   and in Figs. \ref{fig:Model3:BM1}, \ref{fig:Model3:BM2}  for Model 1, 2 
and 3, respectively, with  constraints due to $K^0-\bar K^0$ mixing, cosmology and invisible pion decays 
discussed in the respective sections.  
The paper ends with conclusions in Section \ref{sec:conclusions}, while  details on  decay rates and integral conventions 
are deferred to two short appendices.

\section{The EFT analysis}
\label{sec:EFT:analysis}

We first perform an Effective Field Theory (EFT) based analysis, assuming that the SM is supplemented by a 
single light scalar, $\varphi$, while any other NP states are heavy and  integrated out. The light scalar has flavor violating couplings and is created in the $K^0\to \pi^0 \varphi$ decay.  The effective Lagrangian inducing this transition 
is given by
\beq
\label{eq:Leff}
{\cal L}_{\rm eff}=c^{(4)} \big(\bar s d\big)\varphi+\sum_i \frac{c_i^{(7)}}{\Lambda^3} \big(\bar s \Gamma_id\big)\big(\bar d \Gamma_i' d\big)\varphi+\cdots\;,
\eeq
where we only keep the parity-even operators of lowest dimension  and work in the quark mass basis. There is a single dimension 4 operator, and the sum runs over the  dimension 7 operators, where $\Gamma_i, \Gamma_i'$ include both Dirac and color structures.  In \eqref{eq:Leff}  solely  parity even operators, relevant  the $K\to \pi$ decay, are displayed. 
Moreover,  ${\cal O}^{(5)} = (\bar s \gamma_\mu d) \partial_\mu \varphi$ can be traded for the dimension 4 operator ${\cal O}^{(4)} = (m_s-m_d) (\bar s d) \varphi$, by use of the equations of motion (EOMs). 
Similarly, the EOM  $ \bar s   \overset{\leftrightarrow}{D}_\mu  d =  -   \partial^\nu (\bar s i \sigma_{\mu \nu}  d)   
   +i  (m_s + m_d) \bar d \gamma_\mu  d $ allows us to replace ${\cal O}^{(6)} = (\bar s \overset{\leftrightarrow}{D}_\mu d) \partial^\mu \varphi$  with the same operator $(m_s^2 - m_d^2) \bar s d \varphi$. 
This leaves  the dimension 4 and dimension 7 operators in \eqref{eq:Leff} as  operators of lowest dimension.

\begin{figure}[t!]
\begin{center}
\includegraphics[width=14cm]{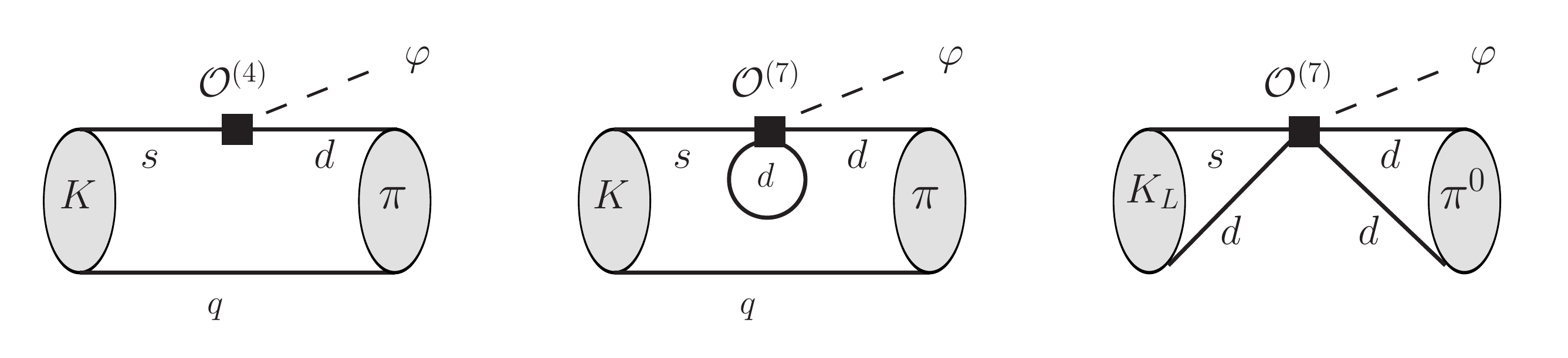}
  \end{center}
\caption{Contributions from dimension 4 (first diagram) and dimension 7 (2nd and 3rd diagrams) EFT operators to the $K\to \pi\varphi$ transition.  The last diagram contributes to $K_L \to \pi^0 \varphi$ only.  
The quark-loop diagram in the middle corresponds to the $S,E$ classes of diagrams  and the weak annihilation 
diagram 
on the right to the $W,C$ classes of diagrams in the lattice computation of Ref.~\cite{Christ:2016mmq}.
\label{fig:EFT}}
\end{figure}

At the quark level the dimension 4 operator induces the $s\to d \varphi$ transition and thus contributes equally to $K^0\to \pi^0 \varphi$ and $K^+\to \pi^+ \varphi$ decays, see the first diagram in  Fig. \ref{fig:EFT}. The resulting matrix elements for the $K_L\to \pi^0 \varphi$ and $K^+\to \pi^+ \varphi$ decays are
\beq
\{{\cal M}^{(4)}(K_L\to \pi^0 \varphi),{\cal M}^{(4)}(K^+\to \pi^+ \varphi)\}=  \frac{m_K^2 -m_\pi^2}{m_s-m_d} f_+(0) \{ \Im c^{(4)}, c^{(4)}\}.
\eeq
The $K_L\to \pi^0 \varphi$ decay is CP violating and vanishes in the limit of zero weak phases, $\Im c^{(4)}\to 0$.
These contributions therefore obey the Grossman-Nir relation,
\beq
|{\cal M}^{(4)}(K_L\to \pi^0 \varphi)| \leq |{\cal M}^{(4)}(K^+\to \pi^+ \varphi)|.
\eeq
 The dimension 7 operators, on the other hand, contribute to $K^0\to \pi^0 \varphi$ and $K^+\to \pi^+ \varphi$ decays in a qualitatively different way. The $K_L\to \pi^0\varphi$ decay can proceed through the weak annihilation type contractions of valence quarks, i.e., through 
  the third diagram in Fig. \ref{fig:EFT}. The $K^+\to \pi^+ \varphi$ transition requires the $\bar d d$ internal line to close in a loop (cf.  
  the 2nd diagram in Fig. \ref{fig:EFT}).  Such contractions also contribute to $K_L\to \pi^0\varphi$. Using at first  perturbative counting the latter contributions are suppressed, giving parametric estimates
 \beq
 \label{eq:M7}
\{{\cal M}^{(7)}(K_L\to \pi^0 \varphi), {\cal M}^{(7)}(K^+\to \pi^+ \varphi)\}\propto  \frac{m_K^3}{\Lambda^3} \Big\{ \Im c_i^{(7)}, \frac{1}{(4\pi)^2}\Big(\frac{\alpha_s}{4\pi}\big)^nc_i^{(7)}\Big\},
\eeq
 where we neglected $m_\pi$ compared to $m_K$ and do not write factors that are parametrically of the same size but may differ by ${\mathcal O}(1)$, such as different form factors in the two cases. 
 Depending on the Dirac-color structures $\Gamma^{(')}_i$ of the operator one or more gluon exchanges may  be required leading to additional  $(\alpha_s/4\pi)^n$-factors  shown in \eqref{eq:M7}. 
 
 A priori this leaves two classes of   NP models with potentially  sizeable violations of the GN bound. The first possibility is heavy NP, with a suppressed $c^{(4)}$ Wilson coefficient such that dimension 7 operators dominate. The other possibility is  light NP such that the EFT assumption, on which the above analysis is based on, is violated. 
 
Building viable heavy NP models that violate the GN bound faces several obstacles. First of all, $c^{(4)}$ would have to be heavily 
 suppressed, $c^{(4)}\ll m_q/\Lambda$, well below naive expectations. 
If this is not the case, the ``heavy'' NP scale needs to be quite light. For instance, for $c^{(4)}\sim  m_q/\Lambda$, 
 $c_i^{(7)}\sim {\mathcal O}(1)$ the dimension 4 operator contributions dominate over the dimension 7 
 ones already for $\Lambda\gtrsim {\mathcal O}(3{\rm~GeV})$ (see also the discussion in \cite{He:2020jzn}).  Furthermore, even if the hierarchy $c^{(4)}\ll c^{(7)}$ was realised, it is not clear whether the 
GN bound
could  be violated by more than a factor of a few. The scaling estimates in   \eqref{eq:M7} were based on perturbative expansion, while the kaon decays are  in the deep non-perturbative  regime of QCD. 
One can get an idea of the size of the ${\cal M}_{K}^{(7)} \propto  \langle \pi \varphi |  {\cal O}^{(7)} | K \rangle $
 matrix elements by linking them to the ones for $K \to \pi \ell^+ \ell^-$ decays that were explored  in lattice QCD 
 for light quark masses above their physical value ($m_\pi = 430~\MeV$ and $m_K = 625~\MeV$) 
 \cite{Christ:2016mmq}.  
 Figure 5 in Ref. \cite{Christ:2016mmq} indicates that the quark-loop  and weak annihilation contractions, 
   corresponding to the middle and the right diagrams in Fig.~\ref{fig:EFT}, lead to contributions of comparable size, contrary to the perturbative expectations in \eqref{eq:M7}.   
 If these results carry over to $K\to \pi \varphi$ decays, 
 it would seem that the ratio of 
${\cal M}^{(7)}(K_L\to \pi^0 \varphi)/{\cal M}^{(7)}(K^+\to \pi^+ \varphi)$ would not easily exceed a factor of $\sim 2$ in models of heavy NP.  It is unclear, however, whether this qualitative feature, based on the evaluation of the SM
$V-A$ four quark operators   
\cite{Christ:2016mmq}, would carry over to a model with scalar-scalar four quark operators, originating from a scalar mediator. 
For instance, for $V-A$ operators the weak annihilation topology is chirally suppressed  in the factorisation approximation, while this is not the case for scalar operators.

In conclusion, for heavy mediators the GN bound might or might not be violated in the case 
$c^{(4)} \ll c^{(7)}$. In this manuscript we therefore focus on the  second possibility, the possibility of light NP mediators, where
we can use Chiral Perturbation Theory (ChPT) with light NP states as a reliable tool to make predictions.

\section{Model 1 -  scalar model leading to two-body kaon decays}
\label{sec:Model1}
In the first example we introduce two real scalar fields, $\phi_1$ and $\phi_2$. The enhancement of the $K\to \pi$+inv branching ratio over the SM is due to the $K\to \pi \phi_1$ decay, while $K\to \pi \phi_2$ is kinematically forbidden,  i.e., we take $m_{\phi_2}>m_K-m_\pi $. The $\phi_1$ interacts feebly with matter and escapes the detector, resulting in a missing momentum signature\footnote{The $\phi_1$ could also  decay to neutrinos, $\phi_1\to \nu \bar\nu$, so that the final state can even be the same as in the SM, though with 
the $\nu \bar\nu$ pair forming a resonant peak. We do not explore this possibility any further.}. The relevant terms in the Lagrangian are
\beq
\label{eq:Lag:model1}
{\cal L}\supset g_{qq'}^{(i)} (\bar q_L q_R') \phi_i+{\rm h.c.} +\lambda m_S \phi_2^2 \phi_1 \;,
\eeq
where $q, q' =\{u,d,s\}$ and summation over repeated indices is implied. 
The couplings  $g_{qq'}^{(i)}$ are complex, and their imaginary parts trigger the 
$K_L \to \pi^0 \phi_1$ decay.

Large violations of the GN bound arise when there is a large hierarchy among the following couplings, 
\beq\label{eq:g:hierarchy}
g_{sd}^{(1)} \ll g_{sd}^{(2)}\ll g_{dd}^{(2)},
\eeq
while all  other couplings are further suppressed. In our benchmarks these remaining couplings 
as well as $g_{sd}^{(1)}$ will be set to zero. 
Before proceeding to predictions for branching ratios and the numerical analysis, it is instructive to perform a naive dimensional analysis (NDA). This will give us insight into why large violations of 
the GN bound are possible as well as to how large these violations can possibly be. 

\begin{figure}[t!]
\begin{center}
  \includegraphics[width=16cm]{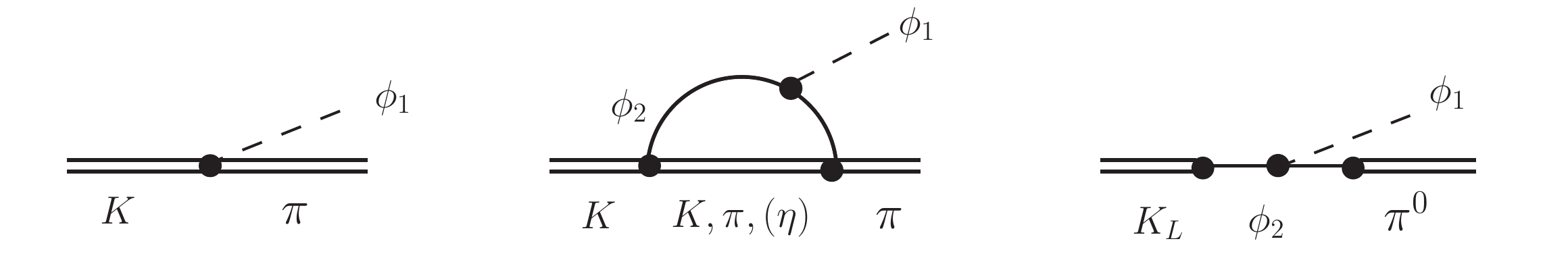} 
  \end{center}
\caption{\small Diagrams for the $K \to \pi \phi_1$ decay in Model 1 with the GN-violating contribution to the very right.  
These diagrams enter the matrix elements in Eqs.~\eqref{eq:K_Lpi0}, \eqref{eq:K+ampl}. Note that the $\eta$ in the loop contributes  to the $K_L$ decay only. 
Diagrams which we neglect, such as the diagrams of ${\mathcal O}(p^4)$ or ${\mathcal O}(g_{qq'}^3)$, are not shown. 
\label{fig:scalar1}}
\end{figure}

Taking $m_S\sim m_{\phi_2}\sim m_K$ the NDA estimate for the two decay amplitudes are, 
\begin{alignat}{2}
\label{eq:KL:NDA}
&{\cal M}(K_L\to \pi^0\phi_1)&\;\propto\;& \Im\ g_{sd}^{(1)} +{\mathcal O}(1)\times \lambda \Re  g_{sd}^{(2)} \Im g_{dd}^{(2)} \;,
\\
\label{eq:K+:NDA}
&{\cal M}(K^+\to \pi^+\phi_1)&\;\propto\;& g_{sd}^{(1)}+{\mathcal O}(1) \times \frac{1}{16\pi^2} \lambda \, g_{sd}^{(2)} g_{dd}^{(2)}\;,
\end{alignat}
where the first term in each line is due to the 1st diagram in Fig. \ref{fig:scalar1}. The second term in \eqref{eq:KL:NDA} is due to the 3rd diagram in Fig. \ref{fig:scalar1}, which is absent in the $K^+\to \pi^+\phi_1$ decay. This is the crucial difference between the two decays and  leads to large violations of the GN bound, 
provided $g_{sd}^{(1)}$  is small. 

However, violations of the GN bound cannot be arbitrarily large. Even if $g_{sd}^{(1)}$ is set to zero, the $K^+\to \pi^+ \phi_1$ transition is generated at the loop level from the 2nd diagram in Fig. \ref{fig:scalar1}, giving the 2nd term in \eqref{eq:K+:NDA}. Without fine-tuning the ratio ${\cal M}(K_L\to \pi^0\phi_1)/{\cal M}(K^+\to \pi^+\phi_1)$ is thus at best as large as the loop factor, $16\pi^2\sim 10^3$. Taking into account the present experimental results, this is more than enough to saturate the present KOTO bound while only marginally modifying the $K^+\to \pi^++$inv decay. 

\begin{figure}[t!]
\begin{center}
  \includegraphics[width=10cm]{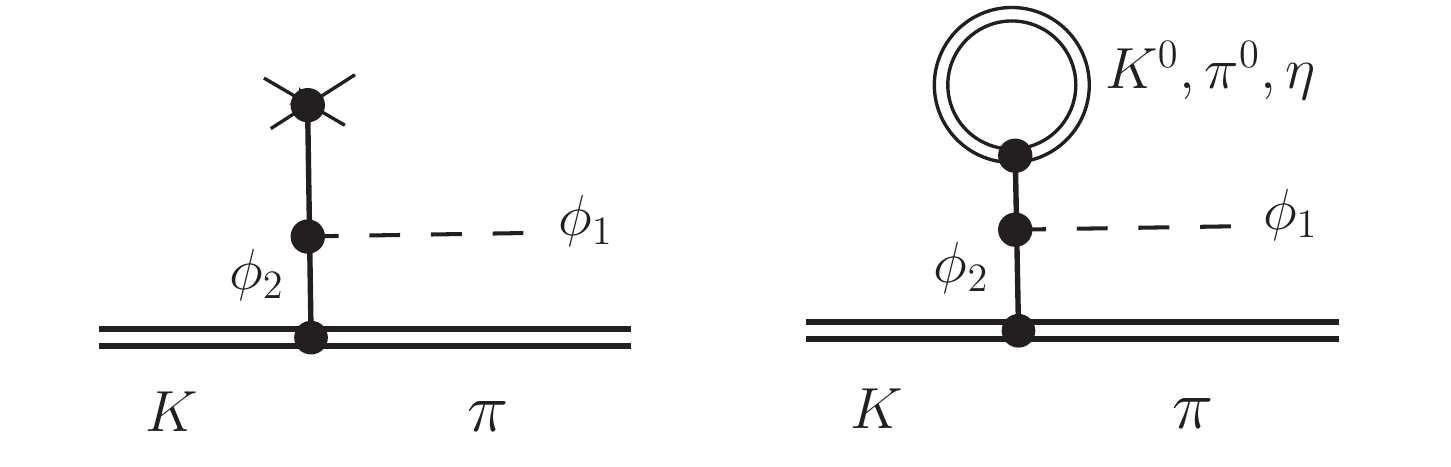} 
  \end{center}
\caption{\small The contributions to the $K\to \pi \phi_1$ decay in Model 1 proportional to $\langle \phi_2 \rangle $ (right), and the related one-loop tadpole diagram (left). 
\label{fig:scalar1vev}}
\end{figure}

In order to simplify the discussion we assume below that the vacuum expectation values (vevs) 
 of the scalar fields $\phi_{1,2}$ vanish, $\langle \phi_1\rangle= \langle \phi_2\rangle=0$. If this is not the case the $K\to \pi\phi_1$ decays receive additional GN-conserving contributions, see Fig. \ref{fig:scalar1vev} (right). More precisely, 
it is the renormalised vevs that are set to zero,  $\langle \phi_1\rangle_{\rm ren}=\langle \phi_2\rangle_{\rm ren}=0$, since we work to one loop order.  That is, we set the sum of the two diagrams in  Fig. \ref{fig:scalar1vev} to be zero. Had we set them instead to their natural value, $\langle \phi_i\rangle_{\rm ren} \sim m_K g_{dd}^{(i)}/16\pi^2$, our results would not change qualitatively. While ${\cal M}(K^+\to \pi^+\phi_1)$ would be modified by an ${\mathcal O}(1)$ factor, in ${\cal M}(K_L\to \pi^0\phi_1)$ such contributions  are always subleading and one would thus still have large violations of the GN bound. 

\subsection{Estimating the transition rates using ChPT}
\label{sec:ChPT}

We use  ChPT to calculate the transition rates. In constructing the ChPT we count $\phi_1\sim \phi_2 \sim {\mathcal O}(p)$.\footnote{That is, we count $m_{\phi_1}$ and $m_{\phi_2}$ both as ${\mathcal O}(p)\sim m_K-m_\pi$, even though $\phi_2$  is  $m_K-m_\pi$ by a factor of a few in large part of the parameter space that we consider.  Hence for heavy $\phi_2$ our ChPT based results should  be taken  as indicative only 
and could receive corrections of ${\mathcal O}(1)$. Since we only wish to demonstrate that large deviations of the GN bound are possible this suffices. However, should an anomalously large $K_L\to \pi^0$+inv rate be experimentally established our results should be revisited, say, for $m_{\phi_2}$ towards and above $1$ GeV.\label{foot:foot}
}
As far as QCD is concerned $\phi_{1,2}$ are external sources and can be treated as spurions ~\cite{Gasser:1984gg,Pich:1995bw} when 
building the low energy effective Lagrangian. 
The QCD Lagrangian, including \eqref{eq:Lag:model1}, can be conveniently rewritten as, 
\beq
\begin{split}
{\cal L}_{{\rm QCD}+\phi}= &\bar q (i \slashed \partial + g_s \slashed G^a T^a)q-\bar q {\cal M}_q q
- \sum_i \phi_i \,  \bar q ( \chi_S^{(i)}- i \chi_P^{(i)}  \gamma_5) q,
\end{split}
\eeq
where we keep only the light quarks,  $q=(u,d,s)$. The diagonal mass matrix is ${\cal M}_q=\diag(m_u,m_d,m_s)$,  while $\chi_{S,P}^{(i)}$ are $3\times3$ Hermitian matrices describing the quark couplings to $\phi_{1,2}$,
\beq
\big[\chi_S^{(i)}\big]_{qq'}=-\frac{1}{2}\big( g^{(i)}_{qq'}+g_{q'q}^{(i)*}\big), \qquad \big[\chi_P^{(i)}\big]_{qq'}=-\frac{i}{2}\big( g^{(i)}_{qq'}-g_{q'q}^{(i)*}\big) \;.
\eeq
Since we set the couplings to the up quark to zero they have the 
following form\footnote{For light $\phi_2$, which is our preferred scenario, 
assuming $ g_{uu}^{(2)} \neq 0$ would not introduce new qualitative features.  
According to chiral counting, $ g_{uu}^{(2)} \neq 0$
induces a $K^+ \pi^- \phi_1$-term at ${\cal O}(p^4)$, and is thus subleading to $K_L\pi^0\phi_1$-terms that we consider. 
Hence we set $g_{uu}^{(2)}$ to zero for simplicity rather than necessity.}
\beq
\label{eq:chiS}
\chi_{S}^{(i)}=- \begin{pmatrix}
0 & 0 & 0
\\
0 &  \Re g_{dd}^{(i)}& \bar g_{ds}^{(i)}
\\
0 &\bar g_{sd}^{(i)} &  \Re g_{ss}^{(i)}
\end{pmatrix},
\qquad
\chi_{P}^{(i)}= \begin{pmatrix}
0& 0 & 0
\\
0 &  \Im g_{dd}^{(i)}& \hat g_{ds}^{(i)}
\\
0 &\hat g_{sd}^{(i)}&  \Im g_{ss}^{(i)}
\end{pmatrix}.
\eeq
The off-diagonal couplings in \eqref{eq:chiS},
\beq
\bar g_{ds}^{(i)}= \bar g_{sd}^{(i)*}=\tfrac{1}{2}\big(g_{ds}^{(i)}+g_{sd}^{(i)*}\big), \qquad \hat g_{ds}^{(i)}=\hat g_{sd}^{(i)*}=-\tfrac{i}{2}\big(g_{ds}^{(i)}-g_{sd}^{(i)*}\big),
\eeq are the origin of the flavor violations. 
 
The Lagrangian for QCD  with the flavor violating $\phi_{1,2}$, ${\cal L}_{{\rm QCD}+\phi}$, is formally invariant under a global $SU(3)_R\times SU(3)_L$ transformation, $q_{R,L}\to g_{R,L} q_{R,L}$, provided $ \chi_{S,P}^{(i)}\phi_i$ and ${\cal M}_q$ are promoted to spurions transforming as 
\begin{align}
s+ip &\to g_R(s+ip) g_L^\dagger,
\end{align}
where $s$ and $p$ stand for
\beq\label{eq:s:p}
 s={\cal M}_q+\sum_i \chi_S^{(i)} \phi_i,\, \qquad p=\sum_i \chi_P^{(i)} \phi_i,
\eeq 
with $\chi_{S,P}^{(i)}$ given in \eqref{eq:chiS}.

The LO ChPT Lagrangian, with $\phi_{1,2}$ included as light degrees of freedom, is given by
\beq\label{eq:coupl:LO}
\begin{split}
{\cal L}_{{\rm ChPT}+\phi}^{(2)} &= \frac{f^2}{4} {\rm Tr}\big(\partial_\mu U \partial^\mu U^\dagger \big)+B_0 \frac{f^2}{2} \Tr\big[ (s-ip) U+ (s+ip) U^\dagger\big]
\\
&+\frac{1}{2}\partial_\mu \phi_i\partial^\mu \phi_i -\frac{m_{\phi_i}^2}{2} \phi_i^2 +\lambda m_S \phi_2^2 \phi_1+   \cdots \; ,
\end{split}
\eeq
where the ellipses stand for additional terms in the scalar potential.
Here $U(x)=\exp(i \lambda^a \pi^a/f)$ is the unitary matrix parametrizing the meson fields~\cite{Gasser:1984gg,Pich:1995bw},  
$B_0$  is a constant related to the quark condensate, $B_0(\mu=2{\rm~GeV})=2.666(57)$ GeV,  $f$ is related to the pion decay constant $f\simeq f_\pi/\sqrt{2}=92.2(1)$ MeV~\cite{Aoki:2019cca}, with normalization $\langle 0| \bar u \gamma_\mu \gamma_5 d(0) |\pi^-(p)\rangle=i p_\mu f_\pi$. 
The kaon decay constant $f_K=155.6 \pm 0.4$ MeV~\cite{Tanabashi:2018oca} accommodates SU(3) breaking at times.

In this paper we work to partial NLO order:  all  LO terms in the chiral expansion ${\mathcal O}(p^2)$ are kept, 
as well as the one loop corrections  which are of order ${\mathcal O}(p^4)$ and all finite.
The complete ${\mathcal O}(p^4)$-expressions for decay amplitudes involves  additional  contact terms (counter-terms or low energy constants),  parametrically of the same size as the one loop corrections.
However, since  $\phi_{1,2}$ are propagating degrees of freedom in our EFT the values of the low energy constants in ${\mathcal O}(p^4)$-ChPT are generally different from the ones in pure QCD and  therefore unknown. 
 The associated error in $K_L\to \pi^0\phi_1$ is small, since the NLO corrections are always subleading, while in $K^+ \to \pi^+ \phi_1$ they could give ${\mathcal O}(1)$ corrections but would not invalidate our conclusions. 
 For simplicity they are set  to zero throughout and we do not discuss them any further.

Next we calculate the $K\to \pi \phi_1$ decay amplitudes. Expanding in the meson fields the ${\mathcal O}(p^2)$ Lagrangian reads 
\beq
\label{eq:ChPT+phi}
\begin{split}
{\cal L}_{{\rm ChPT}+\phi}^{(2)} &\supset B_0 f \sum_i \phi_i  \Big(\sqrt{2} \hat g_{ds}^{(i)} \bar K^0+ \sqrt{2}\hat g_{sd}^{(i)} K^0- \Im g_{dd}^{(i)} \pi^0+ \tfrac{1}{\sqrt{3}}\Im\big( g_{dd}^{(i)} -2 g_{ss}^{(2)}\big)\eta \Big)
\\
&+B_0\sum_i \phi_i \Big\{\Re(g_{dd}^{(i)}+g_{ss}^{(i)}) K^0\bar K^0 +\Re(g_{dd}^{(i)})\big(\tfrac{1}{2}(\pi^{0})^2-\tfrac{1}{\sqrt{3}}\eta\pi^0\big)+
\\
&\qquad+\Re g_{ss}^{(i)}K^+K^-+\Re g_{dd}^{(i)}\pi^+\pi^- +
\\
&\qquad+\Big[\bar g_{sd}^{(i)}\big(-\tfrac{1}{\sqrt{2}}K^0\pi^0+K^+\pi^--\tfrac{1}{\sqrt6}K^0\eta\big)+{\rm h.c.}\Big]+\cdots\Big\}\;,
\end{split}
\eeq
where we only kept   terms relevant for the calculation of the $K\to \pi \phi_1$ transition, and the analysis of experimental bounds on the $\phi_1$-couplings.

The NP contributions to the decay amplitude for the $K_L\to \pi^0 \phi_1$ and $K^+\to \pi^+ \phi_1$ transitions are, see Fig. \ref{fig:scalar1},  
\begin{align}
\label{eq:K_Lpi0}
\begin{split}
{\cal M}(K_L\to \pi^0 \phi_1)_{\rm NP}&= \biggr\{ 2 \Im\hat g_{sd}^{(2)} \Im  g_{dd}^{(2)} 
 \Delta_{\phi_2}(m_K^2)\Delta_{\phi_2}(m_\pi^2) \lambda m_S B_0 f_K f_\pi   \\
&\qquad \quad  + \Im \bar g_{sd}^{(1)}-  \frac{\Im \bar g_{sd}^{(2)}}{8 \pi^2}  \lambda m_S B_0 \FLt(I) \Big\}B_0 
\;,
\end{split}
\\
\label{eq:K+ampl}
{\cal M}(K^+\to \pi^+ \phi_1)_{\rm NP}&= - \Big\{\bar g_{sd}^{(1)} -\frac{\bar g_{sd}^{(2)}}{8 \pi^2}\lambda m_S 
B_0 \Fpt(I) \Big\}B_0 \;,
\end{align}
where $\Delta_X(k^2) \equiv 1/(k^2-m_X^2)$ hereafter and 
\begin{align}
\label{eq:FL}
\FLt(Y) &= \Re g^{(2)}_{ss} Y(m_K) + \Re g^{(2)}_{dd}\big( Y(m_K) + Y(m_\pi) - \tfrac{1}{3}Y(m_\eta)\big) \;, 
\\
\label{eq:Fpl}
\Fpt(Y) &= \Re g^{(2)}_{ss} Y(m_K) + \Re g^{(2)}_{dd} Y(m_\pi)\;,
\end{align}
are structures occurring in all three models. They depend on the loop function 
 $I(m_X)=C_0(m_K^2,m_{\phi_1}^2,m_\pi^2,m_X^2,m_{\phi_2}^2,m_{\phi_2}^2)$, with $C_0$ the standard scalar three-point Passarino-Veltman function (cf. App.~\ref{app:PV}). In the $m_{\phi_2} \gg m_K, m_X$ limit we have $I(m_X) \to -1/m_{\phi_2}^2$. 
 Moreover, the replacement $f^2\to f_\pi f_K/2$  accounts for the main SU(3) breaking effects.

Note that the amplitude vanishes in the limit of no CP violation, $\Im \hat g_{sd}^{(i)}, \Im \bar g_{sd}^{(i)}\to 0$. The first term in \eqref{eq:K_Lpi0}, proportional to $\hat g_{sd}^{(2)}$,  is the ${\mathcal O}(p^2)$ contribution due to the tree level exchange of $\phi_2$, see the 3rd diagram in Fig.~\ref{fig:scalar1}. It is isospin violating since it gives rise to the $K_L \to \pi^0\phi_1$ transition but not to $K^+\to \pi^+\phi_1$. The first term in the second line of Eq. \eqref{eq:K_Lpi0} is the remaining ${\mathcal O}(p^2)$ contribution, due to the emission of $\phi_1$ directly from the meson line, see the 1st diagram in  Fig. \ref{fig:scalar1}. This contribution is isospin conserving -- it is present for both $K_L \to \pi^0\phi_1$ and $K^+\to \pi^+\phi_1$ transitions. It is proportional to $\bar g_{sd}^{(1)}$ and is thus small due to the assumed hierarchy among the couplings, Eq. \eqref{eq:g:hierarchy}.  

The hierarchy of couplings $|g_{sd,ds}^{(2)}| \gg |g_{sd,ds}^{(1)}|$ thus leads to maximal violation of the GN bound by NP contributions. However, this violation cannot be arbitrarily large. Even in the $\bar g_{sd}^{(1)}\to 0$ limit we still have isospin conserving NP contributions generated at one loop, see the 2nd diagram in Fig. \ref{fig:scalar1}, giving the last term 
 in  \eqref{eq:K_Lpi0}.  If $\phi_2$ is heavy and  integrated out these radiative corrections match onto the $\phi_1- K\pi$ vertex, which is then radiatively induced. 
 Moreover the $K_L\to \pi^0\phi_1$ and $K^+\to \pi^+\phi_1$ decays receive contributions from $\pi^0-\phi_1$ mixing where flavor violation comes from the SM $K\to \pi\pi$ transition. For our choices of parameters these contributions are always negligible.

The NP contributions  add coherently to the SM rate, 
\beq
\label{eq:coherent}
\Gamma(K_L\to \pi^0+{\rm inv})=\Gamma(K_L\to \pi^0\nu\bar \nu)_{\rm SM}+\Gamma(K_L\to \pi^0\phi_1)_{\rm NP},
\eeq
and the partial decay width due to NP is
\beq
\label{eq:belowKallen}
\Gamma(K_L\to \pi^0\phi_1)_{\rm NP}=\frac{1}{8\pi} \big|{\cal M}(K_L\to \pi^0 \phi_1)_{\rm NP}|^2\frac{p_\pi}{m_K^2},
\eeq
where $p_\pi = \lambda^{1/2}(m_K^2,m_\pi^2,m_{\phi_1}^2)/(2m_K)$ is the pion's momentum in the 
$K_L$ rest frame and $\lambda(x,y,z) = x^2+y^2+z^2 - 2 xy - 2x z - 2 yz$ the kinematic K\"all\'en function. 
The expressions  for the $K^+\to \pi^++{\rm inv}$ decay is completely analogous. Numerically, this gives (the SM predictions are taken from Refs.~\cite{Buras:2005gr,Brod:2010hi,Buras:2015qea,Mescia:2007kn})
\beq
\begin{split}\label{eq:BrKLpi0MET}
\Br(K_L\to \pi^0+{\rm inv})&=\underbrace{(3.4\pm0.6)\times 10^{-11}}_{{\rm SM}}
\\
&+\underbrace{6.0 \times 10^{-9} \, \biggr(\frac{\Im\hat g_{sd}^{(2)}}{5\cdot 10^{-9}}\biggr)^2
\biggr( \frac{\Im  g_{dd}^{(2)}}{10^{-3}}\biggr)^2 \biggr(\frac{\lambda m_S}{1{\rm~GeV}}\biggr)^2 \biggr( \frac{1{\rm~GeV}}{m_{\phi_2}}\biggr)^8}_{\rm NP},
\end{split}
\eeq
where we kept only the leading term for the NP contribution. The typical values of the inputs parameters for the NP contribution were chosen such that they reproduce roughly the KOTO anomaly (in fact slightly larger, but within $1\sigma$). Note the very high scaling in the $\phi_2$ mass, underscoring that $\phi_2$ needs to be relatively light in order to have large violations of the GN bound. For the charged kaon decay the numerical result is
\beq
\begin{split}
{\rm Br}(K^+\to \pi^++{\rm inv})=&\underbrace{(8.4\pm1.0)\times 10^{-11}}_{{\rm SM}} +\underbrace{5.0 \times 10^{-11} 
\biggr| \frac{\bar g_{sd}^{(1)}}{10^{-13}}\biggr|^2}_{\rm NP},
\end{split}
\eeq
where in the NP contribution we only kept the tree level term and set the value of $\bar g_{sd}^{(1)}$ to be similar to the one-loop threshold correction $\bar g_{sd}^{(1)}\sim \bar g_{sd}^{(2)} g_{dd}^{(2)}/8\pi^2$, cf.~Eq. \eqref{eq:K+ampl}, with the typical values of the later couplings as in \eqref{eq:BrKLpi0MET}. While the correction to $K^+\to \pi^++{\rm inv}$ is ${\mathcal O}(1)$ of the SM branching ratio, the correction to $K_L\to \pi^0+{\rm inv}$ can be orders of magnitude above the SM, giving large violations of the GN bound. Note that  NP in Model 1 contributes  to the 2-body decay $K^+ \to \pi^+ + X^0 $ only, and for massless $X^0$ is subject to the bound  $\Br(K^+\to \pi^+ + X^0) < 0.73 \times 10^{-10}$ from  E949~\cite{Adler:2008zza},  which is slightly stronger than the preliminary NA62 bounds on the 3-body decay $\Br(K^+\to \pi^++{\rm inv})_{\rm exp}<  2.44\times 10^{-10}$ and the 2-body decay $\Br(K^+\to \pi^+ + X^0)_{\rm exp}\lesssim  1.9\times 10^{-10}$ (for massless $X^0$)~\cite{NA62:2020}.

\subsection{Constraints on $\hat{g}_{ds}^{(i)}$ from $K^0-\bar K^0$ mixing}
\label{sec:model1:KKbar}
The $K^0-\bar K^0$ mixing is an important constraint on the model. The contributions to the meson mixing matrix  element are
\beq
\begin{split}
M_{12}=M_{12}^{\rm SM}+M_{12}^{\rm NP}=&- \frac{1}{2 m_K} \langle K^0|\mathcal{L}_{\rm eff}^{\rm SM} (0)|\bar K^0\rangle  - \frac{i}{4 m_K}\times 
\\
&\times 
\int d^4x \langle K^0|T{\mathcal L}_{{\rm ChPT}+\phi}^{(2)}(x), {\mathcal L}_{{\rm ChPT}+\phi}^{(2)}  (0)\}|\bar K^0\rangle+\cdots,
\end{split}
\eeq
where the  tree-level exchanges of $\phi_2$ is
\beq
\begin{split}
\label{eq:M12NP}
M_{12}^{\rm NP}=- \frac{(\hat g_{ds}^{(2)} B_0 f_K)^2}{2m_K(m_{\phi_2}^2-m_K^2)}+\cdots,
\end{split}
\eeq
with the ellipses denoting higher order terms (we also neglect the NP contributions to the absorptive  mixing amplitude since it only enters at one loop). 
The replacement $f \to f_K/\sqrt{2}$ accounts for the SU(3) breaking.

We consider two constraints, $\Delta m_K$ and $\epsilon_K$ which are CP conserving and CP violating respectively.  
Using the relation $\Delta m_K=2{\rm Re} M_{12}$ and conservatively assuming, due to the  relatively uncertain SM predictions of $\Delta m_K$, that the NP saturates the measured $\Delta m_K$, we obtain in the limit $m_{\phi_2}\gg m_K$, 
\beq\label{eq:KKbaraxion}
\begin{split}
\frac{\Delta m_K}{m_K} &\simeq 0.69 \Big| \Re\Big[  \big(\hat g_{ds}^{(2)}\big)^2  \Big]\Big| \biggr(\frac{1{\rm~GeV}}{m_{\phi_2}}\biggr)^2\;,
\end{split}
\eeq
and with the experimental value $\Delta m_K^{\rm expt.}=3.484(6)\times10^{-12}$ MeV~\cite{Tanabashi:2018oca}, this translates to  
\beq\label{eq:gds:DeltamK}
\sqrt{\Big| \Re\Big[  \big(\hat g_{ds}^{(2)}\big)^2  \Big]\Big|}<  1.0 \cdot 10^{-7} \times \biggr(\frac{m_{\phi_2}}{1{\rm~GeV}}\biggr)\;.
\eeq
To obtain the bounds on non-SM CP violating contributions to $K^0-\bar K^0$  mixing we use the normalized quantity
\beq
C_{\varepsilon_K} = \frac{|\epsilon_K^{{\rm SM}+a}|}{|\epsilon_K^{\rm SM}|} \;.
\eeq
 For the theoretical prediction of $\epsilon_K$ we use the expression \cite{Buras:2010pza}
\beq
\epsilon_K=e^{i\phi_\epsilon}\sin\phi_\epsilon\biggr(\frac{\Im M_{12}}{\Delta m_K}+\xi\biggr)\;,
\eeq
where
\beq
\xi\simeq\frac{\Im \Gamma_{12}}{\Delta \Gamma_K}\;.
\eeq
We take the values for $\Delta m_K=m_L-m_S$, $\Delta \Gamma_K=\Gamma_S-\Gamma_L$, and $\phi_\epsilon=\arctan(2\Delta m_K/\Delta \Gamma_K)$ from experiment~\cite{Tanabashi:2018oca}. 
With the SM prediction for $|\epsilon_K|$ from \cite{Brod:2019rzc}, and the NP contribution to $M_{12}$, $\Gamma_{12}$ from Eq.~\eqref{eq:M12NP} we get 
\beq\label{eq:axion:CeK}
\begin{split}
\delta C_{\epsilon_K}=C_{\epsilon_K}-1&=-5.8 \times 10^{16} \,\Im\big[(\hat g_{ds}^{(2)})^2\big] \biggr(\frac{1{\rm~GeV}}{m_{\phi_2}}\biggr)^2,
\end{split}
\eeq
The global CKM fit by the UTFit collaboration results in $0.87< C_{\epsilon_K}<1.39$  at 95\% CL \cite{Bona:2007vi,UTfit}, which translates to the following $1\sigma$ bounds
\beq
\label{eq:gds:epsilonK}
-(2.6 \times 10^{-9})^2 \biggr(\frac{m_{\phi_2}}{1{\rm~GeV}}\biggr)^2< \Im\big[(\hat g_{ds}^{(2)})^2\big]< (1.5\times 10^{-9})^2\biggr(\frac{m_{\phi_2}}{1{\rm~GeV}}\biggr)^2.
\eeq
These bounds will improve in the future, once the improved prediction for $\epsilon_K$~\cite{Brod:2019rzc} is implemented in the global CKM fits. 

\subsection{Constraints from $\epsilon'/\epsilon$}
\label{sec:epspeps}

The tree level exchanges of $\phi_2$ contribute to $K\to \pi\pi$ decays. These contributions can be CP violating and can thus contribute to $\epsilon'/\epsilon$. In general, the matrix elements 
 can be decomposed into isospin amplitudes $A_{I}$ of the final state pions $|(\pi \pi)_I \rangle$. 
 The latter read, 
 with appropriate Clebsch-Gordan coefficients for our 
 chiral Lagrangian  \cite{DAmbrosio:1996lam},
\begin{alignat}{2} 
\label{eq:Miso}
& {\cal M}(K^0 \to \pi^+ \pi^-)  &\;=\;&  A_0 + \frac{1}{\sqrt{2}} A_2 \;, \nonumber \\[0.1cm]
&  {\cal M}(K^0 \to \pi^0 \pi^0)  &\;=\;&  A_0 - {\sqrt{2}} A_2 \;, \nonumber \\[0.1cm]
& {\cal M}(K^+ \to \pi^+ \pi^0)  &\;=\;&  \phantom{A_0 +\;\;\;\;} \frac{3}{2}A_2 \;.
 \end{alignat}
In terms of these amplitudes the real part of ${\epsilon'}/{\epsilon}$ assumes the form
\beq
\Re\left( \frac{\epsilon'}{\epsilon} \right)=-\frac{\omega}{\sqrt{2}|\epsilon_K|}\biggr[\frac{\Im A_0}{\Re A_0}-\frac{\Im A_2}{\Re A_2}\biggr]\;,
\eeq
where $\omega \equiv \Re A_2 /\Re A_0$.
In our model, the isospin amplitudes are easily obtained from \eqref{eq:Miso} through an emission and $s$-channel tree level diagram 
\begin{align}
A_2&=-\frac{2}{3}B_0^2 f \bar g_{sd}^{(2)}\Im g_{dd}^{(2)}\Delta_{\phi_2}(m_\pi^2),
\\
A_0&=\sqrt2 B_0^2 f  \hat g_{sd}^{(2)}\Re g_{dd}^{(2)}\Delta_{\phi_2}(m_K^2) -\frac{1}{\sqrt2}A_2 \;.
\end{align}
Using the measured values, $\Re A_0=27.04(1)\times 10^{-8}$~GeV,  $\Re A_2=1.210(2)\times 10^{-8}$~GeV \cite{Cirigliano:2011ny}, $\omega^{-1}=22.2(1)$ \cite{DAmbrosio:1996lam}, $|\epsilon_K|=(2.10^{+0.27}_{-0.20}) \cdot 10^{-3}$ \cite{Charles:2004jd}, our model then affects the imaginary parts of the isospin amplitudes
and leads to the following shift
\beq\label{eq:epsilon'}
\Re\left( \frac{\epsilon'}{\epsilon} \right)_{\rm BSM}=2\times 10^{-3}\biggr[0.04 \biggr( \frac{\Im \hat g_{sd}^{(2)}}{10^{-9}}\biggr) \biggr(\frac{\Re g_{dd}^{(2)}}{10^{-3}}\biggr)+0.32 \biggr( \frac{\Im \bar g_{sd}^{(2)}}{10^{-9}}\biggr) \biggr(\frac{\Im g_{dd}^{(2)}}{10^{-3}}\biggr)\biggr] \;,
\eeq
with $m_{\phi_2} = 1\; \GeV$ for reference.
For brevity  we used  the central values of the inputs above. This is to be compared with the experimental value $\Re(\epsilon'/\epsilon)_{\rm exp}=(16.6\pm2.3)\times 10^{-4}$ \cite{Tanabashi:2018oca} and the SM prediction from lattice QCD, $\Re(\epsilon'/\epsilon)_{\rm RBC-UKQCD}=(21.7\pm 8.4)\times 10^{-4}$ \cite{Abbott:2020hxn}, which gives the 95 \% C.L. for the positive BSM contributions to be 
$\Re(\epsilon'/\epsilon)_{\rm BSM}<2.2 \times 10^{-3}$ (alternative treatments of lattice QCD inputs as well as isospin breaking effects can lead to somewhat stronger bounds $\Re(\epsilon'/\epsilon)_{\rm BSM}<1.3 (7) \times 10^{-3}$ based on octet (nonet) schemes \cite{Aebischer:2020jto}).


\subsection{Constraints on representative benchmarks}
\label{sec:con-Model1}

To highlight the typical values of couplings that can lead to sizable correction in $K\to \pi+$ inv decays, while passing all other constraints, we form a benchmark 1 (BM1) and a benchmark 2 (BM2),  
\begin{alignat}{3}
\label{eq:BM1}
&{\rm\bf BM~1:} \qquad  g_{dd}^{(2)}&\;=\;&\tfrac{(1+i)}{\sqrt2} g_{dd}\;, \quad  & & \bar g_{sd}^{(2)}\;=\; 
\hat g_{sd}^{(2)}=\tfrac{(1+i)}{\sqrt2} g_{sd} \;,
\\
\label{eq:BM2}
&{\rm\bf BM~2:} \qquad g_{dd}^{(2)} &\;=\;& i g_{dd} \;, \quad  &  & \bar g_{sd}^{(2)}\;=\;0\;, \quad  \hat g_{sd}^{(2)}=i g_{sd}\;.
\end{alignat}
These depend on two real parameters, $g_{dd}$ and $g_{sd}$, parametrizing couplings of $\phi_2$ to quarks. All the remaining couplings of $\phi_2$ to quarks as well as all the direct couplings of $\phi_1$ to quarks are set to zero  in accordance with previous discussions. The triple scalar coupling is fixed to 
$\lambda_S m_S=1{\rm~GeV}$ (and other potentially relevant scalar couplings assumed to be small, see Section \ref{sec:inv:pions}). The mass of $\phi_1$ is taken to be small, $m_{\phi_1}=1$ MeV, while $m_{\phi_2}$ is kept as a free parameter that is varied in the range $m_{\phi_2}\in[0.4,1.5]$ GeV, cf. footnote \ref{foot:foot}.

\begin{figure}[t]
\begin{center}
\includegraphics[width=7cm]{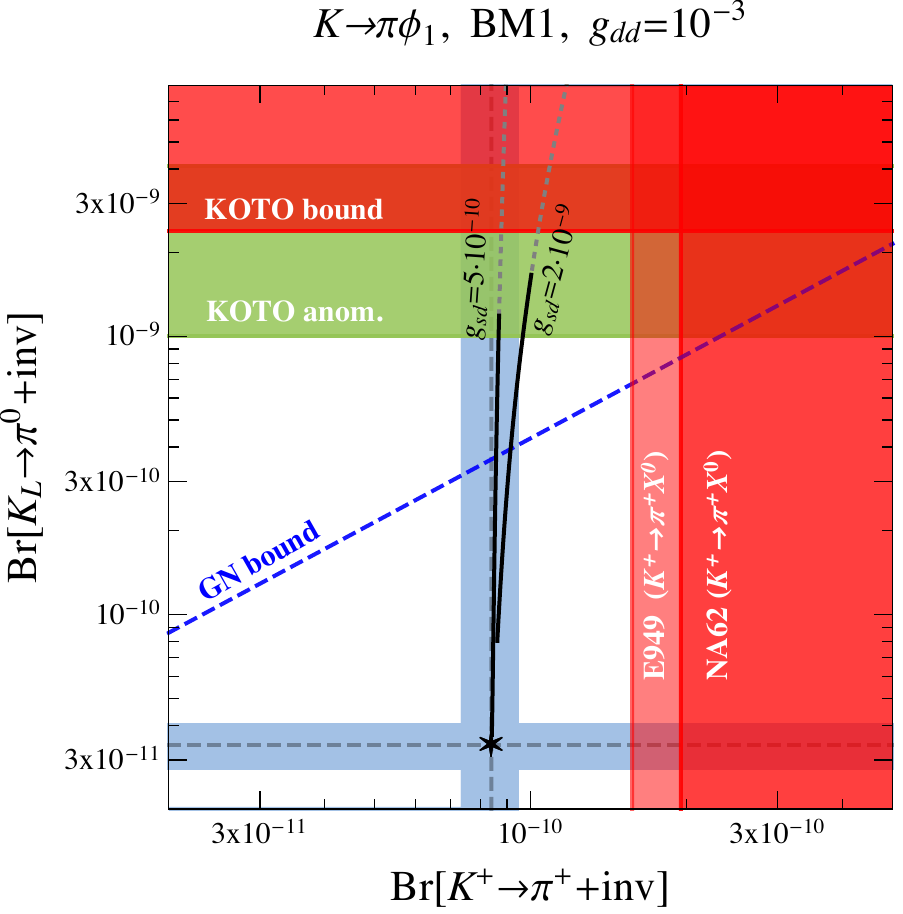}~~~~
\includegraphics[width=7cm]{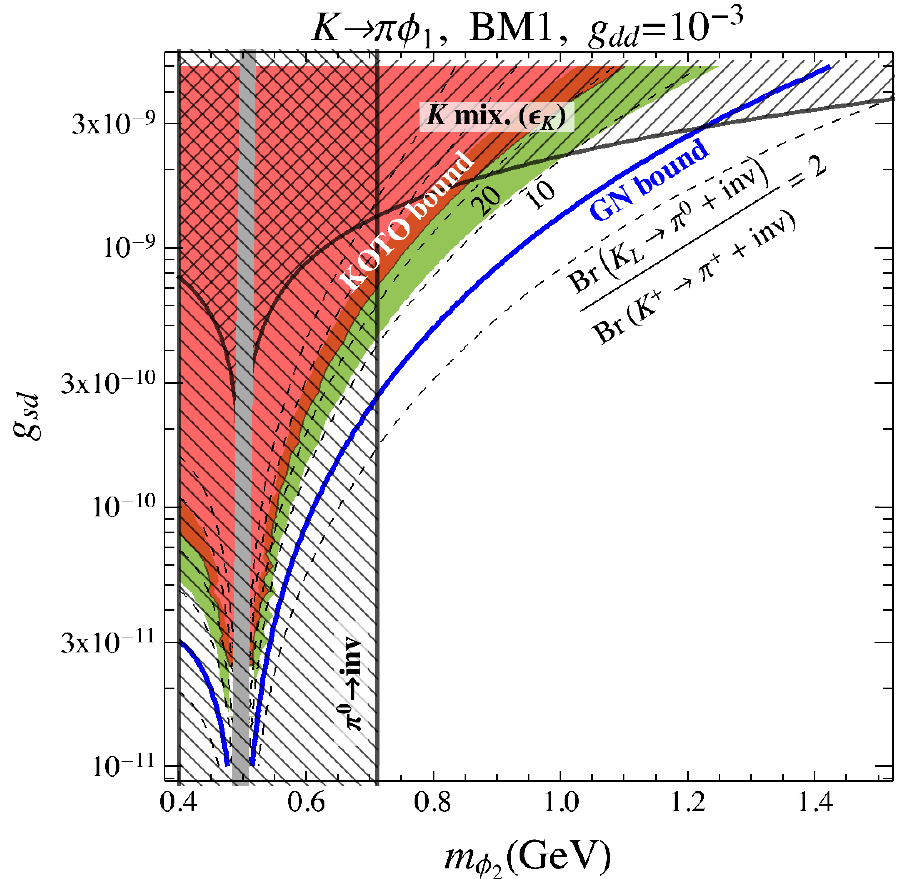}
\end{center}
 \caption{ \label{fig:benchmark1}  The parameter space for Model 1,  BM1,  for $g_{dd}= 10^{-3}$ in \eqref{eq:BM1}. The GN bound is denoted with blue lines, while the green regions give the $1\sigma$ bands corresponding to KOTO anomalous events \cite{Kitahara:2019lws,KOTO:2019}.  Left: the predictions for $\Br(K^+\to \pi^++{\rm inv})$, $\Br(K_L\to \pi^0+{\rm inv})$, varying $m_{\phi_2}\in[0.4,1.5]$ GeV and for two values of $g_{sd}$ (black lines). The values closest to the SM (black cross) are reached for 
 $m_{\phi_2} = 1.5$ GeV.
 Blue regions are the 1$\sigma$ SM prediction bands, with the central values denoted by the dashed lines and a star, red regions are excluded by NA62 \cite{NA62:2019}, E949 \cite{Adler:2008zza} and KOTO \cite{Ahn:2018mvc}. Right: Contours of $\Br(K_L\to \pi^0+{\rm inv})/\Br(K^+\to \pi^++{\rm inv})$ (dashed lines) as functions of $g_{sd}$, $m_{\phi_2}$, with the hatched regions excluded by $K^0-\bar K^0$  mixing and $\pi^0\to inv$ bounds. The region around the kaon mass is masked out (gray region).}
 \end{figure}

The form of couplings in BM1, Eq. \eqref{eq:BM1}, is such that the NP contributions to $\epsilon_K$ are maximized. This benchmark is thus representative of the parameter space that is most constrained. Fixing $g_{dd}= 10^{-3}$ the allowed regions are shown in Fig. \ref{fig:benchmark1}. The red regions are excluded by the NA62 bound on $\Br(K^+\to \pi^+\phi_1)_{\rm exp}\lesssim  1.9\times 10^{-10}$ \cite{NA62:2020}, the E949 bound  $\Br(K^+\to \pi^+\phi_1)_{\rm exp}<  0.73 \times 10^{-10}$~\cite{Adler:2008zza} and by the KOTO bound $\Br(K_L\to \pi^0\phi_1)<2.4 \times  10^{-9}$ \cite{Ahn:2018mvc}.  The E949 and NA62 bounds shown are for massless $\phi_1$, which is a good approximation for our benchmarks, where $m_{\phi_1}=1$ MeV. For heavier masses, above $m_\pi$, the bound is expected to become significantly weaker and completely disappear for $m_{\phi_1}\simeq m_{\pi^0}$, as in \cite{Artamonov:2009sz}.  The green bands denote the $1\sigma$ bands of the branching ratio $\Br(K_L \to \pi^0+{\rm inv})_{\rm KOTO}=\big(2.1^{+2.0}_{-1.1}\big)\times 10^{-9}$ \cite{KOTO:2019,Kitahara:2019lws} that corresponds to the anomalous KOTO events.  The blue line denotes the GN bound, showing that large violations of the GN bound are possible in this model. 

This violation is most apparent in Fig. \ref{fig:benchmark1} (right) which gives the allowed values of $g_{sd}$ as a function of $m_{\phi_2}$,  with the dashed lines denoting contours of the ratio $\Br(K_L \to \pi^0+{\rm inv})/\Br(K^+ \to \pi^++{\rm inv})$. The present KOTO bound is saturated by values for this ratio of around 20,  while still satisfying the   $\epsilon_K$ constraint, Eq. \eqref{eq:gds:epsilonK}, and the $\pi^0\to {\rm inv}$ constrain discussed below, see Eq. \eqref{eq:pi0:inv}. The excluded regions are  shown hatched in Fig. \ref{fig:benchmark1} (right). 
The bound from $\Re(\epsilon'/\epsilon)$, Eq.~\eqref{eq:epsilon'}, is less stringent and  not displayed
as there is already a lot of information in the figure.  It is straightforward to plot this constraint from 
the formulae given in Section\ref{sec:epspeps}.

 The solid black lines in Fig.~\ref{fig:benchmark1} (left) show the values of $\Br(K_L\to \pi^0+{\rm inv})$ and $\Br(K^+\to \pi^++{\rm inv})$ for $g_{sd}=5\cdot 10^{-10}$ and $g_{sd}=2\cdot 10^{-9}$, varying $m_{\phi_2}\in[0.4,1.5]$ GeV, while fixing $g_{dd}=10^{-3}$ (the grey dotted parts of the lines are excluded by a combination of $K^0-\bar K^0$ and $\pi^0\to$ inv constraints).  The SM predictions for the two branching ratios,  $\Br(K^+\to \pi^+\nu \bar \nu)_{\rm SM}=(8.4 \pm 1.0) \times 10^{-11}$ and  $\Br(K_L \to \pi^0\nu \bar \nu)_{\rm SM}=(3.4 \pm 0.6) \times  10^{-11}$~ \cite{Buras:2005gr,Brod:2010hi,Buras:2015qea}, are denoted with blue bands ($1\sigma$ ranges). For the larger value, $g_{sd}=2 \cdot 10^{-9}$, the prediction is still quite far away from the SM for 
$m_{\phi_2}=1.5$ GeV, but would of course tend to the SM for $m_{\phi_2}\to\infty$. For larger values of $g_{sd}$ deviations from the SM prediction for $\Br(K^+\to \pi^++{\rm inv})$ at the level of a few are predicted for this benchmark and subject to the indicated constraints from E949, while for smaller values of $g_{sd}$ the deviations in $\Br(K^+\to \pi^++{\rm inv})$ become negligibly small. That is, it is possible to explain the KOTO anomalous events without having any appreciable NP effects in the charged kaon decay nor in $K^0-\bar K^0$  mixing. 
 
\begin{figure}[t]
 \begin{center}
\includegraphics[width=7cm]{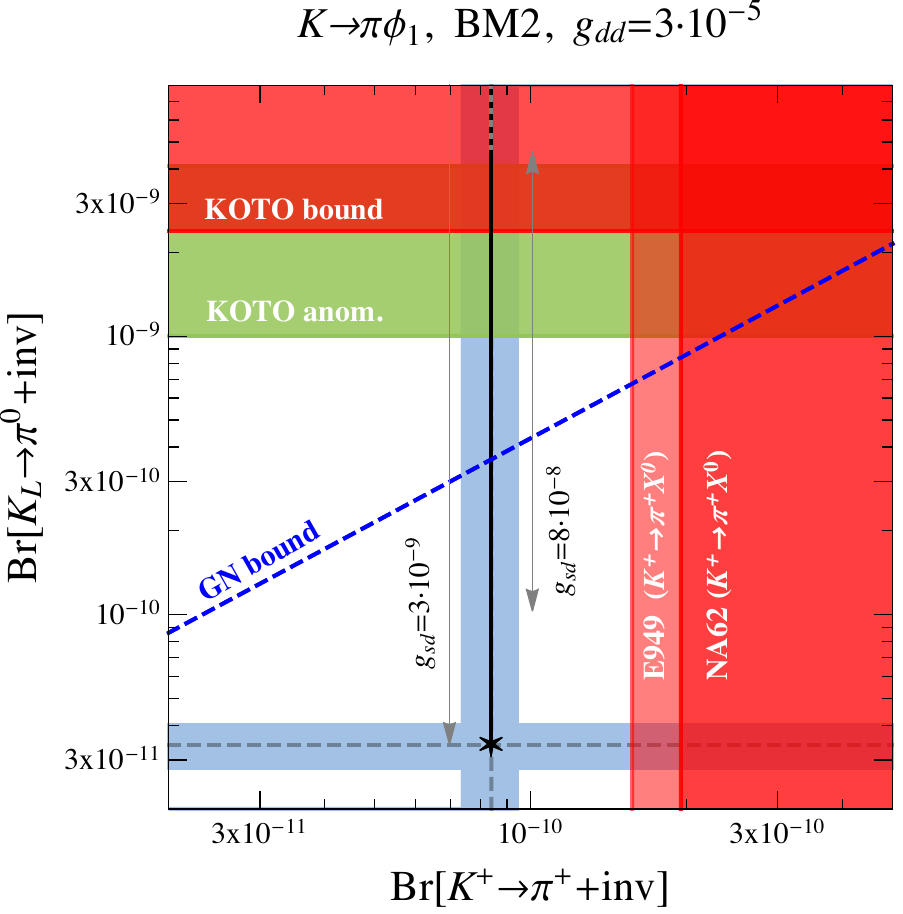}~~~~ 
\includegraphics[width=7cm]{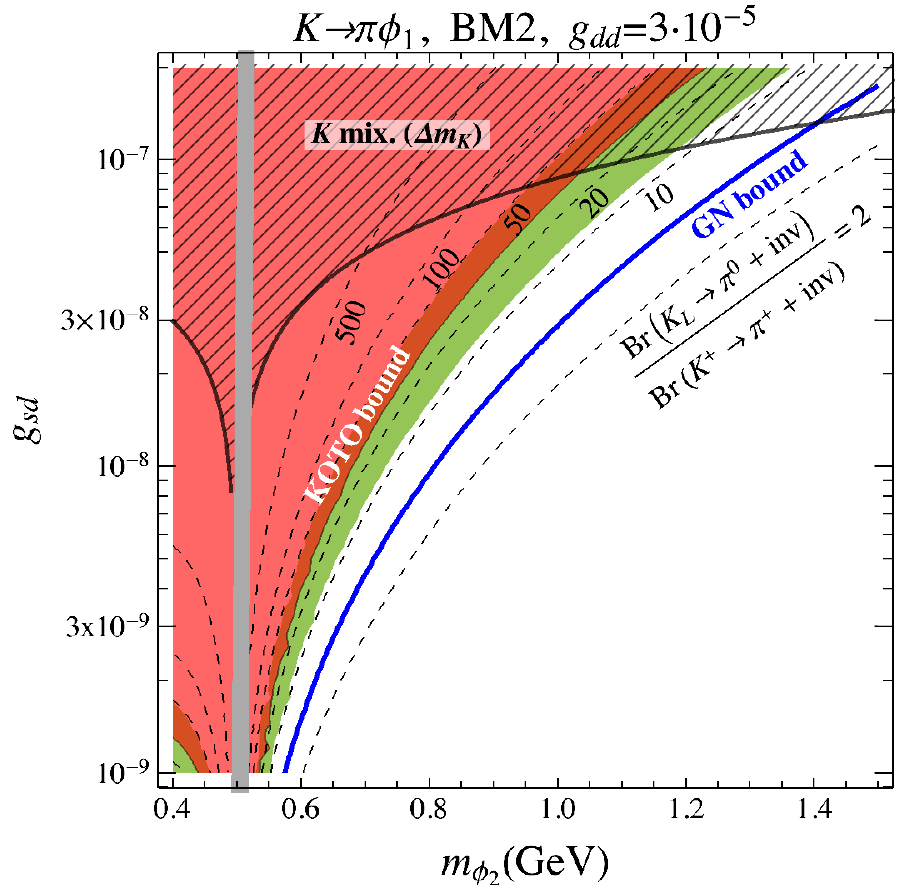}
\end{center}
 \caption{ \label{fig:benchmark2} The preferred parameter space for Model 1, BM2. Color coding is the same as in Fig. \ref{fig:benchmark1}. See end of Section \ref{sec:con-Model1} for comments on these figures.}
 \end{figure}

 We next move  to  BM2. The form of couplings in Eq. \eqref{eq:BM2} was deliberately chosen such that there is no NP CP violation in $K^0-\bar K^0$  mixing, in order to  avoid the  $\epsilon_K$ bound. The bound from $\Delta m_K$, Eq. \eqref{eq:gds:DeltamK}, on CP conserving contributions to $K^0-\bar K^0$  mixing is much weaker, giving the hatched excluded region in Fig. \ref{fig:benchmark2} (right). This means that for the same mass of $\phi_2$ the flavor violating couplings to quarks can be much larger than in BM1. In Fig. (right) \ref{fig:benchmark2} we show the $g_{dd}=3\times 10^{-5}$ slice of the parameter space, in which case $g_{sd}$ can be as large as $10^{-7}$. 
  Furthermore, the form of couplings in  BM2, Eq. \eqref{eq:BM2}, is such that there is no NP effect at all in $\Br(K^+\to \pi^++{\rm inv})$, to the order we are working, and the E949 bound is completely avoided. In contrast, the effect on  $\Br(K_L\to \pi^0+{\rm inv})$ can be very large and easily saturate  KOTO's present  upper bound, as shown for two representative couplings $g_{sd}= 3\times 10^{-9}, 8\times 10^{-8}$ (black lines, with dashed parts of the lines excluded by $\Delta m_K$). BM2 comes with  enhanced symmetry;  $\phi_2$ is a pure pseudoscalar and $\phi_1$ a pure scalar. This has implications for flavor conserving couplings of $\phi_1$, to which we turn next. 
 
\subsection{Constraints on the $\phi_1$-couplings}
So far the scalar mass $\phi_1$ has been  fixed to $1$ MeV.
Next, we show  that for the two benchmarks the radiatively generated couplings of $\phi_1$ to 
pions, nucleons, and photons are all well below the bounds for a large range of $\phi_1$ masses (including  $m_{\phi_1}=1$ MeV). Figs. \ref{fig:benchmark1} and \ref{fig:benchmark2} are thus valid  for a larger set of $\phi_1$ masses, as long as $m_{\phi_1}\ll m_K$.

\subsubsection{Invisible pion decays}
\label{sec:inv:pions}

If kinematically allowed, $\pi^0\to \phi_1\phi_1$ can be an important phenomenological constraint. In Model 1 this decay can proceed through $\phi_2 - \pi^0$ mixing though the loop diagram shown in Fig.~\ref{fig:pi0gagaM1} (left). 
In the  $\mH \gg m_{\pi,\eta}$ limit  the decay amplitude is 
\beq
{\cal M} (\pi^0 \to \phi_1 \phi_1) = \frac{1 }{12 \pi^2} \frac{(\lambda m_S)^2  B_0^2 f}{\mH^4}
\Re g_{dd}^{(2)} \left( 2 \Im g_{dd}^{(2)} -\Im g_{ss}^{(2)} \right)   \, .
\eeq
The corresponding width is given by 
\beq
\label{eq:pi0phi1phi1:dec}
\Gamma (\pi^0 \to \phi_1 \phi_1) = \frac{|{\cal M}|^2}{32 \pi m_\pi} \beta_{\phi_1} \, , 
\eeq
where here $\beta_{\phi_1} \equiv \big(1- 4 m_{\phi_1}^2/m_{\pi}^2\big)^{1/2}$, 
so that in the limit $m_{\phi_1} \ll m_\pi \ll m_{\phi_2}$, one has for the branching ratio (setting $\Im g_{ss}^{(2)} = 0$ for simplicity) 
\begin{align}
\label{eq:pi0:inv}
\Br  (\pi^0 \to \phi_1 \phi_1) & =  1.2 \times 10^{-9} \left( \frac{\Re g_{dd}^{(2)} }{10^{-3}} \right)^2 \left( \frac{\Im g_{dd}^{(2)} }{10^{-3}} \right)^2 \left( \frac{\lambda \, m_S}{\GeV} \right)^4 \left( \frac{\GeV}{m_{\phi_2}} \right)^8 \, .
\end{align}
 The preliminary 90\% C. L. experimental bound reported very recently  by NA62 \cite{NA62:2020} 
\beq
\label{eq:pi0:inv:exp}
 \Br (\pi^0 \to \phi_1 \phi_1) < 4.4 \times 10^{-9}   \, ,
\eeq
improves the   E949 bound of $2.7   \times 10^{-7}$~\cite{Artamonov:2005cu} by almost two orders of magnitude.
BM2 obeys this bound trivially, since $\pi^0\to \phi_1\phi_1$ if forbidden by parity ($\Re g_{dd}^{(2))}=0$). For BM1, on the other hand, the bound on ${\rm Br}(\pi^0\to{\rm inv})$, Eq. \eqref{eq:pi0:inv:exp},   represents a stringent constraint, as shown 
 in Fig. \ref{fig:benchmark1} (right). 
 
 Finally, the $\pi^0\to \phi_1\phi_1$ decay could also proceed at tree level via an additional interaction
 term in \eqref{eq:coupl:LO} of the form
 $\delta {\cal L} =  \lambda' m_S \phi_2 \phi_1^2$.
Whereas, contrary to  Model 2, $\lambda'$ plays no role in the $K\to \pi \phi_1$ decays per se, 
it is potentially dangerous for the invisible pion decay. 
In the absence of a UV completion we may 
choose its initial value to be sufficiently small (zero in practice) to pass the constraint.

\subsubsection{$\phi_1-\pi^0$ mixing}
\begin{figure}[t]
 \begin{center}
\includegraphics[width=7cm]{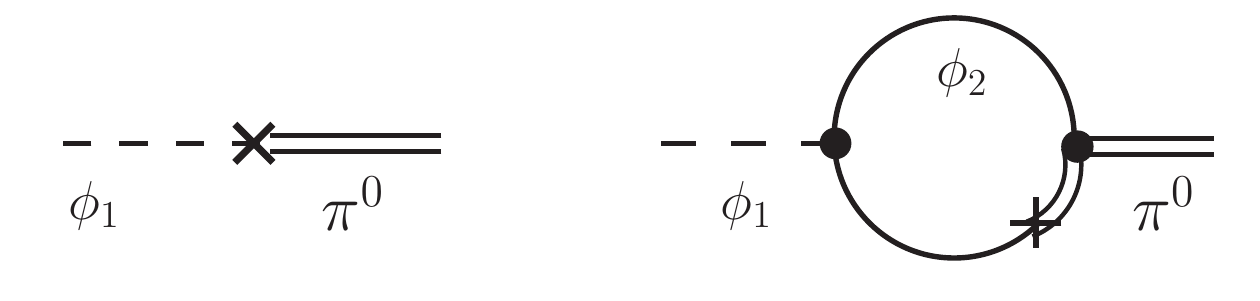}
\end{center}
\vspace{-0.6cm}
 \caption{ The leading order and one loop induced $\phi_1-\pi^0$ mixing.  \label{fig:phi1pi0} }
 \end{figure}

The $\phi_i$ mix with light pseudoscalars through the $g_{qq'}^{(i)}$ couplings, Eq. \eqref{eq:Lag:model1}. The $\phi_1-\pi^0$ part of the mass matrix to one loop receives contributions in Fig.~\ref{fig:phi1pi0}, and 
is parametrized by the Lagrangian, $m_{\phi_2}\gg m_{\phi_1,\pi,\eta}$,  
\beq
\label{eq:Leff:mix}
{\cal L}_{\rm eff}\supset - g_{1 \pi} B_0 f   \phi_1 \pi^0,
\eeq
with the effective $\phi_1-\pi^0$ coupling given by  
\beq
\begin{split}
\label{eq:g1pi}
g_{1\pi}  & =  \Im g_{dd}^{(1)} + \frac{1 }{8 \pi^2 }  \biggr(\frac{\lambda m_S B_0}{m_{\phi_2}^2}\biggr) \biggr\{ \Im g_{dd}^{(2)} \Re g_{dd}^{(2)}  L (m_\pi)  \\ 
& + \frac{1}{3} \left(  \Im g_{dd}^{(2)}- 2 \Im g_{ss}^{(2)}  \right) \Re g_{dd}^{(2)}  L(m_\eta)   +  \left( 
\overline{g}^{(2)}_{ds} \hat{g}^{(2)}_{sd} + {\rm h.c.} \right) L(m_K)\biggr\} 
\; ,
\end{split}
\eeq
where we have exceptionally kept the $g^{(2)}_{ds}$-terms since they are leading in BM2.
The first term is due to tree level mixing, see Fig. \ref{fig:phi1pi0} (left), the second term are the one loop corrections due to diagram in Fig. \ref{fig:phi1pi0} (right).  
The loop function $L (m_X) \equiv - m_{\phi_2}^2 C_0 (0,0,0, m_X^2,m_{\phi_2}^2,m_{\phi_2}^2)$ is normalized such that $L(m_X) \to 1$ for $m_{\phi_2} \gg m_X$. In the following we will take for simplicity this limit, which provides a reasonable approximation for the parameter region of interest, since $L_\pi \simeq 0.8, L_{K,\eta} \simeq 0.4$ for $m_{\phi_2} = 400$  MeV.
In the two benchmarks \eqref{eq:BM1}, \eqref{eq:BM2}, the effective $\phi_1-\pi^0$ couplings are  
\begin{align}
{\bf BM~1}:\qquad g_{1 \pi}^{\rm BM1} & = 2.3 \times 10^{-8} \left( \frac{g_{dd}}{10^{-3}} \right)^2 \left( \frac{\GeV}{m_{\phi_2}} \right)^2 \, , 
\\
{\bf BM~2}:\qquad g_{1 \pi}^{\rm BM2} &= 0 \;.
\end{align}
In BM2 there is no $\phi_1-\pi^0$ mixing $\phi_1$ is 
a pure scalar and parity is conserved. 

Working in the mass insertion approximation for the off-diagonal mass term, Eq. \eqref{eq:Leff:mix}, the $\phi_1 - \pi^0$ mixing angle, $s_\theta \equiv \sin \theta \approx \theta$, between the interaction states $\phi_1$ and the mass eigenstate $\phi_1'\approx \phi_1- s_\theta \pi^0$   is  
\begin{align}
s_{\theta} = \frac{B_0 f}{m_\pi^2 - m_{\phi_1}^2} \,  g_{1 \pi}  \, .
\end{align}
Note that this expression for the mixing angle is only valid for $m_{\phi_1}$ sufficiently far away from $m_\pi$. For the two benchmarks, we have  
\begin{align}
{\bf BM~1}:\qquad s_\theta^{\rm BM1} & = 3.0 \times 10^{-7} \left( \frac{g_{dd}}{ 10^{-3}} \right)^2 \left( \frac{\GeV}{m_{\phi_2}} \right)^2 \, , 
\\
{\bf BM~2}:\qquad s_\theta^{\rm BM2} & =0\;.
\end{align} 
The $\phi_1-\pi^0$ mixing is thus very small in most of the viable parameter space, justifying the use of the 
mass insertion approximation. 

\subsubsection{Couplings of $\phi_1$ to photons}
 The dominant decay channel of $\phi_1$ is to two photons. In the limit $m_{\phi_1}\ll m_\pi$ the interactions with two photons are described by the effective Lagrangian
 \beq\label{eq:phi1:photons}
{\cal L}_{\rm eff} \supset - \frac{1}{8} g_{1 \gamma \gamma} \phi_1 F^{\mu \nu} F^{\rho \sigma} \epsilon_{\mu \nu \rho \sigma} - \frac{1}{4} h_{1 \gamma \gamma} \phi_1 F^{\mu \nu} F_{\mu \nu}\;.
 \eeq
 \begin{figure}[t]
 \begin{center}
\includegraphics[width=10cm]{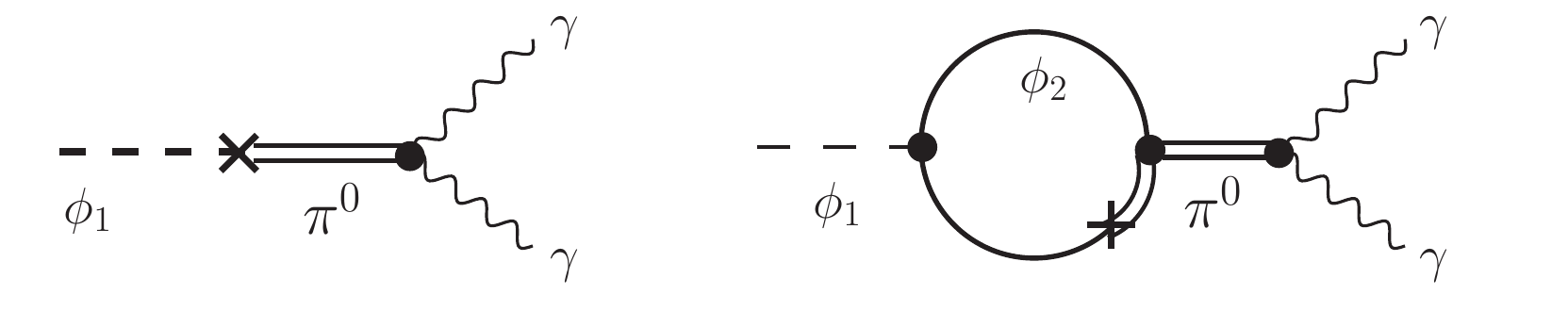}~~~~ 
\end{center}
\vspace{-0.6cm}
 \caption{ \label{fig:phi1:photons:g1} CP violating contributions to   $\phi_1\to \gamma\gamma$, matching onto the  coupling $g_{1\gamma\gamma}$.}
 \end{figure} 
  \begin{figure}[t]
 \begin{center}
\includegraphics[width=14cm]{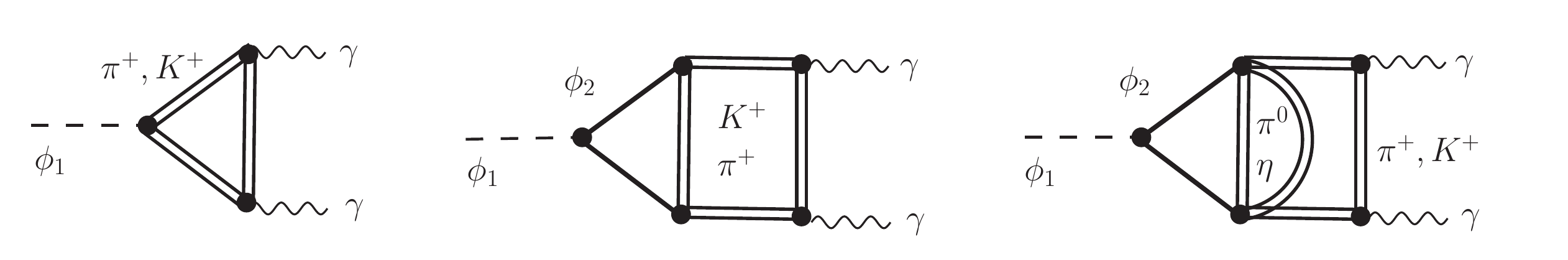}~~~~ 
\end{center}
\vspace{-0.6cm}
 \caption{ \label{fig:phi1:photons:h1} CP conserving contributions to   $\phi_1\to \gamma\gamma$, matching onto the coupling $h_{1\gamma\gamma}$.}
 \end{figure} 
 \begin{figure}[t]
 \begin{center}
\includegraphics[width=9cm]{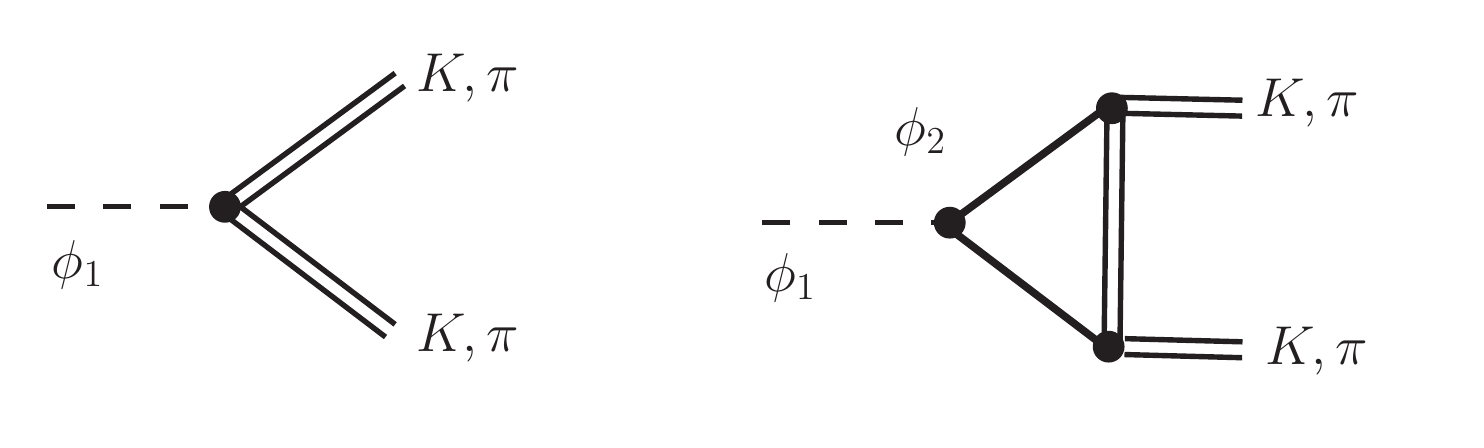}~~~~ 
\end{center}
\vspace{-0.8cm}
 \caption{ \label{fig:phi1:pipi} Tree level and one loop contributions matching onto the effective   
  couplings $g_{1\pi\pi}$/$g_{1KK}$.
 }
 \end{figure} 
The dominant contribution to the CP violating  coupling $g_{1 \gamma \gamma}$ is from the $\pi^0$ anomaly term via the $\phi_1-\pi^0$ mixing, see Fig. \ref{fig:phi1:photons:g1}. Working in the mass insertion approximation for the off-diagonal mass term, Eq. \eqref{eq:Leff:mix}, gives 
\begin{align}
 g_{1 \gamma \gamma} =  s_{\theta} \frac{\sqrt{2} \alpha}{ \pi f_\pi} \approx g_{1 \pi}  \frac{6.2 \times 10^{-3} \, \GeV}{m_\pi^2 - m_{\phi_1}^2}\xrightarrow{m_{\phi_1} \ll m_\pi}  g_{1 \pi}\, 0.34 \GeV^{-1}   \, , 
\end{align}
with $g_{1\pi}$  given in \eqref{eq:g1pi}.
 
 The CP conserving $h_{1 \gamma \gamma}$ coupling receives the first relevant contributions from radiative corrections with $K^+$ and $\pi^+$ running in the loop cf. Fig.~\ref{fig:phi1:photons:h1}.  For our benchmarks the first nonzero contributions arises at two loops, while for BM2 the numerically most important contribution arises at three loops
 \beq
 h_{1 \gamma \gamma}  = h_{1\gamma\gamma}^{\rm 1+2\,loop}+h_{1\gamma\gamma}^{\rm 3\,loop}\,.
 \eeq
 In the $m_{\phi_2} \gg m_K$ ($m_\pi \gg m_{\phi_1}$ by assumption)
 limit the one and two loop contributions, in  Fig.~\ref{fig:phi1:photons:h1},
 assume the form 
\beq
h_{1\gamma\gamma}^{\rm 1+2\,loop}  = \frac{\alpha}{12 \pi} \left( \frac{g_{1 \pi \pi} }{m_\pi^2} + \frac{g_{1 K K} }{m_K^2} \right)\;,
\eeq
whereas the effective couplings of $\phi_1$, ${\cal L}_{\rm eff}  \supset \phi_1 \left(g_{1 \pi \pi} \pi^+ \pi^- + g_{1 K K}  K^+ K^-\right) $,   
to two light charged mesons  evaluate to 
\beq
\label{eq:g1pipi}
g_{1 \pi \pi} = B_0 \biggr[\Re g_{dd}^{(1)}+\frac{ \lambda B_0 m_S}{8 \pi^2 m_{\phi_2}^2}  \left( (\Re g_{dd}^{(2)})^2 + |\overline{g}_{sd}^{(2)}|^2 \right)\biggr]  \;,
\eeq 
and  $g_{1KK} = g_{1 \pi \pi}|_{dd \to ss} $. The first term in \eqref{eq:g1pipi} is the tree level term from \eqref{eq:ChPT+phi}, see Fig.~\ref{fig:phi1:pipi}  (left). In both benchmarks, BM1 and BM2, this contribution was set to zero. The second term in \eqref{eq:g1pipi} is the one loop correction, see Fig.~\ref{fig:phi1:pipi}  (right). We kept the flavor  violating contribution proportional to $\bar g_{sd}$ even though it is numerically negligible. 

For the three loop contribution to $h_{1 \gamma \gamma}$ we resort to a NDA estimate, still in the $m_{\phi_2} \gg m_K$ limit, 
\beq
h_{1\gamma\gamma}^{\rm 3\,loop}\approx \frac{\alpha}{4\pi}\frac{\lambda m_S}{(16\pi^2)^2} \Big(\frac{B_0}{f m_{\phi_2}}\Big)^2 \Big[\big(\Im g_{dd}^{(2)}\big)^2 +{\mathcal O}(1)\times  (\Im g_{ss}^{(2)})^2+ {\mathcal O}(1) \times (\Im g_{dd}^{(2)}) (\Im g_{ss}^{(2)})\Big] \;, \nonumber
\eeq
where  the ${\mathcal O}(1)$ factors are not displayed.

Finally we are in a position to assemble the results for the benchmarks.
Using $g_{sd}\ll g_{dd}$, the  $\phi_1$--photon couplings evaluate to  
\beq
\begin{split}
\label{eq:g1gammagammaBM1}
{\bf BM1:}~~~g_{1 \gamma \gamma}^{\rm BM1}  & \simeq \frac{7.7 \times 10^{-9}}{  \GeV} \left( \frac{g_{dd}}{10^{-3}} \right)^2 \left( \frac{\GeV}{m_{\phi_2}} \right)^2 \, ,
\\
~~~h_{1 \gamma \gamma}^{\rm BM1}  & \simeq \frac{4.8 \times 10^{-10} } {\GeV} \left( \frac{g_{dd}}{10^{-3}} \right)^2 \left( \frac{\GeV}{m_{\phi_2}} \right)^2 \, ,
\end{split}
\eeq
in BM1, while for BM2 they turn out to be 
\beq
\begin{split}
\label{eq:g1gammagammaBM2}
{\bf BM2:}~~~g_{1 \gamma \gamma}^{\rm BM2}  & =0 \;,
\\
~~~h_{1 \gamma \gamma}^{\rm BM2}  & \sim 
\frac{ 2 \times 10^{-14}}{ \GeV} \left( \frac{g_{dd}}{ 3\times 10^{-5}} \right)^2 \left( \frac{\GeV}{m_{\phi_2}} \right)^2 \;,
\end{split}
\eeq
and we remind the reader that $\lambda m_S = 1$ GeV for reference. 
The $g_{1\gamma\gamma}$ coupling vanishes in BM2 since $\phi_1$ is a parity even scalar in that benchmark. The value quoted for $h_{1 \gamma \gamma}^{\rm BM2} $ is the NDA estimate of the flavor conserving 3 loop contribution. 
For representative values of $g_{dd}$ in the two benchmarks we used the values in Figs. \ref{fig:benchmark1} and Fig. \ref{fig:benchmark2} for BM1 and BM2, respectively. 

The above couplings of $\phi_1$ to photons are sufficiently  small that for both benchmarks the $\phi_1$ is stable on collider scales.  More concretely, the $\phi_1\to \gamma\gamma$ partial decay width is given by 
 \begin{align}
\Gamma_{1 \gamma \gamma} & =  \frac{1}{64 \pi} \big(g_{1 \gamma \gamma}^2 + h_{1 \gamma \gamma}^2\big) m_{\phi_1}^3 \, ,
\end{align} 
and this translates to 
\begin{align}
{\bf BM1:~~}c \tau_{1 \gamma \gamma}^{\rm BM1} & = 7  \times 10^{11} \, {\rm m}  \left( \frac{\MeV}{m_{\phi_1}} \right)^3 \left( \frac{ 10^{-3}}{g_{dd}} \right)^{4} \left( \frac{m_{\phi_2}}{\GeV} \right)^{4}, \, 
\\
{\bf BM2:~~}c \tau_{1 \gamma \gamma}^{\rm BM2} & \sim  10^{23} \, {\rm m}  \left( \frac{\MeV}{m_{\phi_1}} \right)^3 \left( \frac{ 3 \times 10^{-5}}{g_{dd}} \right)^{4} \left( \frac{m_{\phi_2}}{\GeV} \right)^{4},
\end{align}
such that $\phi_1$ is stable on solar to cosmological timescales.  For such small couplings the laboratory constraints from, e.g., $\pi^+ \to \phi_1 e^+  \nu$ decays~\cite{Altmannshofer:2019yji} are irrelevant, whereas astrophysical and cosmological constraints are important (cf. figure 1 in Ref.~\cite{Jaeckel:2015jla}) 
and  further discussed in Section \ref{sec:constraints:combo}.

\subsubsection{Couplings of $\phi_1$ to nucleons}

\begin{figure}[t]
 \begin{center}
\includegraphics[width=14cm]{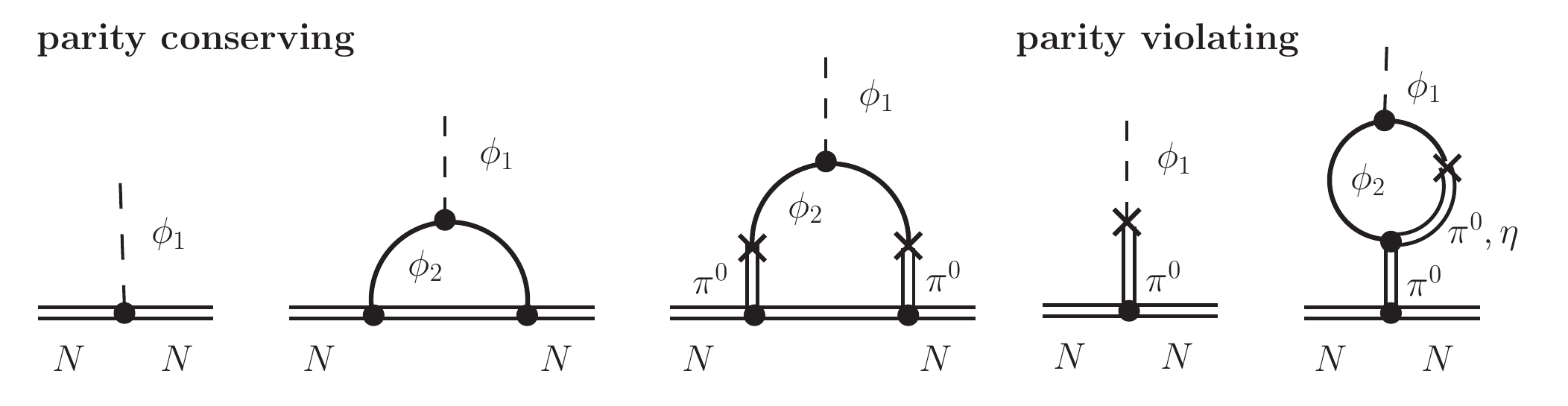}
\vspace{-1,0cm}
\end{center}
 \caption{ The leading order and one loop  induced $\phi_1$-couplings to nucleons grouped 
 into   parity conserving coupling $g_{1NN}$ (even in $g_A$) and parity violating coupling $\tilde{g}_{1NN}$ in \eqref{eq:g1NN} (odd in $g_A$).
  The 3rd diagram is the only non-vanishing contribution to $g_{1NN}$ in BM2. 
 \label{fig:phi1nucleon} }
 \end{figure}

The couplings of $\phi_1$ to protons and neutrons are tree-level and loop-level 
induced by $g_{dd}^{(1)}$ and $g_{dd}^{(2)}$ respectively, cf. Fig. \ref{fig:phi1nucleon}.
One can use Heavy Baryon Chiral Perturbation Theory (HBChPT)~\cite{Jenkins:1990jv}  to organize different contributions. We only keep only the leading terms which are (in  relativistic notation)
\beq
\label{eq:Lnucleon}
{\cal L}=\frac{g_A}{f}(\bar N\gamma^\mu \gamma_5 t^a N)\partial_\mu \pi^a+\sum_i (\bar N Y_i^N N) \phi_i+\cdots,
\eeq
with $t^a=\sigma^a/2$, $a=1,2,3$ and $\sigma^a$ are Pauli matrices, $N=(p,n)$  the isospin doublet of nucleons, and  
\beq
\label{eq:Y}
Y_i^N=
\begin{pmatrix}
\sum_q\Re(g_{qq}^{(i)}) {\sigma_q^p}/{m_q} &0
\\
0 &\sum_q \Re(g_{qq}^{(i)}) {\sigma_q^n}/{m_q}
\end{pmatrix},
\eeq
the coupling between $\phi_i$ and nucleons with  summation over $q=d,s$ (by assumption the couplings of $\phi_{1,2}$ to up quarks are zero). For the matrix elements of the scalar current, $ \sigma_q^N \bar u_N u_N=\langle N |m_q\bar q q|N \rangle$ we use the values from \cite{Bishara:2017pfq}, $\sigma_d^p=(32\pm 10){\rm~MeV},  \sigma_d^n=(36\pm 10){\rm~MeV},  \sigma_s^p=\sigma_s^n=(41.3\pm 7.7){\rm~MeV}$, along with the quark masses at $\mu=2$ GeV, $m_d=4.67(33)$ MeV, $m_s=93(8)$ MeV, while $g_A=1.2723(23)$ \cite{Tanabashi:2018oca}.

In the heavy $\phi_2$ limit the following  effective Lagrangian 
\beq
\begin{split}
{\cal L}_{\rm eff}&= g_{1NN} m_N \phi_1(\bar N N)+2  \tilde g_{1NN} m_N \phi_1 (\bar N i\gamma_5 t^3 N) \;,
\end{split}
\label{eq:g1NN}
\eeq
provides a  good description of the $\phi_1$-nucleon system. 
Assuming  $m_{\phi_2, N} \gg m_{\phi_1} $,
the diagrams in Fig. \ref{fig:phi1nucleon},  evaluate to   
\begin{align}
\label{eq:g1NN} 
g_{1NN}&=\frac{1}{m_N}\biggr[Y_1^N-\frac{\lambda m_S}{8\pi^2 m_N} \biggr\{ (Y_2^N)^2 F(r)-\Big(g_A  \Im g_{dd}^{(2)} \frac{B_0}{m_N}\Big)^2\tilde F(r)\biggr\}\biggr]\;,
\\
\label{eq:tildeg1NN}
\tilde g_{1NN}&=  \frac{ g_A B_0}{m_\pi^2-m_{\phi_1}^2}\biggr[\Im g_{dd}^{(1)} +\frac{\lambda m_S}{12\pi^2}\biggr(\frac{  B_0}{m_{\phi_2}^2}\biggr)\Big(2 \Im g_{dd}^{(2)} -\Im g_{ss}^{(2)}\Big)\Re g_{dd}^{(2)}
\biggr]\;,
\end{align}
where $Y_1^N$ stands for the nucleon-nucleon entries in \eqref{eq:Y}. In the $\tilde F(r)$ term in \eqref{eq:g1NN}  we in addition assumed the $m_\pi \gg m_{\phi_1}$ limit. 
The real-valued loop functions  $F(r)$, $\tilde F(r)$, with $r=m_{\phi_2}^2/m_N^2$,  are given 
by\,\footnote{It is noted that the 3rd diagram, in Fig.~\ref{fig:phi1nucleon}, 
does not introduce any infrared (IR) divergences in the limit $m_\pi \to 0$. This is a consequence of  
the derivative couplings of pions, cf. Eq. \eqref{eq:Lnucleon}. We note in passing  that for a double insertion of this  interaction term
one cannot use the naive EOM and replace 
$g_A (\bar N\gamma^\mu \gamma_5 t^a N)\partial_\mu \pi^a \to  -2 m_N g_A \bar N \gamma_5 t^a N \pi^a$. 
For a concise technical discussion we refer the reader to Ref. \cite{Simma:1993ky}. 
Use of the naive EOM leads to the IR divergence that is linked to the absence of the derivative coupling in that case. The same applies to the single insertion of the $g_A$-term in 4th and 5th diagram.}
\begin{align}
\label{eq:F}
F(r)&=\frac{(r-3)}{2}\log r -1+ (1-r)\sqrt{1-4/r}\log\biggr[\frac{1}{2}\biggr(\sqrt{r-4}+
\sqrt{r}  \biggr)\biggr] \;,
\\
\tilde F(r)&=\frac{1}{r^2\sqrt{1-{4}/{r}}}\log\biggr[\frac{1}{2}\biggr(\sqrt{r-4}+
\sqrt{r}  \biggr)\biggr] \;. 
\end{align}
 In the limit $m_{\phi_2}\gg m_N$ we have $F(r)\to -3/(2r)$, $\tilde F(r)\to 
  \ln r / (2r^2)$.
 For $m_{\phi_2}\in [0.5,1.5]$ GeV the loop functions take values in the intervals $F(r)\in  [-2.7, -0.46]$, $\tilde F(r)\in [4.5, 0.13]$.
The first term in \eqref{eq:g1NN} is due to the 1st diagram, while the one loop corrections are due to the 2nd and 5rd diagram in Fig. \ref{fig:phi1nucleon}. 
For the pseudoscalar coupling to nucleons, $\tilde g_{1NN}$, 
we keep   the pion exchange term (dropping the $\eta$-exchange) in the 4th and the 5th diagram in Fig. \ref{fig:phi1nucleon} resulting in the tree level and one loop terms in \eqref{eq:tildeg1NN}. To simplify the expressions we show the one loop contribution in \eqref{eq:tildeg1NN} only in the heavy $m_{\phi_2}$ limit.

Numerically,  we have for BM1, setting $m_{\phi_2}=1$ GeV, 
\begin{align}
\label{eq:g1NN:BM1}
g_{1 NN}^{\rm BM1} &\simeq 3.5 (4.3) \times 10^{-7} \,  \GeV^{-1} \left( \frac{g_{dd}}{10^{-3}} \right)^2  \, , 
\\
\label{eq:tildeg1NN:BM1}
\tilde g_{1 NN}^{\rm BM1} &\simeq \phantom{-} 4  \times 10^{-6} \,  \GeV^{-1} \left( \frac{g_{dd}}{10^{-3}} \right)^2   \, , 
\end{align}  
where the $g_{1 NN}^{\rm BM1}$ central value refers  to protons (neutrons), while for BM2, 
\begin{align}
\label{eq:g1NN:BM2}
g_{1 NN}^{\rm BM2} &\simeq 8 \times 10^{-11} \,  \GeV^{-1} \left( \frac{g_{dd}}{3 \times 10^{-5}} \right)^2  \, , 
\\
\label{eq:tildeg1NN:BM2}
\tilde g_{1 NN}^{\rm BM2} &= 0  \; .
\end{align}  
Below we analyse the combined constraints from the previous two subsections.

   \begin{figure}[t]
 \begin{center}
\includegraphics[width=14cm]{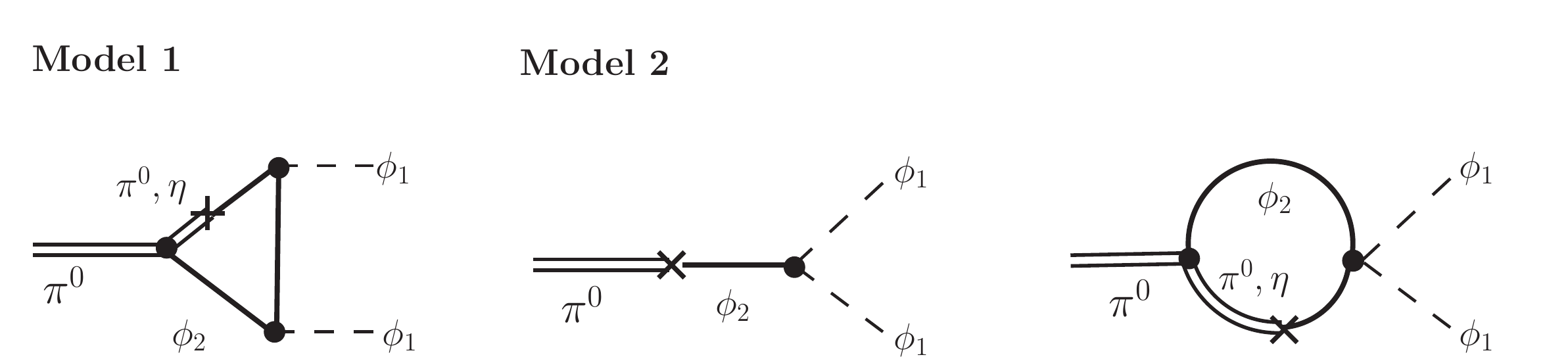}
\end{center}
 \caption{Diagrams for the invisible pion decay, $\pi^0 \to \phi_1 \phi_1$, in Model 1 (left) and Model 2 (middle 
 and right). The diagram for Model 3 are analogous to Model 2 with the difference that the graph on the right 
 needs an extra $\psi_2$ propagator as in Fig.~\ref{fig:fermion}.
 \label{fig:pi0gagaM1} }
 \end{figure} 

\begin{figure}[t]
\centering
\includegraphics[width=7cm]{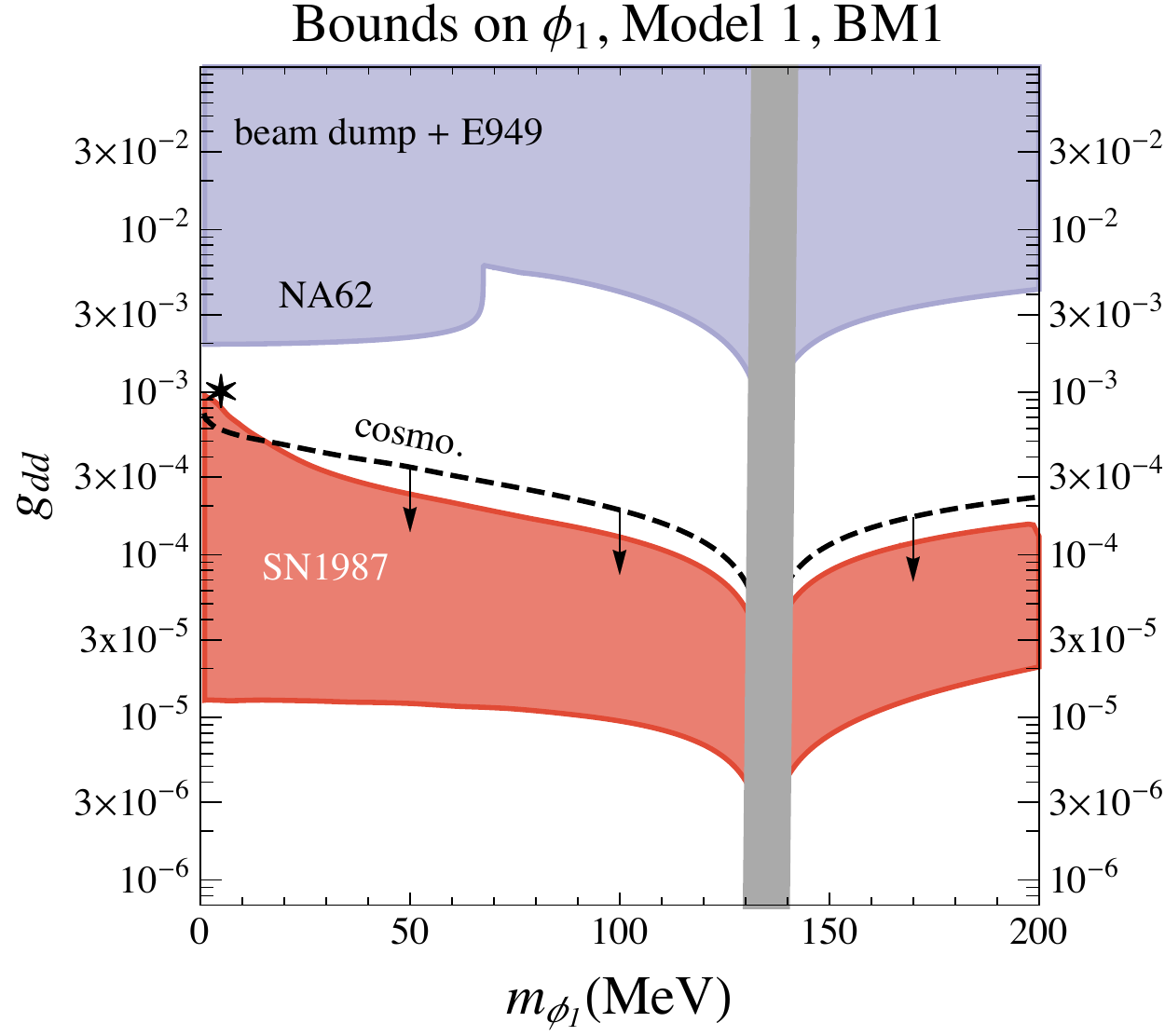}~~~
\includegraphics[width=7cm]{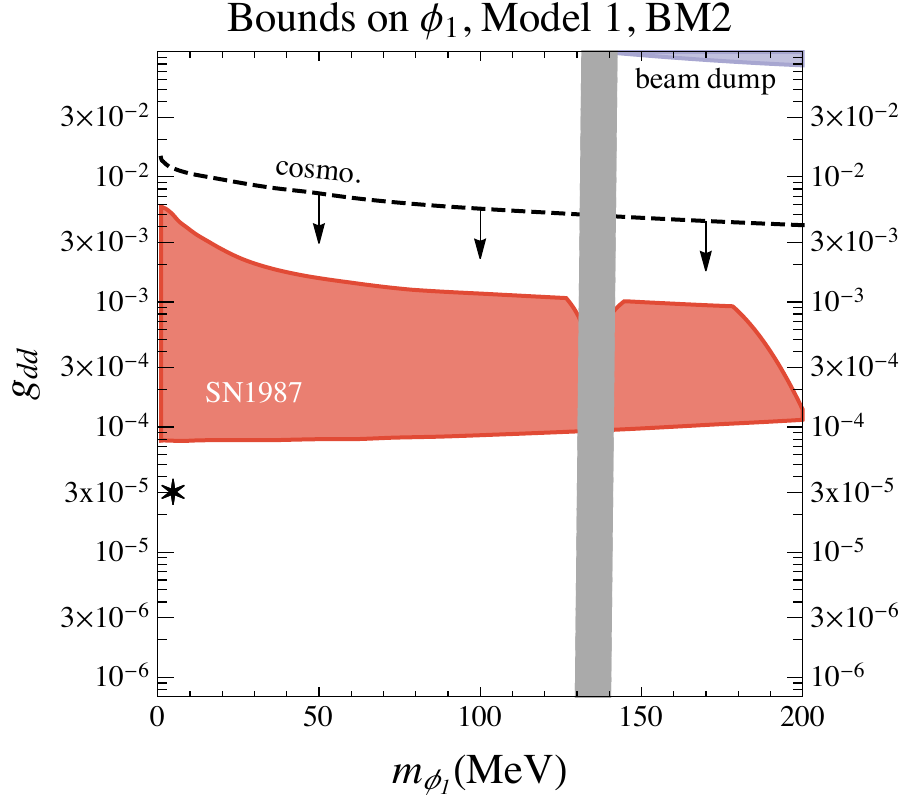} 
 \caption{ The constraints on the $g_{dd}$ coupling in BM1 (left) and BM2 (right) due to couplings of $\phi_1$ to photons and nucleons as a function of the $\phi_1$ mass. The purple regions are excluded by beam dump searches, E949 ($K^+ \to \pi^+ X$) and NA62($\pi^0 \to {\rm inv}$), the red region by SN1987, while the dashed line shows the upper bound from cosmology in the absence of any other light states or $\phi_1$-couplings. The star denotes the values of $g_{dd}$ and $m_{\phi_1}$ in Fig. \ref{fig:benchmark1} (Fig. \ref{fig:benchmark2}) for BM1 
(BM2). The region around $m_{\phi_1}\simeq m_{\pi^0}$ is masked out (gray region). 
  \label{fig:gddBM1} 
} 
 \end{figure} 

 \subsubsection{Combined analysis of $\phi_1$-constraints}
 \label{sec:constraints:combo}
 The most important constraint on the $\phi_1$-couplings comes from the neutrino burst duration observed in 
 the supernova SN1987A. The interactions of $\phi_1$ with matter inside an 
 exploding supernova are dominated by its couplings to nucleons. For $m_{\phi_1} = 1$ MeV, used in our benchmarks, the SN1987A observations exclude $g_{1NN}^{\rm eff}\equiv (\tilde g_{1NN}^2+(3/2) g_{1NN}^2)^{1/2}$ in the range $7 \cdot 10^{-10} \, \GeV^{-1} \lesssim g_{1 NN}^{\rm eff} \lesssim 4 \cdot  10^{-6} \, \GeV^{-1}$~\cite{Lee:2018lcj}. For larger values of $g_{1 NN}^{\rm eff}$ the $\phi_1$ gets trapped inside the proto-neutron star (PNS) and does not contribute to the cooling. This is the case for BM1, see Eqs. \eqref{eq:g1NN:BM1}, \eqref{eq:tildeg1NN:BM1}. For smaller values of $g_{1 NN}^{\rm eff}$ the emission of  $\phi_1$ is suppressed sufficiently that it again does not contribute appreciably to the cooling of PNS. BM2 falls in this regime, see Eqs. \eqref{eq:g1NN:BM2}, \eqref{eq:tildeg1NN:BM2}. 
 
 The photon couplings of $\phi_1$ are less relevant for SN1987A since the Primakoff emission of $\phi_1$ is always subdominant relative to the emission of $\phi_1$ in nucleon-nucleon scattering. This is best illustrated by the fact that SN1987A would exclude the range $10^{-8} \GeV^{-1}  \lesssim g_{1\gamma\gamma}, h_{1\gamma\gamma} \lesssim 10^{-5} \GeV^{-1}$, if $\phi_1$ were to 
 coupled to photons only. The induced couplings of $\phi_1$ to photons are at the lower edge of this range for BM1 and well below for BM2, cf. 
 Eqs.~\eqref{eq:g1gammagammaBM1} and  \eqref{eq:g1gammagammaBM2}  respectively. 
This should be contrasted with nucleon couplings which for BM1 traps $\phi_1$ inside the PNS as it is above 
 and not below the exclusion window.
 
 The constraints from the SN1987A neutrino burst duration are shown for a range of  ${\phi_1}$ masses for benchmarks BM1 and BM2 in Fig.~\ref{fig:gddBM1} (left) and (right) as red regions, respectively.  According to the analysis of Ref.~\cite{Lee:2018lcj}, the bounds are relevant all the way up to $m_{\phi_1}\lesssim 300$ MeV, though we truncate the plots at 200 MeV. These bounds may however depend on the details of the SN1987A explosion, and may even be absent if this was due to a  collapse-induced thermonuclear explosion \cite{Bar:2019ifz}. 
 
 In addition, Fig.~\ref{fig:gddBM1} shows with purple shading the constraints  from beam dump experiments (we use the combined limit as quoted in~\cite{Lee:2018lcj}), and from the invisible pion decay by NA62~\cite{NA62:2020}. The $\phi_1-\pi^0$ mixing angle $s_\theta$  needs to be smaller than about $2 \times 10^{-5}$ in order to satisfy the $K^+\to \pi^+ X$ constraints  from E949~\cite{Adler:2008zza} and NA62~\cite{NA62:2020}. This imposes a constraint on $g_{dd}$ that is comparable but slightly less stringent than the beam dump limit. 
 The upper bound  from cosmology, i.e., the impact of $\phi_1$ decays on big bang nucleosynthesis and distortions of cosmic microwave background, are shown with a dashed line \cite{Cadamuro:2011fd}. This bound is very sensitive to the details of the model. For instance, if the $\phi_1$ decays predominantly to neutrinos these bounds would be drastically  modified and thus potentially 
 irrelevant.
 \section{Model 2 - scalar model leading to the three-body kaon decays}
\label{sec:Model2}
\begin{figure}[t!]
\begin{center}
  \includegraphics[width=12cm]{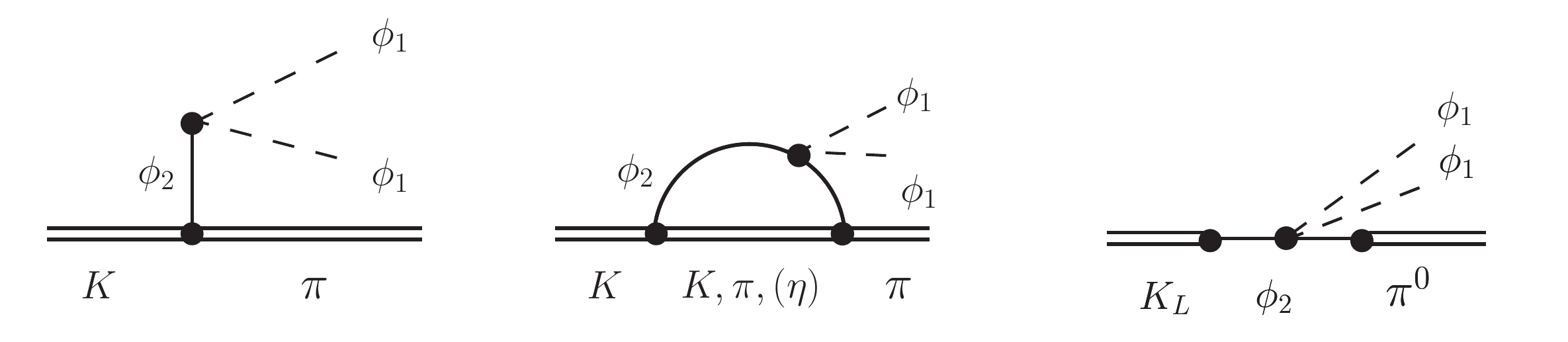} 
  \end{center}
\caption{\small The diagrams inducing the $K\to \pi \phi_1 \phi_1$ decays in Model 2,  
with the matrix elements shown in Eqs.~\eqref{eq:K_Lpi0:model2} and \eqref{eq:K+ampl:model2}.  The 3rd diagram violates  the GN bound.  
\label{fig:scalar2}}
\end{figure}

Model 2 has the same field content as Model 1, except that we impose a $Z_2$ symmetry under which the scalar $\phi_1$ is odd, $\phi_1\to -\phi_1$. The relevant terms in the Lagrangian are 
\beq
{\cal L}\supset g_{qq'}^{(2)} (\bar q_L q_R') \phi_2+\rm{h.c.} +\lambda_4 \phi_2^2 \phi_1^2+ \lambda' m_S \phi_2 \phi_1^2+\lambda'' m_S \phi_2^3 + \cdots.
\eeq
Note that the coupling $(\bar q_L q_R') \phi_1$ is forbidden by the $Z_2$-parity. 
Because of the $Z_2$ parity the $\phi_1$ always appears in pairs in the final state 
and  we thus focus on the  $K\to \pi \phi_1\phi_1$ transitions with leading  diagrams shown 
in Fig. \ref{fig:scalar2}. 

The 1st diagram in Fig. \ref{fig:scalar2},  proportional to the trilinear coupling $\lambda'$, gives 
the same contribution to both $K^+\to \pi^+\phi_1\phi_1$  and $K^0\to \pi^0 \phi_1\phi_1$ transitions 
in accordance with isospin. 
Since we are interested in  violations of the GN bound, we  impose the hierarchy
\beq
\label{eq:lambda}
\lambda', \lambda''\ll \lambda_4 \;, 
\eeq
and assume $m_S = {\mathcal O} (m_K)$. For simplicity we further assume that $\phi_{1,2}$ do not have vevs, or that they are negligibly small (cf. related discussion for Model 1 in Section \ref{sec:Model1}).

 Keeping the leading diagrams in the $\lambda'$ and $\lambda_4 g_{dd}^{(2)}$ expansion, i.e., the diagrams in Fig. \ref{fig:scalar2},  the $K_L\to \pi^0\phi_1\phi_1$ decay amplitude reads
\beq
\label{eq:K_Lpi0:model2}
\begin{split}
{\cal M}(K_L\to \pi^0 \phi_1& \phi_1)_{\rm NP}= i \biggr\{ 4 \Im\hat g_{sd}^{(2)} \Im  g_{dd}^{(2)} \lambda_4 
\Delta_{\phi_2}(m_K^2) \Delta_{\phi_2}(m_\pi^2) B_0 f_K f_\pi \\
&\qquad- 2  \Im\bar g_{sd}^{(2)} \lambda' m_S \Delta_{\phi_2}(q^2) - \frac{\Im \bar g_{sd}^{(2)}}{4 \pi^2} \lambda_4 \FLt(\tilde I)  B_0 \Big\}B_0 
\;,
\end{split}
\eeq
with $\FLt$ given in \eqref{eq:FL}, 
while  the $K^+\to \pi^+ \phi_1 \phi_1$ decay amplitude is
\beq
\label{eq:K+ampl:model2}
\begin{split}
{\cal M}(K^+\to \pi^+ \phi_1\phi_1)_{\rm NP}=  \Big\{&2 \bar g_{sd}^{(2)} \lambda' m_S \Delta_{\phi_2}(q^2)
+\frac{\bar g_{sd}^{(2)}}{4 \pi^2}\lambda_4 B_0 \Fpt(\tilde{I}) \Big\}B_0 
\;,
\end{split}
\eeq
with $\Fpt$ defined in \eqref{eq:Fpl}, $\tilde I(m_M)=C_0(m_K^2,q^2,m_\pi^2,m_M^2,m_{\phi_2}^2,m_{\phi_2}^2)$, 
and $q^2=(p_1+p_2)^2$ is the invariant mass squared of the $\phi_1\phi_1$ final state system. 
As for Model 1,  $f^2\to f_\pi f_K/2$ in order to account for the main SU(3) breaking effect.

The structure of the two decay amplitudes is reminiscent of the results in Model 1  in 
Eqs. \eqref{eq:K_Lpi0}, \eqref{eq:K+ampl}. The main difference is that there is no direct coupling of $\phi_1$ to quarks due to the $Z_2$ symmetry. The $\phi_1\phi_1$ pair couples to $d\to s$ current instead through the off-shell tree level exchange of $\phi_2$, see the 1st diagram in Fig. \ref{fig:scalar2}. This leads to isospin symmetric contributions to $K^+\to \pi^+ \phi_1\phi_1$ and $K_L\to \pi^0 \phi_1\phi_1$, proportional to the  trilinear $\lambda'$ coupling.  
Hence, in the $\lambda'\to 0$ limit,  the $K^+\to \pi^+ \phi_1\phi_1$ transition only receives loop
 contributions, and the GN bound is maximally violated. Note that $\lambda'$ cannot be arbitrarily small, since it is generated at one loop through $\phi_2$ loop, $\lambda'\sim \lambda_4 \lambda''/(16\pi^2)$, and at two loops with $\phi_2$ and $\pi^0,\eta$ running in the loop: $\lambda'\sim \lambda_4 (g_{dd}^{(2)})^3/(16\pi^2)^2$. For our benchmarks this gives a vanishingly small $\lambda'$ and thus this contribution can be safely ignored in our analysis provided the bare value of $\lambda', \lambda''$ are set to zero. In this limit the first isospin conserving contribution is at one loop due to the 2nd diagram in Fig. \ref{fig:scalar2}. The GN-violating contribution instead arises at tree level, see the 3rd diagram in Fig. \ref{fig:scalar2} and the first term in \eqref{eq:K_Lpi0:model2}.

The total rate of $K_L \to \pi^0 \phi_1 \phi_1$ adds coherently to the SM $K_L \to \pi^0 \nu\bar \nu$ rate. The differential rate for $K_L \to \pi^0 \phi_1 \phi_1$ is given by 
\begin{align}
\frac{d \Gamma}{d E_\pi} = \frac{\left| {\cal M}\right|^2}{128 \pi^3 m_K } p_\pi \beta_{\phi_1}  \;,
\end{align}
where  
$p_\pi = \sqrt{E_\pi^2 - m_\pi^2}$ and 
$E_\pi = (m_K^2 + m_\pi^2 - q^2)/(2m_K)$ are the pion's momentum and energy in the $K_L$ rest frame, 
while $\beta_{\phi_1} = (1 -  4 \mL^2/q^2)^{1/2} $.

\subsection{Benchmarks for Model 2}
\label{sec:BM:Model2}
\begin{figure}[t]
\begin{center}
\includegraphics[width=7cm]{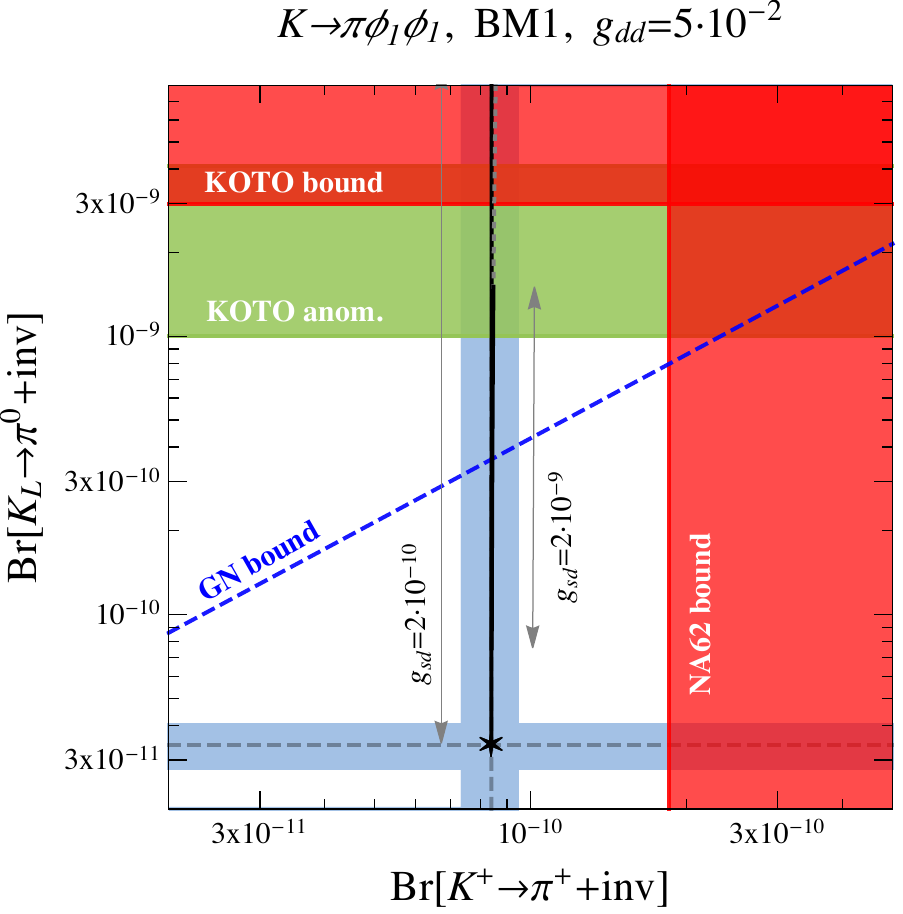}~~~~
\includegraphics[width=7cm]{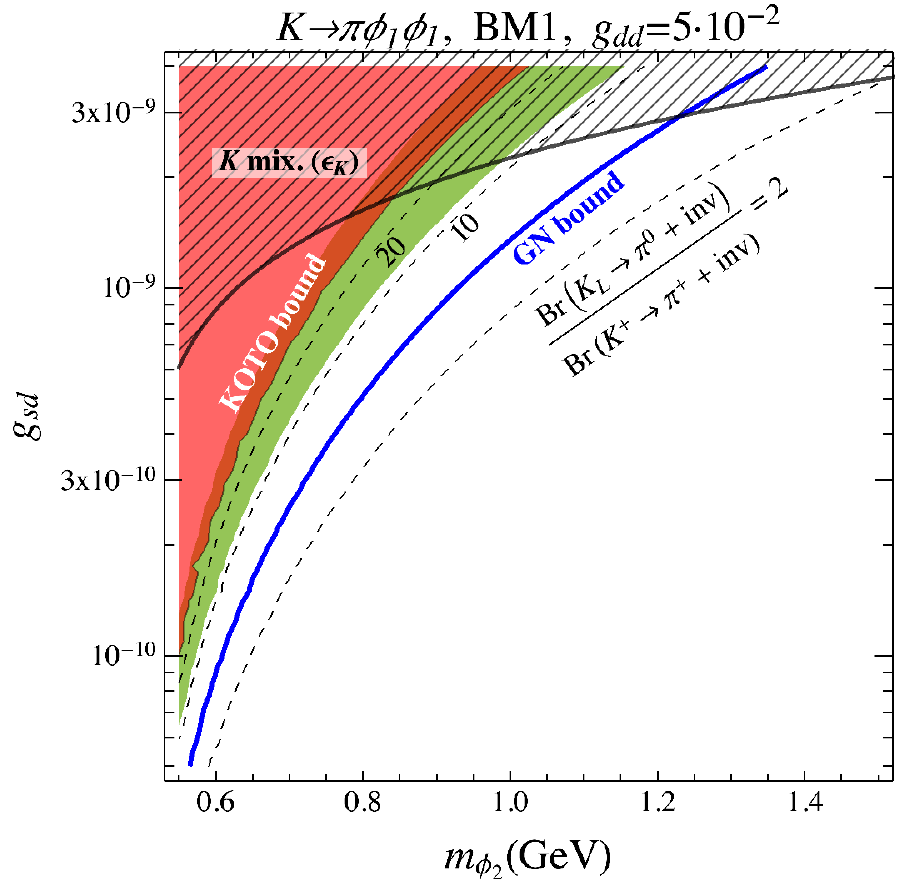}
\end{center}
 \caption{ \label{fig:Model2:BM1} The parameter space for Model 2, BM1, Eq. \eqref{eq:BM1:Model2}.
 The color coding is the same as in Fig. \ref{fig:benchmark1}. In the predictions for $\Br(K^+\to \pi^++{\rm inv})$, $\Br(K_L\to \pi^0+{\rm inv})$ on the left plot (black lines) we vary $m_{\phi_2}\in[0.55,1.5]$ GeV for two values of $g_{sd}=2\times 10^{-10}, 2\times 10^{-9}$.}
 \end{figure}

\begin{figure}[t]
 \begin{center}
\includegraphics[width=7cm]{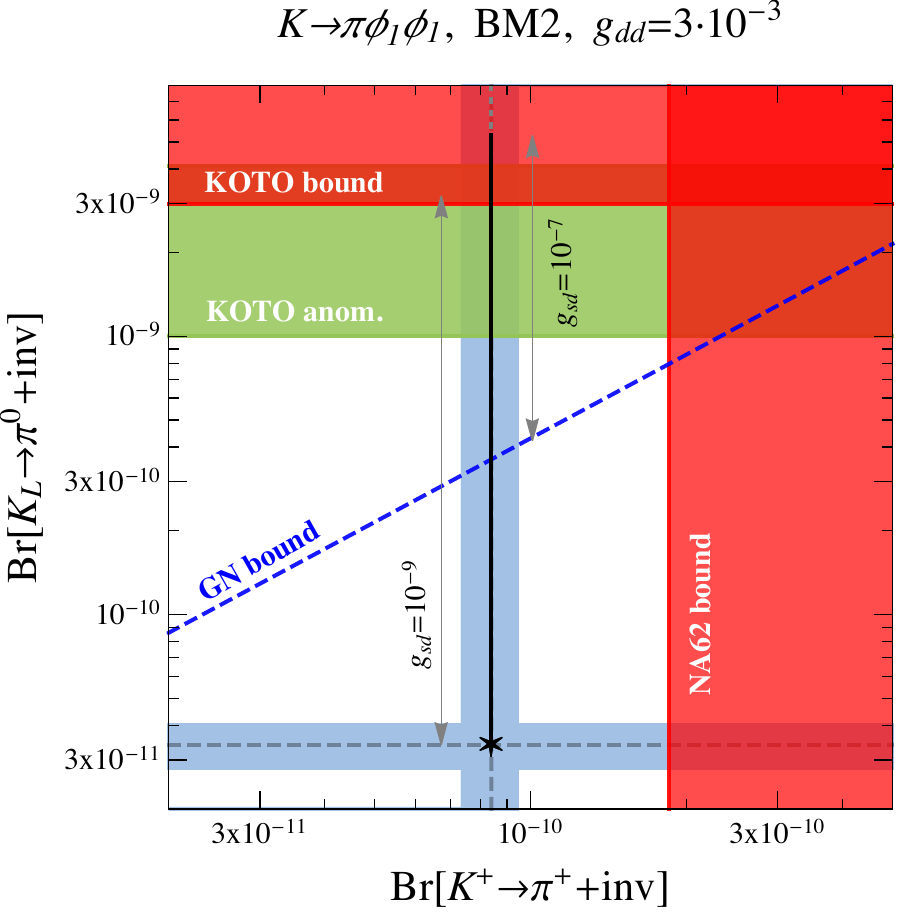}~~~~ 
\includegraphics[width=7cm]{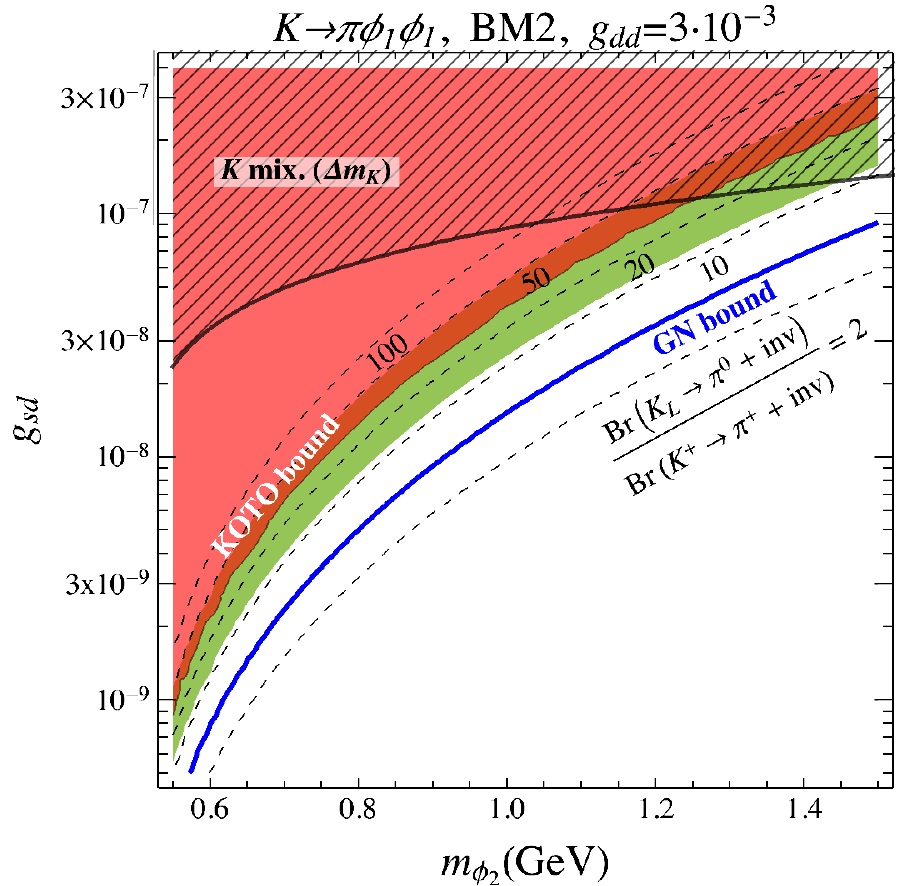}
\end{center}
 \caption{ \label{fig:Model2:BM2} The preferred parameter space for Model 2, BM2. The color coding is the same as in Fig. \ref{fig:Model2:BM1}. For the model predictions (black lines) in the left panel we set $g_{sd}=10^{-9}, 10^{-7}$ and vary $m_{\phi_2}\in [0.55,1.5]$ GeV.}
 \end{figure}

The bounds from $K^0-\bar K^0$  mixing on flavor violating $\phi_2$-coupling $\hat g_{sd}^{(2)}$ are exactly the same as for Model 1, Section \ref{sec:model1:KKbar}. To illustrate the available parameter space we therefore use the same two benchmarks for the $\phi_2$-couplings, Eqs. \eqref{eq:BM1}, \eqref{eq:BM2}, with  results shown in Figs. \ref{fig:Model2:BM1}, \ref{fig:Model2:BM2}. In both cases we set $\lambda_4=1$ and all the other couplings, apart from the ones  in Eqs. (\ref{eq:BM1}, \ref{eq:BM2}), to zero (including $\lambda'$). 
In summary, the two benchmarks for Model 2 are thus
\begin{align}
\label{eq:BM1:Model2}
&{\rm\bf Model~2,~BM~1:}&\text{Eq. \eqref{eq:BM1} and~}m_{\phi_1}=100{\rm~MeV}, \lambda_4=1,\, \lambda'=\lambda''=0\;,
\\
\label{eq:BM2:Model2}
&{\rm\bf Model~2,~BM~2:}&\text{Eq. \eqref{eq:BM2} and~}m_{\phi_1}=100{\rm~MeV}, \lambda_4=1,\, \lambda'=\lambda''=0\;,
\end{align}
while $m_{\phi_2}$ is kept as a free parameter.  The results in Figs. \ref{fig:Model2:BM1}, \ref{fig:Model2:BM2} are fairly independent of the  $\phi_1$ mass as long as it is taken to be small, $m_{\phi_1}\ll m_K$, and thus does not modify the final phase space. The choice of benchmark value  $m_{\phi_1}=100{\rm~MeV}$ is driven by the constraints of the invisible pion decays, see Section \ref{sec:phi1:Model2}. BM1 and BM2 thus have three free parameters: $g_{dd}, g_{sd}$ and $m_{\phi_2}$. 

BM1, shown in Fig.~\ref{fig:Model2:BM1}, has a well restricted $\{g_{sd}, m_{\phi_2}\}$ parameter space, since the tree level exchanges of $\phi_2$ contributes a new CP violating source in $K^0-\bar K^0$  mixing. This then restricts $g_{sd}$ to be below the hatched region in Fig. \ref{fig:Model2:BM1} (right), see also Eq. \eqref{eq:gds:epsilonK}. However, large enhancements of $\Br(K_L\to \pi^0+{\rm inv} )$ over $\Br(K^+ \to \pi^++{\rm inv} )$ are still possible in significant parts of the parameter space. For instance, setting $g_{dd}=5\cdot 10^{-2}$, the KOTO upper bound $\Br(K_L \to \pi^0 \nu \bar \nu)_{\rm exp}<3.0 \times  10^{-9}$ \cite{Ahn:2018mvc}   (red region in Fig. \ref{fig:Model2:BM1}) are obtained for $g_{sd}\lesssim {\mathcal O}(10^{-9})$ and $m_{\phi_2}\lesssim {\mathcal O}(1{\rm~GeV})$.  Fig. \ref{fig:Model2:BM1} (left) shows that in the relevant region of  parameter space the deviations in $\Br(K^+ \to \pi^++{\rm inv} )$ from the SM prediction are negligible, while $\Br(K_L\to \pi^0+{\rm inv} )$ can be enhanced well above the GN bound (blue line). For illustration we vary $m_{\phi_2}\in [0.55, 1.5]$ GeV, the same range as is shown in Fig. \ref{fig:Model2:BM1} (right), fix $g_{dd}=5\cdot 10^{-2}$ and show predictions for two choices of $g_{sd}=2\cdot 10^{-10}, 2\cdot 10^{-9}$ (black lines). The resulting range in $\Br(K_L\to \pi^0+{\rm inv} )$ is denoted with arrows. For $g_{sd}=2\cdot 10^{-9}$  the $\epsilon_K$ bound is reached, and the exclusion range is shown with gray dotted lines.
 For both choices of $g_{sd}$ the enhancements can  easily be in the range of the KOTO anomaly (green band) without violating any other bounds.

 For BM2 the allowed $\{g_{sd}, m_{\phi_2}\}$ parameter space is much larger, since in this case $\phi_2$ exchanges only induce CP conserving contributions to $K^0-\bar K^0$  mixing. This gives the bound in \eqref{eq:gds:DeltamK}, denoted in Fig. \ref{fig:Model2:BM2} (right) with the hatched region. Very large violations of the GN bound (blue line) are thus possible without violating $K^0-\bar K^0$  mixing constraints. For instance, for $g_{dd}=3\cdot 10^{-3}$ the KOTO upper bound is reached for $g_{sd}\sim$ few$\times 10^{-8}$ and $m_{\phi_2}\sim1$ GeV. Fig. \ref{fig:Model2:BM2} (left) shows predictions for $\Br(K^+\to \pi^++{\rm inv})$, $\Br(K_L\to \pi^0+{\rm inv})$, for two choices of $g_{sd}=10^{-9}, 10^{-7}$ (black lines, gray dotted line excluded by $K^0-\bar K^0$  mixing) when varying $m_{\phi_2}\in [0.55,1.5]$ GeV, setting $g_{dd}=3\cdot 10^{-3}$. The deviations in $\Br(K^+\to \pi^++{\rm inv})=1.0\cdot 10^{-8}$ vanish in BM2, while over a large region of $\{g_{dd}, g_{sd}, m_{\phi_2}\}$ the KOTO upper limits on $\Br(K_L\to \pi^0+{\rm inv})$ are saturated, while avoiding any other constraints. 

\subsection{Constraints on the $\phi_1$-couplings}
\label{sec:phi1:Model2}

   The most stringent constraints on the couplings of $\phi_1$ to quarks are due to the invisible $\pi^0$ decay. 
  In the $m_{\phi_2} \gg m_\pi$ limit, the $\pi^0\to \phi_1\phi_1$ decay amplitude is  given by, 
\begin{align}
{\cal M} (\pi^0 \to \phi_1 \phi_1) & =\frac{  B_0 f }{m_{\phi_2}^2}\biggr[2 \lambda^\prime m_S  \Im g_{dd}^{(2)} +  \frac{  \lambda_4 B_0 }{6 \pi^2} \Re g_{dd}^{(2)} \left( 2 \Im g_{dd}^{(2)} - \Im g_{ss}^{(2)} \right) 
\biggr] \, ,
\end{align}
where the first term in the parenthesis  originates from the tree level  exchange of the $\phi_2$, Fig.~\ref{fig:pi0gagaM1} (middle), the second from the one loop contribution shown in Fig.~\ref{fig:pi0gagaM1} (right).

Using Eq. \eqref{eq:pi0phi1phi1:dec} for the $\pi^0\to \phi_1\phi_1$ partial decay width, the corresponding branching ratio in the $m_{\phi_1} \ll m_\pi \ll m_{\phi_2}$ limit are, setting $g_{ss}^{(2)}=0$, 
\beq
\Br  (\pi^0 \to \phi_1 \phi_1)  = 
 \left\{
 \begin{array}{ll}
 3.8 \times 10^{-3} \, \lambda_4^2 \left( \frac{\Re g_{dd}^{(2)}  }{3 \times  10^{-2}} \right)^2 \left( \frac{\Im g_{dd}^{(2)} }{3 \times  10^{-2}} \right)^2 \left( \frac{ \GeV}{m_{\phi_2}} \right)^4, &{\rm~for~}\lambda^\prime=0\;, \, 
  \\
2.1 \times 10^{-7} \left( \frac{ \lambda^\prime m_S }{10^{-5} \, \GeV} \right)^2 \left( \frac{\Im g_{dd}^{(2)} }{3 \times  10^{-2}} \right)^2 \left( \frac{ \GeV}{m_{\phi_2}} \right)^4, &{\rm~for~}\lambda_4=0\;.
  \end{array}
  \right.
\eeq
These should be compared with the experimental bound $\Br (\pi^0 \to \phi_1 \phi_1) < 4.4 \times 10^{-9}$~\cite{NA62:2020}.
For $\lambda_4\sim {\mathcal O}(1)$, which is required for large violations of the GN bound, this excludes $\phi_1$ masses $m_{\phi_1}\lesssim m_{\pi}/2$ for BM1. In BM2 the $\pi^0\to \phi_1\phi_1$ is forbidden due to parity, so that $\phi_1$ can be light, as long as the parity breaking term  $\lambda' m_S$ is sufficiently small.

The beam dump and SN constraints in BM1 and BM2 are very similar to the ones shown in Fig. \ref{fig:gddBM1}  for Model 1, but with rough identification $g_{dd}\big|_{\rm Model~1} \to  1/(4\pi) \times g_{dd}\big|_{\rm Model~2}$ and $m_{\phi_1}\big|_{\rm Model~1} \to  2 m_{\phi_1}\big|_{\rm Model~2}$, since the transitions now involve two $\phi_1$ particles in the final state. In particular, for the choices of parameters in Figs. \ref{fig:Model2:BM1} and \ref{fig:Model2:BM2} the collider and SN constraints are presumably satisfied. 

\subsection{$\phi_1$ as a dark matter candidate}
Since $\phi_1$ is  odd under the $Z_2$-parity, 
it is absolutely stable and could be a dark matter  (DM) candidate. If $m_{\phi_1}>m_{\pi^0}$ the 
$\phi_1\phi_1\to \pi^0\pi^0$ annihilation  channel is open. 
Below we shall see that 
the annihilation cross section is large enough, in part of the parameter space, such  that the correct DM relic abundance is obtained. Note however, that this restricts $\phi_1$ to a rather narrow mass range, $m_{\pi^0}\leq m_{\phi_1}\leq (m_{K_L}-m_{\pi^0})/2$, or numerically, $135~{\rm MeV}\lesssim m_{\phi_1}\lesssim 181$ MeV.

\begin{figure}[t!]
\begin{center}
  \includegraphics[width=15.4cm]{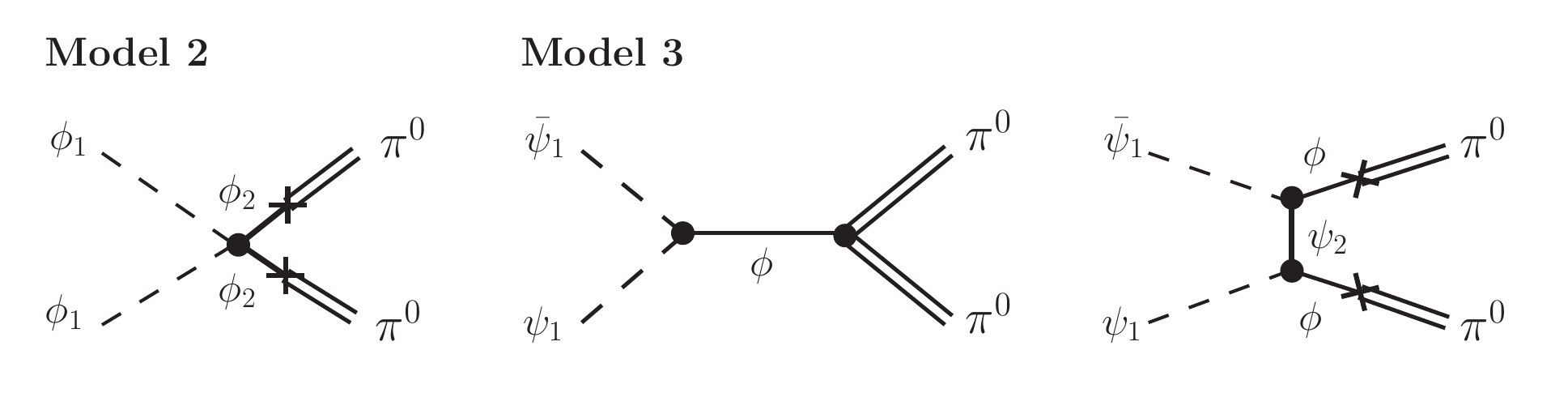} 
  \end{center}
  \vspace{-0.6cm}
\caption{\small Graph dominating the annihilation cross section of $\phi_1 \phi_1 \to \pi^0 \pi^0$ and 
$ \psi_1 \bar \psi_1 \to \pi^0 \pi^0$ in Model 2 and Model 3 respectively. 
\label{fig:annihilation}
}
\end{figure}

The annihilation cross section for $\phi_1\phi_1\to \pi^0\pi^0$ process 
is dominated by the $\lambda_4$ vertex and $\phi_2-\pi^0$-conversion, while the 
$\phi_2$ $s$-channel resonance contribution is subleading, since $\lambda' \ll \lambda_4$, cf. \eqref{eq:lambda} and Fig.~\ref{fig:annihilation}. Assuming a non-relativistic $\phi_1$, as is the case at the time of  freeze-out, the leading thermally averaged
cross section is given by
\beq
\langle \sigma v\rangle =
\frac{1}{16\pi} \lambda_4^2 \big(\Im g_{dd}^{(2)}\big)^4 \biggr(\frac{B_0 f_\pi}{m_{\phi_2}^2}\biggr)^4 \frac{p_\pi}{m_{\phi_1}^3}\;,
\eeq
where in this approximation $p_\pi = (m_{\psi_1}^2- m_\pi^2)^{1/2}$.
Taking $m_{\phi_1}=160$ MeV as a representative value gives 
 \beq
\langle \sigma v\rangle\simeq 3 \cdot 10^{-26}\frac{{\rm cm}^3}{{\rm s}}\times\lambda_4^2 \, \biggr(\frac{\Im{g_{dd}^{(2)}}}{2.56\times 10^{-2}}\biggr)^4\, \biggr(\frac{1{\rm~GeV}}{m_{\phi_2}}\biggr)^4\;,
\eeq
which is of the right size to get the correct DM relic abundance ($3 \cdot 10^{-26} {\rm cm}^3/{\rm s} \approx 1 {\rm \, pb} $). 

For $m_{\phi_1}<m_{\pi^0}$ the $\phi_1\phi_1\to \pi^0\pi^0$ annihilation cross section is kinematically forbidden. In that case the dominant annihilation channel becomes $\phi_1\phi_1\to \gamma\gamma$. The resulting annihilation cross section is so small, that if this were the only annihilation channel, the $\phi_1$ would 
overclose the universe \cite{Bertone:2004pz}. This means that $\phi_1$ should also couple to other light states. For instance, $\phi_1$ could annihilate into light SM particles, e.g. $\phi_1\phi_1\to e^+ e^-$ or $\phi_1\phi_1\to \nu\bar \nu$. 
Alternatively it could annihilate away to other light dark sector particles or dark photons $\phi_1\phi_1\to \gamma_D \gamma_D$
(if $\phi_1$ was gauged under a dark $U(1)$). Since none of these couplings are related to $K\to \pi \phi_1\phi_1$ decays we do not explore the related phenomenology 
any further, beyond stating the obvious -- that $\phi_1$ could well be a thermal relic for appropriate values of these additional couplings.

\section{Model 3 - light dark sector fermions}
\label{sec:Model3}

In this model we introduce a real scalar, $\phi$, of mass $m_\phi$, and two Dirac fermions, $\psi_1, \psi_2$,  with masses $m_{\psi_1, \psi_2}$, where the couplings relevant for the $K\to \pi$+inv decay are 
\beq
{\cal L}\supset  g^{(\phi)}_{qq'} (\bar q_L q_R') \phi + y_{ij} \phi \bar \psi_{L,i} \psi_{R,j} + {\rm h.c.} \; .
\eeq
The fermion $\psi_2$  is massive enough such that the  decays of $K\to \pi \psi_2 \bar \psi_2$ and $K\to \pi \psi_1 \bar \psi_2$  are kinematically forbidden. 
In contrast and crucially, the decay $K\to \pi \psi_1 \bar \psi_1$  is assumed to be kinematically allowed. 
 The couplings of $\phi$ to the quarks are assumed to have a hierarchical flavor structure
\beq
\label{eq:g}
g^{(\phi)}_{sd,ds}\ll g^{(\phi)}_{dd,ss}   \;,
\eeq
reflecting the suppression of flavor changing neutral currents of the SM,  
whereas the Yukawa couplings of $\phi$ to $\psi_{1,2}$ are assumed to favor off-diagonal transitions,
\beq
\label{eq:y}
y_{11,22}  \ll  y_{12,21}  \;.
\eeq
\begin{figure}[t!]
\begin{center}
  \includegraphics[width=14cm]{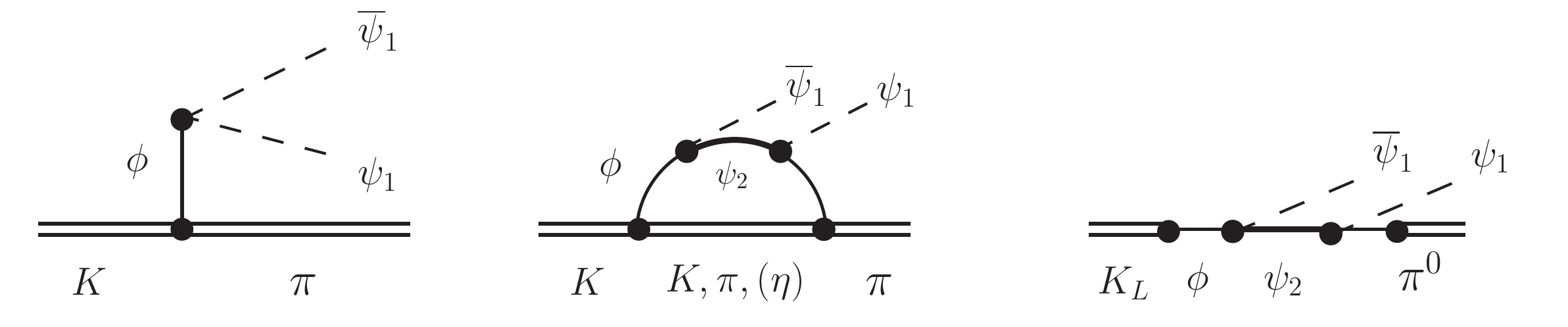} 
  \end{center}
\caption{\small The $K \to \pi \psi_1\bar   \psi_1$ in the fermion Model 3 with contribution evaluated in Eqs.~\eqref{eq:K_Lpi0:model2} and \eqref{eq:K+ampl:model3} respectively. 
The third diagram only contributes to $K_L \to \pi^0 \psi_1 \bar \psi_1$, as the notation suggests, 
and is therefore responsible for potential violation of the GN bound in Model 3. 
\label{fig:fermion}}
\end{figure}
While we do not attempt to build a full flavor model we remark in passing that such flavor structures can 
easily be realised within  Froggatt-Nielsen (FN) type models \cite{Froggatt:1978nt}. Choosing for instance the $U(1)_{\rm FN}$  charges  to be $[\psi_{L,2}]=[\psi_{R,2}]=0$, $[\psi_{L,1}]=-[\psi_{R,1}]=[\phi]=1$
and with $\epsilon=\langle \phi_{\rm FN}\rangle/M_{\rm FN}$  the FN spurion carrying the charge $[\epsilon]=-1$,  the Yukawa and mass matrices take the form
\beq
y_{ij}\sim 
\begin{pmatrix}
\epsilon^* & 1
\\
1 & \epsilon
\end{pmatrix} \;, \quad 
{\cal M}_\psi \sim m_0
\begin{pmatrix}
\epsilon^2 & \epsilon
\\
\epsilon & 1
\end{pmatrix} \;, 
\eeq
where the ``$\sim$'' sign denotes equality up to ${\mathcal O}(1)$ factors. Similarly, if $[d_{L,R}]\gg [s_{R,L}]$, the $g^{(\phi)}_{sd,ds}$ can be arbitrarily suppressed in accordance with \eqref{eq:g}.

Keeping the leading diagrams in  $y_{ij}$ and $g^{(\phi)}_{dd}$, shown 
in Fig~\ref{fig:fermion}, gives  the following $K_L\to \pi^0  \psi_1\bar \psi_1$ decay amplitude
\begin{alignat}{1}
\label{eq:K_Lpi0:model3}
 {\cal M}(K_L &\to \pi^0  \psi_1\bar\psi_1)_{\rm NP} =
 \nonumber \\ 
  - i \biggr\{&
 \Im \hat{g}^{(\phi)}_{sd}  \Im {g}^{(\phi)}_{dd}  \Delta_{\phi}(m_K^2) \Delta_{\phi}(m_\pi^2)  \Big[  m_{\psi_2} y_{12} y_{21} \big(\bar u P_R v\big) \, \bar \Delta +
 |y_{12}|^2 \big( \bar  u \gamma_\mu P_L v \big)  \bar \Delta^\mu \Big]  B_0 f_K f_\pi \,  
 \nonumber  \\[0.1cm]
-&   \frac{\Im \bar g^{(\phi)}_{sd}}{16 \pi^2}  \Big[  m_{\psi_2} y_{12} y_{21} \big(\bar u P_R v\big) \, 
 \FL^{(\phi)}( \bar{I}_4) +   |y_{12}|^2 \big(\bar  u \gamma_\mu P_L v \big)
 \FL^{(\phi)}(\bar I^\mu_4)  \Big]  B_0 \,  \nonumber  \\[0.1cm]
+&  \Im \bar g^{(\phi)}_{sd}  \,
y_{11} \big(\bar u P_R v\big)    \Delta_{\phi}(q^2) \biggl\}B_0  + 
\big\{y_{ij},\gamma_5  \leftrightarrow y^*_{ji}, - \gamma_5 \big\} 
\;, 
\end{alignat}
where $2 P_{R,L} \equiv 1 \pm \gamma_5$, 
we have shortened $\bar u\equiv\bar u(\bar p)$, $v\equiv v(p)$,  
while  $q^2 = (p + \bar{p})^2$  is the invariant four momentum of the  fermion pair. The 
$\bar \Delta^{(\mu)}$  stands for combinations of fermion propagators  
\begin{alignat}{2}
\label{eq:barDelta}
&\bar{\Delta}  &\;=\;&  \Big[ \Delta_{\psi_2}( 
(p_\pi \!+\!\bar p)^2)  + \{\bar p \leftrightarrow   p \} \Big] \;, \nonumber  \\[0.1cm]
&\bar{\Delta}^\mu 
&\;=\; &   \Big[ (p_\pi + \bar p)^\mu \Delta_{\psi_2}( 
(p_\pi \!+\!\bar p)^2)  - \{\bar p \leftrightarrow   p \}   \Big] \;,
\end{alignat}
where $\Delta_X(k^2)$ is defined below \eqref{eq:K_Lpi0}, 
while   $\FL^{(\phi)}(Z)= \FLt(Z)|_{g^{(2)}_{qq} \to g^{(\phi)}_{qq}}$, with the latter defined in \eqref{eq:FL}. Its arguments are given in terms of loop integrals, 
 \begin{align}
  \bar{I}_4(m_M)  
  &\;=\; I_4 (m_M,\bar p)  + \big\{\bar p \to   p \big\}  \;, \nonumber  \\
   \bar{I}^\mu _4(m_M) &\;=\;         \big\{ I^\mu_4(m_M,\bar p) + ( \bar p + p_\pi)^\mu I_4(m_M,\bar p) \big\}
 -   \big\{\bar p \to   p \big\} \;,
 \end{align}
 where
 $I_4(m_M,P) \equiv D_0(m_\pi^2,(p_\pi \!+\!P)^2, (p_K \!-\! p_\pi \!-\! P)^2, (p_K\!-\! p_\pi )^2,P^2, m_K^2,
m_\phi^2 , m_M^2, m_{\psi_2}^2, m_\phi^2)  $  (cf.~Appendix~\ref{app:PV})
and $I^\mu_4$ is the same integral with an additional 
Lorentz-vector $k^\mu$ in the integrand.

The decay amplitude $K^+\to \pi^+   \psi_1\bar\psi_1$ is analogous, but without the 3rd diagram in 
Fig.~\ref{fig:fermion}. This gives 
\beq
\begin{split}
\label{eq:K+ampl:model3}
 {\cal M}(K^+\to \pi^+  \psi_1\bar\psi_1)_{\rm NP} &=  
 \\ 
- \biggr\{   \frac{ \bar g^{(\phi)}_{sd} }{16 \pi^2} & \Big[  m_{\psi_2} y_{12} y_{21} \big(\bar u P_R v\big) \, 
 \Fp^{(\phi)}( \bar{I}_4) +   |y_{12}|^2 \big(\bar  u \gamma_\mu P_L v\big)  \Fp^{(\phi)}(\bar I^\mu_4) \Big]  B_0 \,    \\[0.1cm]
&  - \bar{g}^{(\phi)}_{sd} 
y_{11} \big(\bar u P_R v \big) \Delta_\phi(q^2)  \biggr\} B_0 \! +\! 
\big\{y_{ij},\gamma_5  \leftrightarrow y^*_{ji}, - \gamma_5 \big\}  
\;,  
\end{split}
\eeq
where $\Fp^{(\phi)}(Z)= \Fpt(Z)|_{g^{(2)}_{qq} \to g^{(\phi)}_{qq}}$ with the later defined in \eqref{eq:Fpl}. A formula for the rate, in differential form, is given in Appendix \ref{app:diff-rate}. 

\subsection{Benchmarks for Model 3}
\label{sec:BM:Model3}

The new elements of Model 3 are the Yukawa couplings between $\phi$ and $\psi_{1,2}$, as well as the absence of the light-scalar $\phi_1$. 
In order to ease comparisons with  Model~1 and Model~2, we use $g^{(2)}_{qq'} \to g^{(\phi)}_{qq'}$ 
Eqs. \eqref{eq:BM1}, \eqref{eq:BM2}
\begin{alignat}{4}
\label{eq:Model3:BM1}
&{\rm\bf Model~3,~BM~1:} \quad  g_{dd}^{(\phi)}&\;=\;&\tfrac{(1+i)}{\sqrt2} g_{dd}\;, \quad  & & \bar g_{sd}^{(\phi)}\;=\; 
\hat g_{sd}^{(\phi)}=\tfrac{(1+i)}{\sqrt2} g_{sd}, && \quad  y_{12}=y_{21}=1 \;,
\\
\label{eq:Model3:BM2}
&{\rm\bf Model~3,~BM~2:} \quad g_{dd}^{(\phi)} &\;=\;& i g_{dd} \;, \quad  &  & \bar g_{sd}^{(\phi)}\;=\;0\;, \quad  \hat g_{sd}^{(\phi)}=i g_{sd}, && \quad  y_{12}=y_{21}=1\;,
\end{alignat}
while all the other couplings are set to zero. In particular, the only nonzero Yukawa couplings of $\psi_i$ fermions for $\phi$ are the flavor violating ones, $y_{12,21}$, while the diagonal ones are assumed to be vanishingly small, and set to $y_{11,22}=0$. The mass of the lightest fermion is set to $m_{\psi_1}=100$~MeV.  The benchmarks are thus described by four continuous variables: the masses $m_{\phi}, m_{\psi_2}$ and the real 
parameters $g_{dd}, g_{sd}$.
\begin{figure}[t]
\begin{center}
\includegraphics[width=5cm]{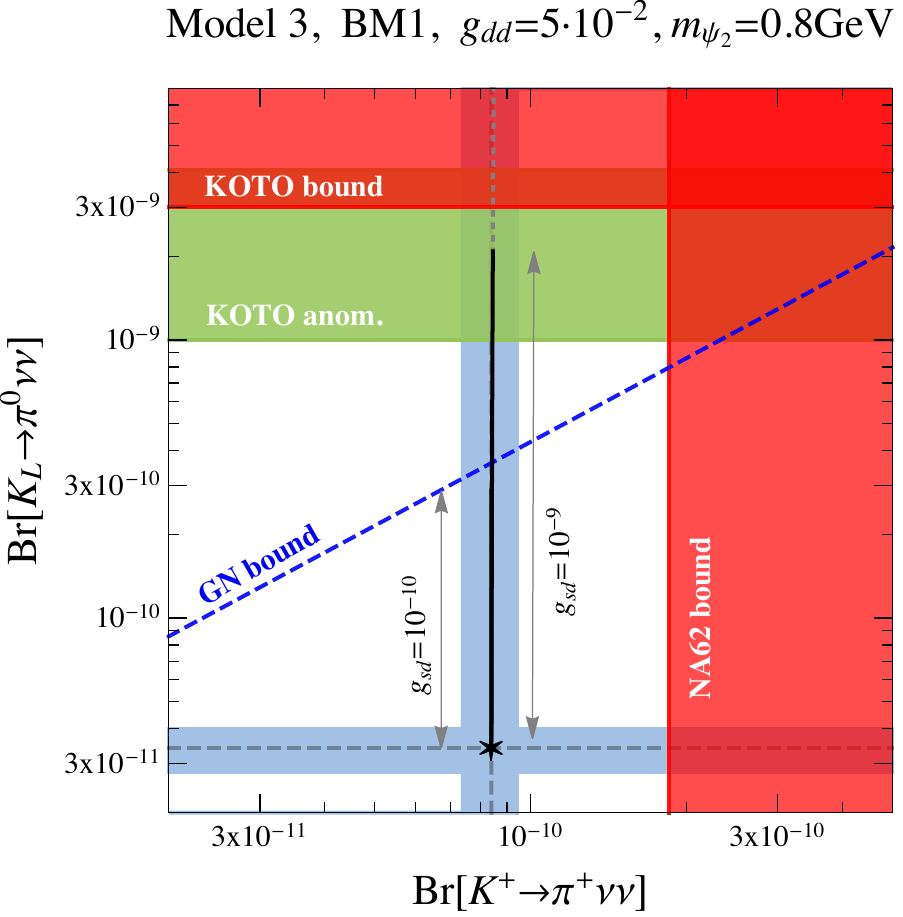}~
\includegraphics[width=5cm]{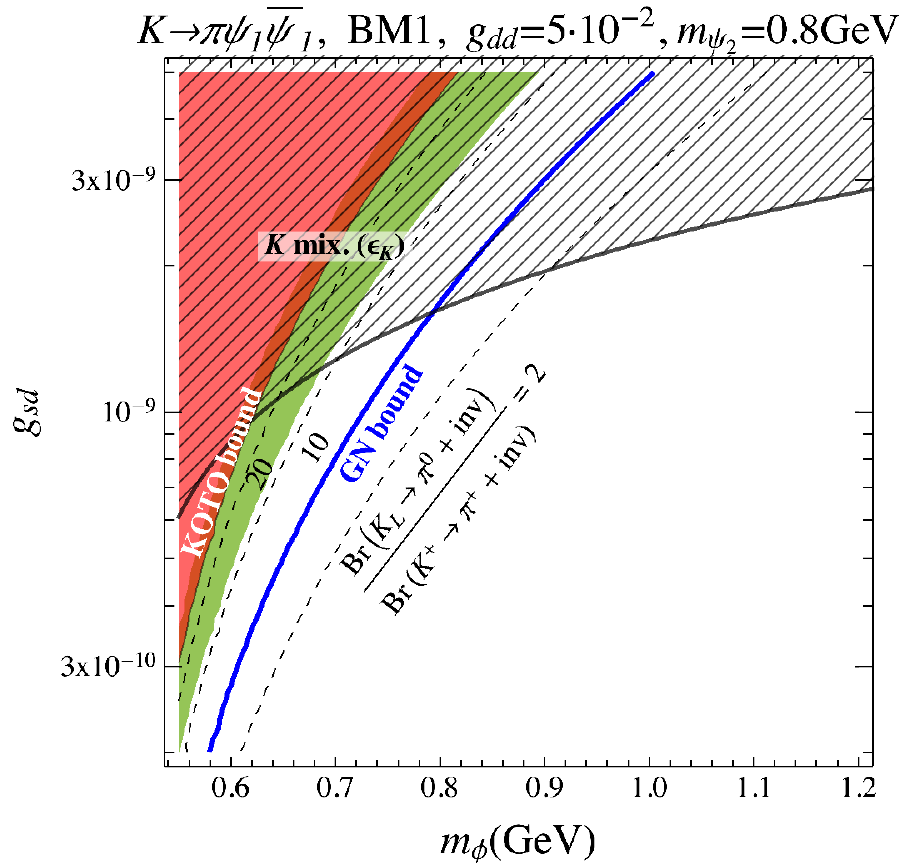}~
\includegraphics[width=5cm]{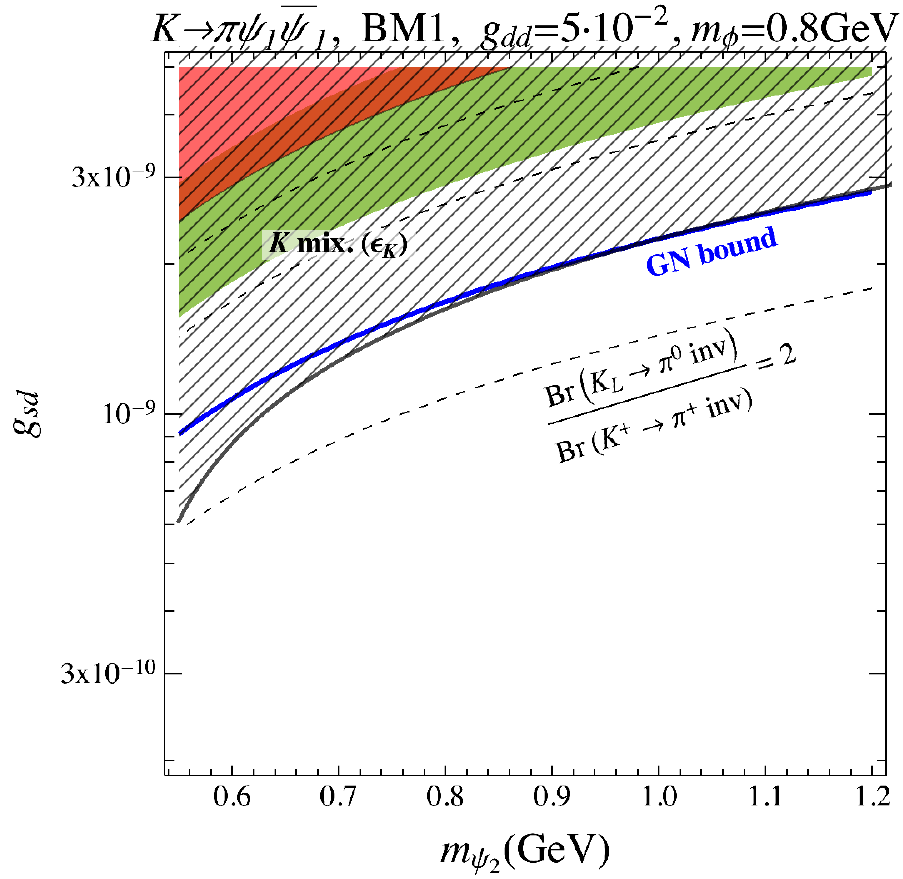}
\end{center}
 \caption{ \label{fig:Model3:BM1} 
 The parameter space for Model 3, BM1, Eq. \eqref{eq:Model3:BM1}.
 The color coding is the same as in Fig. \ref{fig:benchmark1}. In the predictions for $\Br(K^+\to \pi^++{\rm inv})$, $\Br(K_L\to \pi^0+{\rm inv})$ in the left plot (black lines) we vary $m_{\phi}\in[0.55,1.2]$ GeV for two values of $g_{sd}=10^{-10}, 10^{-9}$ and fix $m_{\psi_2}=0.8$ GeV. Right (middle) panels show the parameter space as functions of $m_{\psi_2} (m_{\phi})$, fixing $m_{\phi(\psi_2)}=0.8$~GeV. }
 \end{figure} 
 
 The flavor violating coupling,  $\hat g_{sd}^{(\phi)}$, is constrained by $K^0-\bar K^0$  mixing. The bounds are the same as for $\phi_2$ in  Model 1, Section \ref{sec:model1:KKbar}, and are thus obtained from   Eqs. \eqref{eq:gds:epsilonK}, \eqref{eq:gds:DeltamK} through the $\hat g_{sd}^{(2)}\to \hat g_{sd}^{(\phi)}$, $m_{\phi_2}\to m_\phi$ replacements. BM1 is severely constrained by $\epsilon_K$ since tree level  
 exchange of $\phi$ induces  a new CP violating contribution to $K^0-\bar K^0$ mixing. Fig. \ref{fig:Model3:BM1} (middle) and (right) show that large enhancements of $\Br(K_L\to \pi^0+{\rm inv})/\Br(K^+\to \pi^++{\rm inv})$ are possible only for small values of $m_{\phi}$ and $m_{\psi_2}$, comparable to the kaon mass. Still, such light NP states are not excluded experimentally and can saturate the present KOTO bound. Fig. \ref{fig:Model3:BM1} (left) shows that in this regime it is possible to have values for this ratio well above the GN bound, 
 in the range of the anomalous KOTO  events (green band). 
 
 \begin{figure}[t]
\begin{center}
\includegraphics[width=5cm]{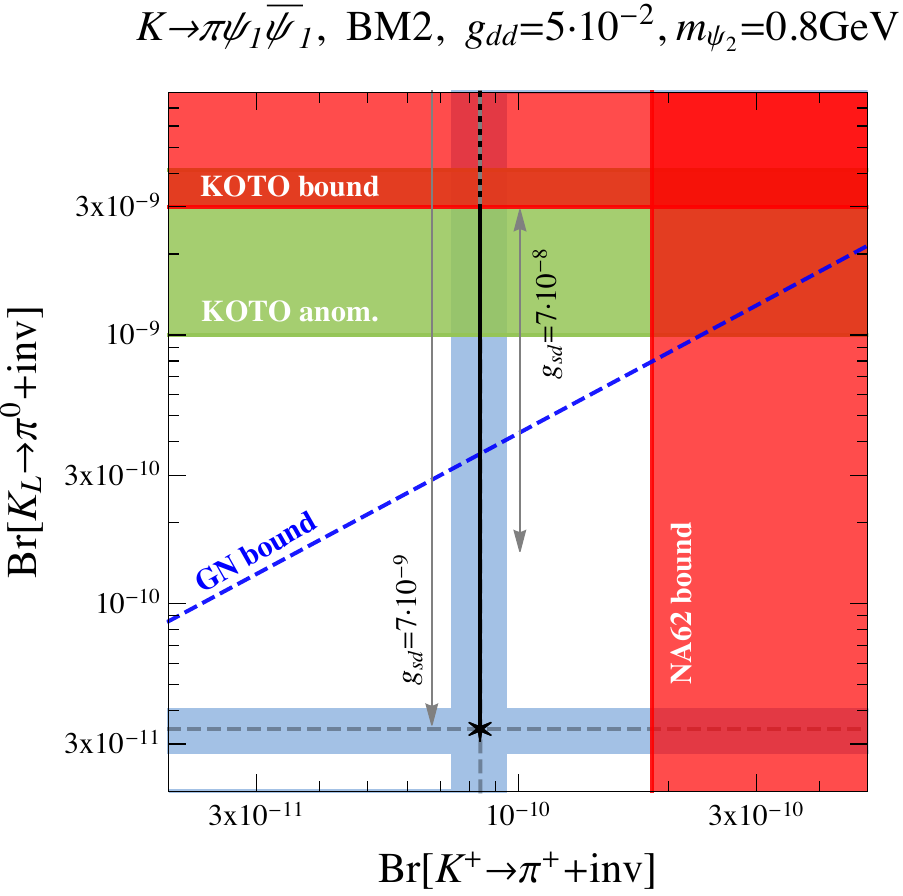}~
\includegraphics[width=5cm]{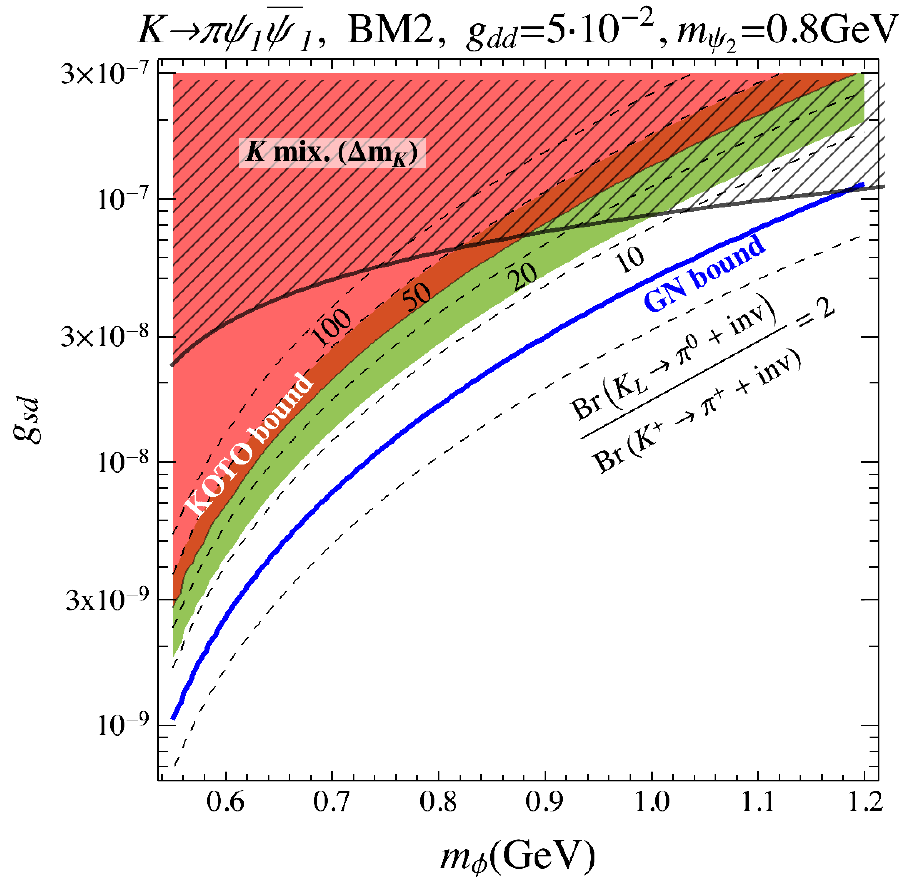}~
\includegraphics[width=5cm]{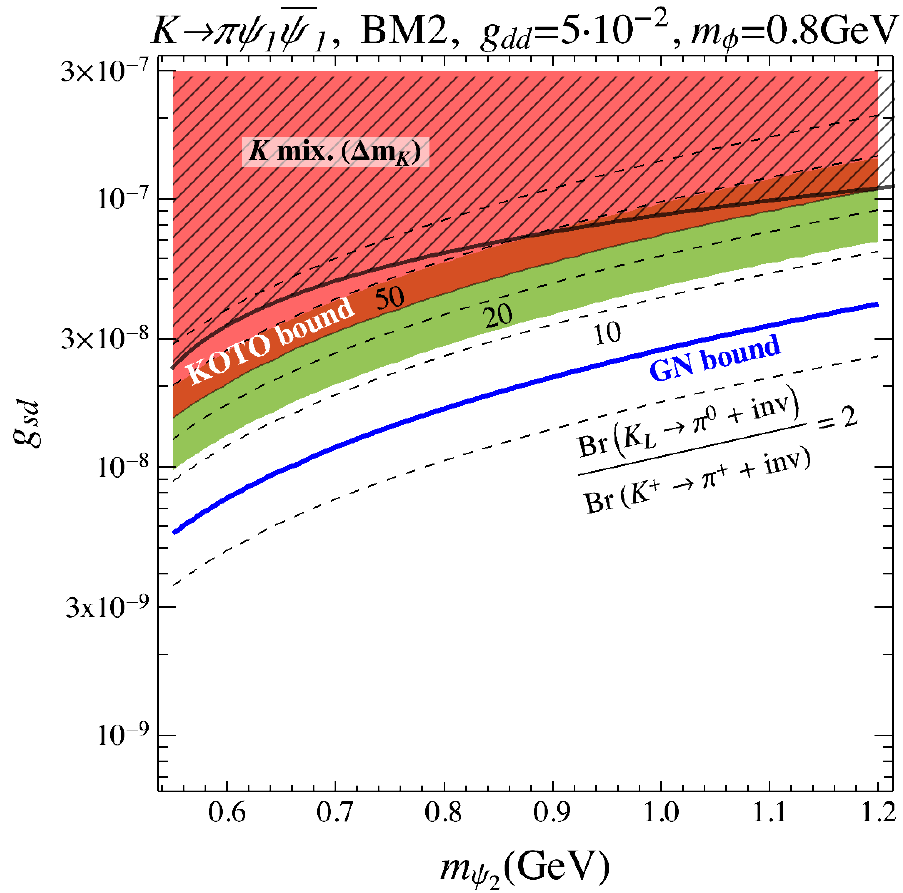}
\end{center}
 \caption{ \label{fig:Model3:BM2} Same as Fig. \ref{fig:Model3:BM1} but for Model 3 BM2.  In the predictions for $\Br(K^+\to \pi^++{\rm inv})$, $\Br(K_L\to \pi^0+{\rm inv})$ in the left plot (black lines) we vary $m_{\phi}\in[0.55,1.2]$ GeV for two values of $g_{sd}=7\cdot 10^{-9}, 7\cdot 10^{-7}$ and fix $m_{\psi_2}=0.8$ GeV. }
 \end{figure} 

BM2, on the other hand, does not lead to tree level contributions to $\epsilon_K$. The constraints from $K^0-\bar K^0$  mixing are therefore relaxed compared to BM1 as they are only due to $\Delta m_K$. As shown in Fig. \ref{fig:Model3:BM2}, it is thus possible to saturate the present KOTO upper bounds over a much larger set of parameter space, with masses of $m_\phi$ and $m_{\psi_2}$ up to $\sim 1$ GeV for $g_{dd}=5\cdot 10^{-2}$.  
Next, we discuss  the constraints on the $\psi_1$-couplings.

\subsection{Constraints on the $\psi_1$-couplings}

The leading diagrams for $\pi^0 \to \psi_1 \bar \psi_1$, relevant to the invisible pion constraint,
are analogous to the Model 2 ones shown in Fig.~\ref{fig:pi0gagaM1} with 
$\phi_2 \to \phi$, $\phi_1 \to \psi_1$ with a $\psi_2$ inserted in between the final state pair in the 
loop diagram.
Assuming $m_{\phi,\psi_2} \gg m_{\pi,\eta}$,  the corresponding matrix element reads 
\begin{equation}
{\cal M}(\pi^0 \to   \psi_1\bar\psi_1) = \big(\bar u P_R v\big) \frac{ B_0 f }{m_\phi^2} \hat{\cal M} 
 +  \big\{y_{ij},\gamma_5  \leftrightarrow y^*_{ji}, - \gamma_5 \big\} + {\cal O}( m_\pi^2/m_{\phi,\psi_2}^2) \;, 
\end{equation}
where $\hat{\cal M}$ is a shorthand for
\begin{equation}
\hat{\cal M} =  \Im g^{(\phi)}_{dd} y_{11} +  \Re g^{(\phi)}_{dd} y_{12}y_{21} \frac{1}{12 \pi^2} \frac{B_0}{ m_{\psi_2}}  \left( 2 \Im g_{dd}^{(\phi)} -  \Im g_{ss}^{(\phi)} \right) H (x) \;,
\end{equation}
where $x = m_\phi^2/m_{\psi_2}^2$, $H(x) =  (1 + x(\ln x -1))/(1-x)^2$ quoting $H(0)=1$ and $H(1) = 0.5$
as representatitve  values. 
The total rate is easily obtained from the matrix element squared given in \eqref{eq:Msq-generic} and
the $1 \to 2$  decay rate \eqref{eq:pi0phi1phi1:dec} (without the symmetry factor $1/2$)
\begin{equation}
\Gamma(\pi^0 \to   \psi_1\bar\psi_1) = 
\frac{( B_0f_\pi)^2}{ m_\phi^4} \left( \beta_{\psi_1}^2 
(\Re \hat{\cal M} )^2  +  (\Im \hat{\cal M} )^2    \right)\, \frac{m_\pi}{16 \pi} \beta_{\psi_1}   \;,
\end{equation}
replacing $f \to f_\pi/\sqrt{2}$ and adapting $\beta_{\psi_1} = ( 1- 4 m_{\psi_1}^2/m_\pi^2)^{1/2}$. Assuming 
 $ \Im g^{(\phi)}_{ss} =0$, 
$m_{\psi_1} = 0$ and  $H(x) \to 1$ one gets values
\begin{align}
\Br  (\pi^0 \to \psi_1 \bar \psi_1)  = 
 \begin{cases}
 2.6 \cdot 10^{-9} \,   \left( \frac{ |y_{11}|  }{5 \cdot 10^{-6}} \right)^2 \left( \frac{\Im g_{dd}^{(2)} }{5 \cdot 10^{-2}} \right)^2 \left( \frac{ \GeV}{m_\phi} \right)^4\,,  &  \hspace{-0.5cm} y_{12} y_{21} =0  \, ,
  \\
3.3 \cdot 10^{-9}   \left( \frac{|y_{12} y_{21}|}{10^{-2}}\right) ^2 \left( \frac{\Re g_{dd}^{(2)} }{5 \cdot 10^{-2}} \frac{\Im g_{dd}^{(2)} }{5 \cdot 10^{-2}} \right)^2
 \left( \frac{ \GeV}{m_\phi} \right)^4 \left( \frac{4  \, \GeV}{m_{\psi_2}} \right)^2 ,
  &  y_{11}  =0   \;,
  \end{cases}
\end{align}
which are close to the upper experimental bound $\Br (\pi^0 \to \phi_1 \phi_1) < 4.4 \times 10^{-9}  $  
\cite{NA62:2020}.
Clearly the tree graph is leading and imposes a
constrain  $|y_{11}| \lesssim {\mathcal O}(5 \cdot 10^{-6})$ on the Yukawa couplings to the light fermion for the two benchmarks in Figs. \ref{fig:Model3:BM1}  and \ref{fig:Model3:BM2}. We observe that, for both BM1 and BM2 the lightest fermion is required to be heavy enough that invisible pion decay is kinematically forbidden, $m_{\psi_1}\gtrsim m_{\pi^0}/2$. Reducing somewhat the value of $|y_{12}y_{21}| \sim {\mathcal O}(10^{-2})$, 
very light $\psi_1$ are possible. Even in this case the KOTO bounds could be saturated (at least for the BM2 flavor structure of the couplings).

\subsection{$\psi_1$ as a dark matter candidate}

In the minimal version of Model 3, presented in this work,  $\psi_1$ and $\psi_2$ are odd under the $Z_2$-parity. The lightest fermion, $\psi_1$ can thus be a DM candidate.  The situation is similar to Model 2. For $\psi_1$ in the mass range $135~{\rm MeV}\lesssim m_{\psi_1}\lesssim 181$ MeV,  $\psi_1 \bar \psi_1\to \pi^0 \pi^0$ is kinematically allowed, and can lead to the correct relic DM abundance. 
For lighter $\psi_1$ only the $\psi_1 \bar \psi_1\to \gamma\gamma$ annihilation  is allowed. However, if $\phi$ were to couple  to electrons or neutrinos, the resulting annihilation cross sections can be large enough such that $\psi_1$ can be the DM. 

For now, let us assume that $\psi_1 \bar \psi_1\to \pi^0 \pi^0$ is kinematically allowed. 
Then at leading order there are two relevant diagrams as shown in Fig.~\ref{fig:annihilation}. 
The corresponding matrix element reads 
\begin{alignat}{2}
\label{eq:MAnn}
& {\cal M}(\psi_1 \bar \psi_1 \to \pi^0\pi^0) &\;=\;&    \left(B_0 f  \Im g_{dd}^{(\phi)} \Delta_{\phi}(m_\pi^2)\right)^2 
\Big[  m_{\psi_2} y_{12} y_{21}\tilde \Delta  \big(\bar v P_R u\big)  +
 |y_{12}|^2 \tilde \Delta^\mu  \big( \bar  v \gamma_\mu P_L u \big)  \Big]
\nonumber \\[0.1cm] &  &\;\phantom{=}\;& + B_0 \Re g_{dd}^{(\phi)}   \Big[ y_{11}  \Delta_{\phi}(s) \bar v P_R u \Big]   + 
\big\{y_{ij},\gamma_5  \leftrightarrow y^*_{ji}, - \gamma_5 \big\} \;,
\end{alignat}
where $\bar v \equiv \bar v( p)$, $u \equiv u (\bar p)$,    $s \equiv q^2 = (p + \bar p)^2 $,
and by crossing symmetry from the right diagram in Fig.~\ref{fig:fermion}: 
$\tilde \Delta^{(\mu)}  = \bar \Delta^{(\mu)}|_{p,\bar p \to - \bar p,- p}$ in \eqref{eq:barDelta}.
The cross section is obtained from the spin-averaged squared matrix element (including a symmetry factor for identical final states)
\begin{align}
\frac{d \sigma}{d \Omega} =   \frac{ \left| {\cal \overline{M}} \right|^2
}{512 \pi^2 s} \frac{\beta_\pi}{\beta_{\psi_1}} \; , 
\end{align}
where $\beta_{\pi,\psi_1} = (1 - 4 m_{\pi,\psi_1}^2/s)^{1/2}$ are the respective velocities in the centre of mass frame.
The thermally  averaged cross section at leading order in the non-relativistic expansion 
is given by 
\begin{alignat}{1}
\langle \sigma v \rangle =  \frac{p_\pi}{32 \pi m_{\psi_1}}  |P|^2 \;,
\end{alignat}
where in this approximation $p_\pi = (m_{\psi_1}^2- m_\pi^2)^{1/2}$, and
\begin{alignat}{1}
 P   = i \{ & B_0 \re g_{dd}^{(\phi)}  \Delta_\phi( 4 m_{\psi_1}^2) \Im y_{11}  \,+ \nonumber  \\[0.1cm]
  & \left(B_0 f_\pi \Im g_{dd} \Delta_\phi(m_\pi^2)  \right)^2 \Delta_{\psi_2}(m_\pi^2 \!-\! m_{\psi_1}^2)  m_{\psi_2}   \Im y_{21} y_{12}    \}  \;,
\end{alignat}
is the pseudoscalar part in \eqref{eq:dec}. It is easily obtained from \eqref{eq:MAnn} taking into account 
that  the role of $u$ and $v$ are interchanged. All other contributions, such as $|S|^2$, 
vanish in the non-relativistic approximation. 
And for values  of input parameters, 
 $y_{11} =0$, $m_{\psi_1} = 160~\MeV$, and $m_\phi = 1~\GeV$  
one obtains a total cross section 
\begin{equation}
\label{eq:crossModel3}
\langle \sigma v \rangle = 3 \times 10^{-26}\frac{{\rm cm}^3}{{\rm s}} \left(\frac{ \Im g^{(\phi)}_{dd} }{7.5 \times 10^{-2}} \right)^4
 \left(\frac{ \Im y_{21} y_{12}}{1}\right)^2 \left( \frac{1 ~\GeV}{m_{\psi_2}}\right)^2   \;,
\end{equation}
which is of the right order of magnitude to produce the  required relic abundance ($1 {\rm \, pb} \approx 3 \cdot 10^{-26} {\rm cm}^{3} s^{-1}$). 
In quoting the dependences in \eqref{eq:crossModel3}, 
we have neglected terms of ${\mathcal O }(m_{\pi,\psi_1}^2/m_{\psi_2}^2)$.

\section{Conclusions}
\label{sec:conclusions}
We have presented three models that can lead to large deviations in $\Br(K_L\to \pi^0+{\rm inv})$, while leaving $\Br(K^+\to \pi^0+{\rm inv})$ virtually unchanged from the SM expectation. The three models are: Model 1 where the invisible decay is the two body transition $K_L\to \pi^0\phi_1$, Model 2 with $K_L\to \pi^0\phi_1\phi_1$ and Model 3 with $K_L\to \pi^0\psi_1 \bar \psi_1$ three body transitions, can be viewed as representatives of a larger class of models. The scalar $\phi_1$ or fermion $\psi_1$ that escape the detector could be replaced by a dark gauge boson, or more complicated dark sector final states, without affecting our main conclusions. 

Common to all these possibilities is that in addition to the invisible final state particles (in our case $\phi_1$ and $\psi_1$), there has to be at least one additional light mediator with a ${\mathcal O}(1~{\rm GeV})$-mass  in order to have large violations of the Grossman-Nir bound. In Models 1 and 2 the mediator is another scalar, $\phi_2$, while in Model 3 there are two mediators, the fermion $\psi_2$ and scalar $\phi$. The scalar mediators mix with $K_L$ and $\pi^0$, which then leads  to  enhanced $\Br(K_L\to \pi^0+{\rm inv})$ rates. The required mixings are small, and thus for large parts of parameter space the most stringent constraints are due to the present KOTO upper bound on $\Br(K_L\to \pi^0+{\rm inv})$. If the anomalous events seen by KOTO turn out to be a  true signal of new physics, then these models are natural candidates for their explanation. 

In Models 2 and 3 the lightest states, $\phi_1$ and $\psi_1$, can be dark matter candidates.
For the restricted mass range $m_\pi \leq m_{\phi_1, \psi_1}\leq (m_K-m_\pi)/2$ and suitable parameter ranges these particles can be the thermal relic.
For lighter $\phi_1$ or $\psi_1$ new annihilation channels are required. 
For example the mediators could couple to either electrons or neutrinos, in addition to the couplings to quarks. 
 
In the numerics we followed the principle of minimality and switched on the minimal set of couplings required for large violations of the GN bound. We took great care to ensure that the radiative corrections do not modify the assumed flavor structure and potentially invalidate our conclusions. In the future, it would be interesting to  revisit our simplified models in more complete flavor models which fix all the couplings to quarks.  An even more ambitious possible research direction could be to explore whether the light mediators could be tied to the SM flavor puzzle itself, e.g. along the lines of Ref. \cite{Smolkovic:2019jow}. We leave this and related open questions for future investigations. \\[0mm]


{\bf Acknowledgements.}
We gratefully acknowledge the paradise conference in Tenerife, the 2nd Workshop on Hadronic Contributions to New Physics searches (HC2NP 2019), where this collaboration started.
We are grateful to Alex Kagan, Antonin Portelli, and Diego Redigolo for useful discussions. 
JZ acknowledges support in part by the DOE grant de-sc0011784. JZ thanks the Higgs Centre for Theoretical Physics at The University of Edinburgh for the hospitality during the collaboration on this project. 
RZw is supported by an STFC Consolidated Grant, ST/P0000630/1. The work of RZi  is supported by project C3b of the DFG-funded Collaborative Research Center TRR 257, ``Particle Physics
Phenomenology after the Higgs Discovery".

\appendix

\section{The $K \to \pi \psi_1 \bar \psi_1$ decay rate}
\label{app:diff-rate}

For completeness we give here the explicit expression for the $K \to \pi \psi_1 \bar \psi_1$ differential rate in Model 3. The expressions become rather involved 
because of the presence of the fermions in the final states. 
For instance, the analytic expression for the rate can only be given as a double differential rate since both variables enter the loop 
diagram, cf. Fig.~\ref{fig:fermion}, in a non-trivial way. 
At the end of the appendix we also comment on how this decay defies the helicity formalism used for semileptonic 
and flavor changing neutral currents.

The generic double differential rate in terms of Dalitz plot variables is given by \cite{PDG}
\begin{equation}
\frac{d^2  }{dq^2 d Q^2} \Gamma(K \to \pi \psi_1 \bar \psi_1) = 
\frac{1}{( 2\pi)^3} \frac{1}{ 32 m_K^3} |{\cal M}(K \to \pi \psi_1 \bar \psi_1)|^2 \;,
\end{equation}
where $q^2 \equiv (p + \bar p)^2$, $Q^2 \equiv ( p + p_\pi)^2$  
are the kinematic variables 
with ranges  $4 m_{\psi_1}^2 < q^2 < (m_K - m_\pi)^2$ 
and $ Q^2_- <  Q^2 < Q^2_+$, with
\begin{equation}
Q^2_{\mp} = (E_{\psi_1}+ E_\pi)^2 - \left(  p_{\psi_1} \pm 
p_{\pi}  \right)^2 \;,
\end{equation}
 $ E_{\psi_1} = q/2$,
 $E_\pi = (m_K^2- q^2 - m_\pi^2)/2q$ and $p_i  =  (E_i^2 - m_i^2)^{1/2}$.
 
 Decomposing the matrix element ${\cal M}$ in terms of fermion bilinears 
 \begin{equation}
 \label{eq:dec}
{\cal M} = S \bar u v + P \bar u \gamma_5 v + V^\mu \bar u \gamma_\mu v + A^\mu \bar u \gamma_\mu \gamma_5  v +  T^{\mu\nu} \bar u \sigma_{\mu\nu}  v  \;,
\end{equation}
 the generic matrix element squared, summing over fermion polarizations, reads 
\begin{alignat}{1}
\label{eq:Msq-generic}
 |{\cal M}|^2   &=    2 q^2  \big\{  |P|^2 +  \beta_{\psi_1}^2 |S|^2 + \frac{2 m_{\psi_1}}{q^2}
  ( (p + \bar p) \! \cdot \! A P^* + 
(p - \bar p) \! \cdot \! V S^ * + \textrm{h.c.} 
)   \\ 
& +A^\mu (m_{\mu\mu'} - \beta_{\psi_1}^2 g_{\mu\mu'}   ) A^{\mu'*} 
+ V^\mu (m_{\mu\mu'} - g_{\mu\mu'}   ) V^{\mu'*}   \nonumber \\ 
& + 2 T^{\mu\nu} g_{\mu\mu'}(g_{\nu\nu'} - 2 m_{\nu\nu'} ) T^{\mu'\nu'*}  +  \frac{4  }{q^2}  \left(i T^{\mu\nu} ( 
m_{\psi_1}( \bar p - p)_{\mu} A^*_{\nu} +  \bar p_\mu  p_\nu P^*) + \textrm{h.c.} \right)    \big\} \;, \nonumber
\end{alignat}
where $\sigma_{\mu\nu} = i/2 [ \gamma_\mu,\gamma_\nu]$, 
$\beta_{\psi_1} = ( 1- 4 m_{\psi_1}^2/q^2)^{1/2}$ and $m_{\mu\nu} \equiv  2 (\bar p_\mu  p_\nu + 
p_\mu  \bar  p_\nu)/q^2$. For completeness we have included the tensor current in \eqref{eq:dec} even though 
it does not appear in our models. 

The conversion from a form ${\cal M} = L \bar u P_L v + R \bar u P_R v + 
L^\mu \bar u \gamma_\mu P_L v +R^\mu \bar u \gamma_\mu P_R v $, used in  \eqref{eq:K_Lpi0:model2}, 
to the form in \eqref{eq:dec} proceeds via: $S[P] = 1/2(R\pm L)$ and $V[A]^\mu = 1/2(R \pm L)^\mu$. 
In particular for  $K_L \to \pi^0 \psi_1 \bar \psi_1$
\begin{align}
S[P]_L &\;=\;  -i  B_0 \left(  B_0 m_{\psi_2} \Re[i \Im] (y_{12}y_{ 21}) X_L + 
\Re[i \Im] y_{11} \Im \bar g^{(\phi)}_{sd} \Delta_{\phi}(q^2) \right)  \;, 
\\
 V[A]_L^\mu  &\;=\; -i B_0^2/2 \left( |y_{21}|^2 \pm |y_{12}|^2 \right) X_L^\mu \;,
\end{align}
with 
$X_L \equiv   \Im \hat{g}_{sd} \bar \Delta f_K f_\pi - \frac{\Im \bar{g}_{sd}}{16 \pi^2} \FL^{(\phi)} ( \bar{I}_4) $
and $X_L^\mu = X_L|_{\bar{I}_4 , \bar{\Delta} \to \bar{I}_4^\mu , \bar{\Delta}^\mu}$, whereas 
for  $K^+ \to \pi^+ \psi_1 \bar \psi_1$ \eqref{eq:K+ampl:model3} the decomposition reads 
\begin{align}
S[P]_+ &\;=\;   B_0 \left(  B_0 m_{\psi_2} \Re[i \Im] (y_{12}y_{ 21}) X_+ + 
\Re[i \Im] y_{11}  \bar g^{(\phi)}_{sd} \Delta_{\phi}(q^2) \right)  \;, \\
 V[A]_+^\mu  &\;=\;  B_0^2/2 \left( |y_{21}|^2 \pm |y_{12}|^2 \right) X_+^\mu \;,
\end{align}
with 
$X_+ \equiv   - \frac{ \bar{g}_{sd}}{16 \pi^2} \Fp^{(\phi)} ( \bar{I}_4) $
and $X_+^\mu = X_+|_{\bar{I}_4   \to \bar{I}_4^\mu }$.

It seems worthwhile to point out that this decay cannot be cast in the Jacob-Wick helicity formalism, 
since it does not corresponds to a chain of $1 \to 2$ decays. This also applies, e.g., to the 
generalisation of the formalism  to effective theories used for $B \to K \ell^+\ell^-$~\cite{Gratrex:2015hna}. The issue is that in Model 3 the $\phi$ particle breaks factorization of the fermion part 
and the rest in the same way as  the photon does between the lepton and the quarks, cf. Section 5.3. in Ref.~\cite{Gratrex:2015hna}. 
On a technical level, this can easily be seen from the decomposition of the
vector matrix element 
\begin{equation}
V^\mu = V^{(p)} p^\mu  + V^{(\bar p)} \bar p^\mu + V^{(p_\pi)} p_\pi^\mu \;,
\end{equation}
 which necessitates all independent momenta of the decay.
In the $B \to K \ell^+ \ell^-$ case, induced by the standard dimension six effective Hamiltonian and no QED or electroweak corrections,  the amplitude only depends on $p_\pi$,  $p+ \bar p$ but  not the difference 
$p - \bar p$. 
However, using such a decomposition in the expressions given above does allow in practice  for a fast numerical evaluation of the differential rate thereby retaining one of the main advantages of the helicity amplitude formalism.

\section{Integral conventions}
\label{app:PV}

For convenience and clarity we collect here the conventions of the Passarino-Veltman functions 
\cite{Passarino:1978jh,Hahn:1998yk,Patel:2015tea,Hahn:1998yk} used in this work. The conventions are   
 equivalent to those of LoopTools \cite{Hahn:1998yk} and  FeynCalc \cite{FeynCalc1,FeynCalc2}.
The loop function used are the triangle and box integrals defined by  
 \beq
C_0(p_1^2,p_2^2,(p_1 \!+ \!p_2)^2, m_1^2,m_2^2,m_3^2)  \equiv
 \int_k  \frac{1}{ (k^2 - m_1^2)((k\!+\!p_1)^2 -m_2^2),((k\!+\!p_1\!+\!p_2)^2 -m_2^2)} \;, 
 \eeq
and
\beq
 \begin{split}
&  D_0(m_\pi^2,(p_\pi \!+\!P)^2, (p_K \!-\! p_\pi \!-\! P)^2, (p_K\!-\! p_\pi )^2,P^2, m_K^2,
m_\phi^2 , m_M^2, m_{\psi_2}^2, m_\phi^2)  \equiv  \\  
&\qquad\qquad    \int_k      \frac{1 } {((k \!+\! p_K)^2\!-\! m_{\phi}^2)( (k\!+\!p_\pi)^2\!-\!m_{\phi}^2)((k\!+\! p_\pi\!+\!P)^2\!-\!m_{\psi_2}^2) (k^2 \!-\!m_{M}^2) }\;,
\end{split}
\eeq
respectively, 
with  $i0$-prescription  suppressed and   $\int_k \equiv (2 \pi \mu)^{4 \!-\!  d}/( i \pi^2)  \int d^dk $.
The arguments of the $D_0$ function are those appearing in Model 3 in section \ref{sec:Model3}.
 It seems worthwhile to mention that the two-point Passarino-Veltman function does not appear in this paper
 and the symbol $B_0$ is used for a quantity related to quark condensate as stated at the beginning 
 of Section \ref{sec:ChPT}.

\bibliographystyle{apsrev4-1.bst}

\bibliography{zzzbiblio}

\end{document}